\documentclass{aa}
\usepackage[varg]{txfonts}
\usepackage{graphicx}

\usepackage{color}
\begin{document}

   \title{The third realization of the International Celestial Reference Frame by very long baseline interferometry}

   \author{P. Charlot
          \inst{1}
          \and
          C. S. Jacobs
          \inst{2}
          \and
          D. Gordon
          \inst{3}
          \and
          S. Lambert
          \inst{4}
          \and
          A. de Witt
          \inst{5}
          \and
          J. B\"ohm
          \inst{6}
          \and
          A. L. Fey
          \inst{7}
          \and
          R. Heinkelmann
          \inst{8}
          \and
          E.~Skurikhina
          \inst{9}
          \and
          O. Titov
          \inst{10}
          \and
          E. F. Arias
          \inst{4}
          \and
          S. Bolotin
          \inst{3}
          \and
          G. Bourda
          \inst{1}
          \and
          C. Ma
          \inst{11}\thanks{Retired}
          \and
          Z. Malkin
          \inst{12,13}
          \and
          A. Nothnagel
          \inst{14}\thanks{Now at Technische Universit\"at Wien, Vienna, Austria}
          \and
          D.~Mayer
          \inst{6}\thanks{Now at Federal Office of Metrology and Surveying, Vienna, Austria}
          \and
          D. S. MacMillan
          \inst{3}
          \and
          T. Nilsson
          \inst{8}\thanks{Now at Lantm\"ateriet -- The Swedish mapping, cadastral and land registration authority, Geodetic Infrastructure, G\"avle, Sweden}
          \and
          R.~Gaume
          \inst{15}
          }

   \institute{Laboratoire d'astrophysique de Bordeaux, Univ. Bordeaux, CNRS,
              B18N, All\'ee Geoffroy Saint-Hilaire, 33615 Pessac, France \\
              \email{patrick.charlot@u-bordeaux.fr}
         \and
             Jet Propulsion Laboratory, California Institute of Technology,
             4800 Oak Grove Drive, Pasadena, CA 91109-8099, USA
         \and
             NVI Inc. at NASA Goddard Space Flight Center,
             Code 61A.1, Greenbelt, MD 20771, USA
         \and
             SYRTE, Observatoire de Paris, Université PSL, CNRS, Sorbonne Université, LNE,
             61 Av. de l'Observatoire, 75014 Paris, France
         \and
             Hartebeesthoek Radio Astronomy Observatory,
             PO Box 443, Krugersdorp 1740, South Africa
         \and
             Department of Geodesy and Geoinformation, Technische Universit\"at Wien,
             Wiedner Hauptstra\ss e 8-10, 1040 Vienna, Austria
         \and
             U.S. Naval Observatory,
             3450 Massachusetts Avenue NW, Washington, DC 20392-5420, USA
         \and
             Helmholtz Centre Potsdam, German Research Centre for Geosciences, Telegrafenberg, A17, D-14473 Potsdam, Germany
         \and
             Institute of Applied Astronomy, Russian Academy of Sciences,
             Nab. Kutuzova 10, St. Petersburg 191187, Russia
         \and
             Geoscience Australia,
             P.O. Box 378, Canberra, ACT 2601, Australia
         \and
             NASA Goddard Space Flight Center,
             Code 61A.1, Greenbelt, MD 20771, USA
         \and
             Pulkovo Observatory,
             St. Petersburg 196140, Russia
         \and
             Kazan Federal University,
             Kazan 420000, Russia
         \and
             Institut für Geod\"asie und Geoinformation, Univers\"at Bonn, Nu\ss allee 17, D-53115 Bonn, Germany
         \and
             National Science Foundation,
             2415 Eisenhower Avenue, Alexandria, Virginia 22314, USA
             }

   \date{Received 7 May 2020 / Accepted 31 July 2020}


  \abstract
   {A new realization of the International Celestial Reference Frame (ICRF) is presented based on the work achieved by a working group of the International Astronomical Union (IAU) mandated for this purpose. This new realization follows the initial realization of the ICRF completed in 1997 and its successor, ICRF2, adopted as a replacement in 2009. The new frame, referred to as ICRF3, is based on nearly 40~years of data acquired by very long baseline interferometry at the standard geodetic and astrometric radio frequencies (8.4~and 2.3~GHz), supplemented with data collected at higher radio frequencies (24~GHz and dual-frequency 32 and 8.4~GHz) over the past 15~years. State-of-the-art astronomical and geophysical modeling has been used to analyze these data and derive source positions. The modeling integrates, for the first time, the effect of the galactocentric acceleration of the solar system (directly estimated from the data) which, if not considered, induces significant deformation of the frame due to the data span. The new frame includes positions at 8.4~GHz for 4536~extragalactic sources. Of these, 303~sources, uniformly distributed on the sky, are identified as ``defining sources'' and as such serve to define the axes of the frame. Positions at 8.4~GHz are supplemented with positions at 24~GHz for 824~sources and at 32~GHz for 678~sources. In all, ICRF3 comprises 4588~sources, with three-frequency positions available for 600~of these. Source positions have been determined independently at each of the frequencies in order to preserve the underlying astrophysical content behind such positions. They are reported for epoch~2015.0 and must be propagated for observations at other epochs for the most accurate needs, accounting for the acceleration toward the Galactic center, which results in a dipolar proper motion field of amplitude 0.0058~milliarcsecond/yr (mas/yr). The frame is aligned onto the International Celestial Reference System to within the accuracy of ICRF2 and shows a median positional uncertainty of about 0.1~mas in right ascension and 0.2~mas in declination, with a noise floor of 0.03~mas in the individual source coordinates. A subset of 500 sources is found to have extremely accurate positions, in the range of 0.03 to 0.06~mas, at the traditional 8.4~GHz frequency. Comparing ICRF3 with the recently released Gaia Celestial Reference Frame 2 in the optical domain, there is no evidence for deformations larger than 0.03 mas between the two frames, in agreement with the ICRF3 noise level. Significant positional offsets between the three ICRF3 frequencies are detected for about 5\% of the sources. Moreover, a notable fraction (22\%) of the sources shows optical and radio positions that are significantly offset. There are indications that these positional offsets may be the manifestation of extended source structures. This third realization of the ICRF was adopted by the IAU at its 30th~General Assembly in August 2018 and replaced the previous realization, ICRF2, on January~1,~2019.}


   \titlerunning{The third realization of the International Celestial Reference Frame}
   \maketitle
%

\section{Introduction}
\label{Sec:intro}
   The International Celestial Reference Frame (ICRF) and its successor, ICRF2, have been the basis for high-accuracy astrometry for more than two decades. Both frames have drawn on simultaneous 8.4~GHz and 2.3~GHz observations of compact extragalactic radio sources acquired by very long baseline interferometry (VLBI), starting from the end of the 1970s. The primary frequency in this observing scheme is 8.4~GHz (X~band), while 2.3~GHz (S~band) is only used for ionosphere calibration. Hereafter, such dual-frequency VLBI observations are referred to as~S/X~band, following the standard designation.

   The ICRF \citep{Ma1998} was the first all-sky realization of an extragalactic frame with milliarcsecond (mas) position accuracy and the first realization of the International Celestial Reference System (ICRS) \citep{Arias1995}. It was adopted by the International Astronomical Union (IAU) at its 23rd~General Assembly in 1997, replacing the Fifth Fundamental Catalog of stars (FK5) \citep{Fricke1988} as the fundamental celestial reference frame as of January 1, 1998. Unlike previously (e.g., for the FK5 or other stellar frames), the definition of the frame axes was no longer related to the equinox and equator but relied on coordinates of so-called defining sources. The ICRF included 212~such defining sources out of a total of 608~objects for which positions were reported. The orientation of the ICRF axes was deemed to be accurate to 20~microarcseconds (${\rm\mu}$as) while source coordinates had a noise floor of~250~${\rm\mu}$as. As a consequence of the~ICRS definition, all source positions in the ICRF were independent of epoch. The release of ICRF was a major step forward and as such it became the required passage to link other celestial reference frames to the ICRS, among which were the dynamical frame \citep{Standish1998} and the Hipparcos stellar frame \citep{Kovalevsky1997}. The ICRF has also allowed for many advances in other fields, such as geodesy and Earth orientation studies, or related to practical applications like deep space navigation.

   As time passed, VLBI observing networks and technology improved and further sources were observed. On the geodesy side, observations became organized in a formal way under the umbrella of the International VLBI Service for Geodesy and Astrometry (IVS), established in 1999 \citep{Schluter2007}, permitting more resources to be pooled together. Two ICRF extensions, ICRF-Ext.1 and ICRF-Ext.2, were constructed in 2000 and 2002, adding another 109 sources to the frame \citep{Fey2004}. At the same time, systematic surveys of the VLBI sky were initiated using the Very Long Baseline Array (VLBA)\footnote{The VLBA is a facility of the National Science Foundation operated under cooperative agreement by Associated Universities, Inc.}, enlarging the pool of VLBI sources with milliarcsecond-accurate positions to more than a thousand \citep{Beasley2002}. By the mid 2000s, it was realized that the amount of data, their accuracy, and the denser sky coverage, along with modeling improvements since ICRF was delivered, would justify the building of a new reference frame to get the full potential of the available data sets. This prompted the construction of the second realization of the ICRF, named ICRF2, which was completed in 2009 and adopted by the IAU at its 27th General Assembly in the same year \citep{Fey2015}. As a result, ICRF2 replaced ICRF on January 1, 2010. The new realization comprises 3414~sources, of which 295 are defining sources. The orientation of its axes is known to 10~${\rm\mu}$as, while source coordinates have a noise floor of 40~${\rm\mu}$as. The large increase in the number of sources from ICRF to ICRF2 (a factor of six) was largely due to the inclusion of observations from a series of VLBA Calibrator Survey (VCS) astrometric campaigns carried out between 1994 and 2007 \citep{Beasley2002,Fomalont2003,Petrov2005,Petrov2006,Kovalev2007,Petrov2008}. The goal of these was to expand the pool of calibrators available for VLBI observations in phase-referencing mode \citep{Beasley1995}. Such campaigns added nearly 2200~sources exclusively observed by the VLBA to the catalog of sources derived from IVS sessions. The VCS sources, still, had position uncertainties typically five times larger than the other sources due to being observed in survey mode and generally only in a single session. For this reason, they were categorized separately from the IVS sources in ICRF2.

   Since the release of ICRF2, the VLBI database has continued to expand thanks to ongoing observing programs run by the IVS but also through specific projects carried out independently, notably by using the VLBA. The latter includes a complete re-observation of all VCS sources in 2014--2015, which has brought an overall factor of five improvement in coordinate uncertainties, hence bringing position uncertainties for the VCS sources closer to those for the rest of the ICRF2 sources \citep{Gordon2016}. A specific effort has also been made to strengthen observations of optically bright ICRF2 sources within IVS programs to facilitate the alignment of the optical reference frame which is being built by the Gaia space mission \citep{LeBail2016}. At the same time, VLBI observations at higher radio frequencies began to develop, namely at 24~GHz (K~band) and 32~GHz (Ka~band), the latter with simultaneous X~band observations for ionosphere calibration (hence the usual X/Ka band designation for this dual-frequency observing scheme). An initial catalog of 268~sources, together with VLBI images of the sources, was produced using the VLBA at K~band \citep{Lanyi2010,Charlot2010}, while positions for 482~sources were reported at X/Ka band from observations with the Deep Space Network (DSN) \citep{Jacobs2012}. Despite the limited data sets, both catalogs showed an overall agreement with ICRF2 at the 300~${\rm\mu}$as level when comparing individual source coordinates, hence revealing the value of such high-frequency observations. By 2012, the wealth of the additional VLBI data already acquired, or foreseen, together with the need to have a state-of-the-art VLBI frame to align as well as possible the future Gaia optical frame onto the ICRS, created the necessary motivation for generating a new realization of the ICRF. Shortly after the 28th General Assembly of the IAU in Beijing, a working group under IAU Division A was assembled to this end. The mandate of the working group was to generate the third realization of the ICRF by 2018, for adoption at the 30th IAU General Assembly to take place that year.

   The work accomplished toward the generation of the third realization of the ICRF, hereafter referred to as ICRF3, was organized along several lines. One such line was aimed at acquiring new appropriate data to correct deficiencies of ICRF2. In this respect, the focus was placed not so much on trying to increase the number of sources but rather on improving uniformity and internal consistency. The campaign to re-observe all VCS sources \citep{Gordon2016} falls into this line. As noted above, this campaign vastly improved coordinate uncertainties for the relevant 2200 such sources, resulting in an overall distribution of source position uncertainties that is now much more uniform. Specific efforts were also targeted to strengthen observations in the far south (i.e., for declinations below $-45\degr$). However, the limited number of VLBI telescopes in the Southern Hemisphere remains as a major bottleneck to reach the same quality in that area of the sky (in terms of source density and position accuracy) as that available further north. The full data set used to generate ICRF3 is described in Sect.~\ref{Sec:data}. Compared to ICRF and ICRF2, a new feature is the inclusion of data at K~band and X/Ka band in addition to those at the standard S/X~band geodetic frequencies. The resulting three-frequency positions (at X, K, and Ka~band) are herewith reported as part of ICRF3 without being combined in order to preserve the underlying astrophysical information.

   Another major activity of the working group consisted in generating ICRF3 prototype realizations at several stages of the work. These allowed the group to study the impact of data sets, astronomical and geophysical modeling, analysis configuration, software packages, and individuals that analyze the data on the resulting frame. In practice, such prototype realizations were produced at eight different institutions\footnote{Geoscience Australia (Australia), Technische Universit\"at Wien (Austria), Observatoire de Paris (France), Helmholtz Centre Potsdam (Germany), Institute of Applied Astronomy St. Petersburg (Russia), NASA Goddard Space Flight Center (USA), Jet Propulsion Laboratory (USA), U.S. Naval Observatory (USA).} in Australia, Austria, France, Germany, Russia, and USA using five different software packages. Of interest is that one such ICRF3 prototype realization was delivered to the Gaia science team in July~2017 and used in the process to generate the catalog for the Gaia Data Release~2 (Gaia DR2) in order to align the Gaia frame with the ICRS \citep{Lindegren2018}. The final solution for ICRF3 incorporates data up to spring~2018 and was produced in July~2018. Section~\ref{Sec:analysis} below reviews the adopted modeling and analysis configuration, while Sect.~\ref{Sec:errors} describes the alternate analyses that have been conducted to assess errors in ICRF3. A newly added feature in the modeling is the galactocentric acceleration of the solar system, long sought and first detected by \citet{Titov2011}. With a magnitude of about 5~${\rm\mu}$as/yr, this effect produces detectable apparent drifts in the source positions, especially over the almost 40-year span now reached by the S/X band data. Unlike previously (i.e., for ICRF and ICRF2), the source coordinates in ICRF3 are referred to a specific epoch and hence should be properly propagated for epochs away from that reference epoch, accounting for Galactic acceleration, for the most demanding needs.

   Details of ICRF3 are given in Sect.~\ref{Sec:ICRF3}, including source categorization and tables of positions reported separately at X~band, K~band, and Ka~band. The frame has a new set of defining sources selected in a way to be uniformly distributed on the sky. Recommendations to future users on how to use the cataloged ICRF3 positions, depending on their needs, are also provided. Section~\ref{Sec:discussion} makes an assessment of the alignment of ICRF3 onto ICRS, emphasizes the improvement and benefits of the new realization over ICRF2, and compares ICRF3 with the Gaia DR2 celestial reference frame (Gaia-CRF2), which is the first extragalactic frame ever built in the optical domain~\citep{Mignard2018}. Consistency between multi-frequency radio positions and between radio and optical positions is also addressed as part of that section. The final sections outline the adoption process by the IAU and the future evolution of the ICRF.

\section{Observations}
\label{Sec:data}

\begin{figure*}
   \centering
   \includegraphics[trim=0 90 0 88, clip, width=1.0\textwidth]{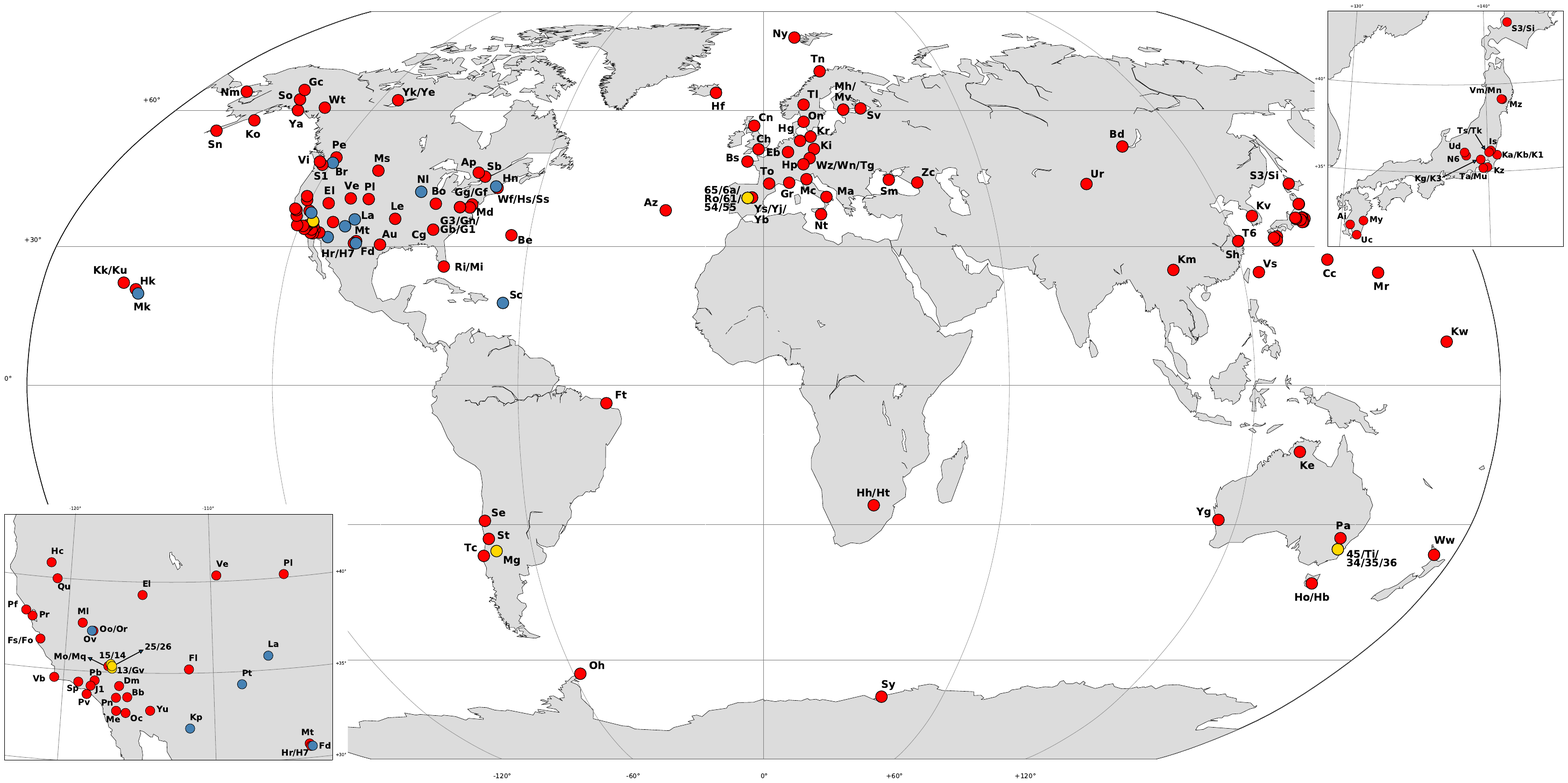}
   \caption{World map showing the geographical location of the 167 antennas (situated on 126 different sites) that participated in the observations used for ICRF3. The red dots show the antennas from the IVS network (and pre-existing adhoc VLBI arrays that observed at S/X band), the blue ones those from the VLBA, and the yellow ones those from the DSN and ESA. The two-character codes printed near each dot correspond to the short names of the antennas, as defined in the IVS nomenclature. The two insets show enlargements of western US and Japan where a large number of antennas (including mobile VLBI stations) have been used to collect geodetic VLBI data over the years due to the seismic nature of these regions.}
   \label{Fig:ICRF3_network_map}
   \end{figure*}

\begin{figure}
   \includegraphics[width=1.0\columnwidth, trim=15 10 37 33, clip]{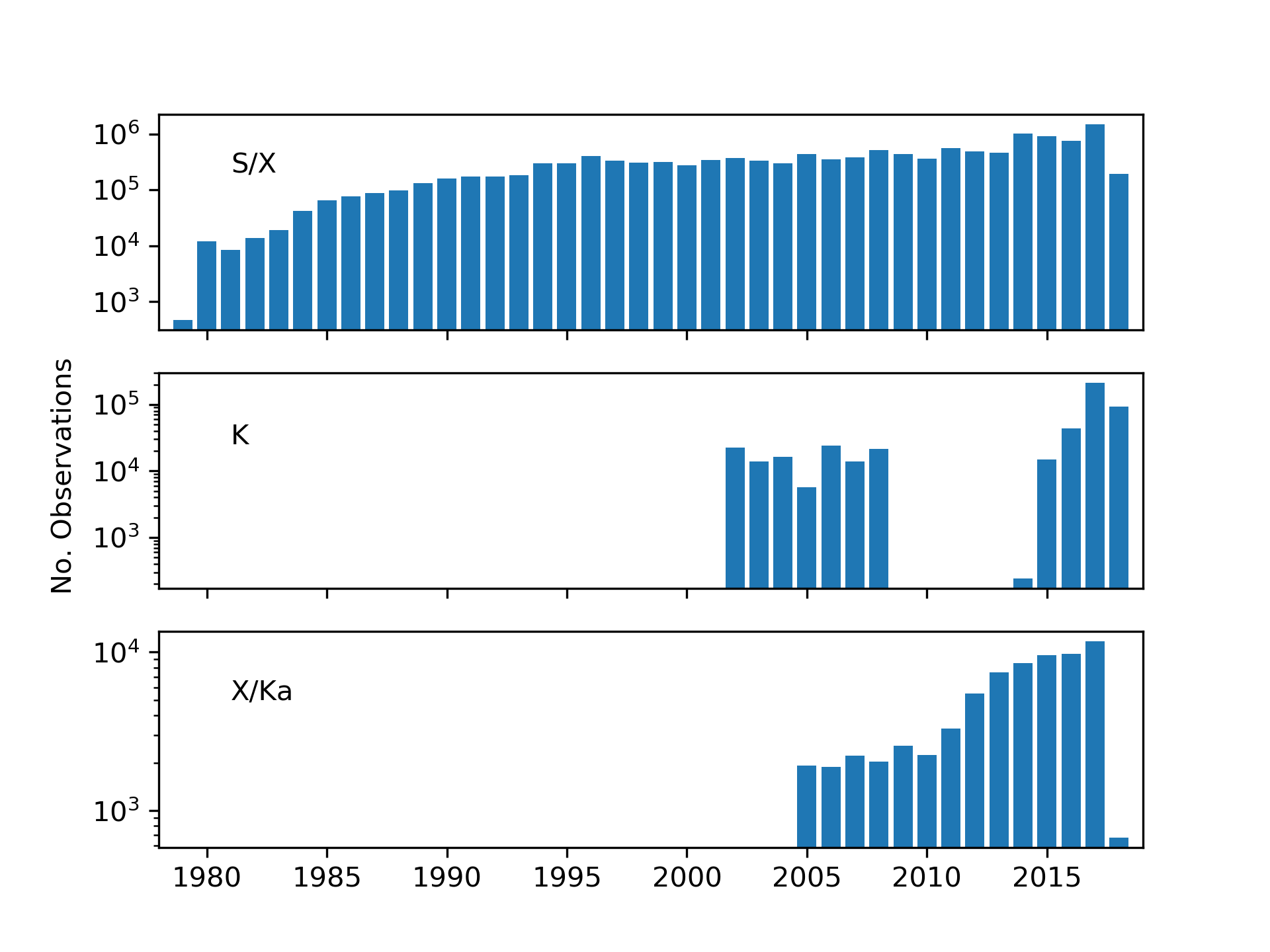}
   \centering
   \caption{Distribution of the observations used for ICRF3. The three histograms show the number of VLBI delays per year at S/X~band (upper panel), K~band (middle panel), and X/Ka~band (lower panel). For ease of reading, the number of observations is plotted with a logarithmic scale.}
   \label{Fig:Data_distribution}
   \end{figure}

   The VLBI data used to build ICRF3 were acquired by arrays of 2~to 20 radiotelescopes organized in their vast majority under the umbrella of the~IVS, the VLBA, and the DSN. Over the years, a total of 167 telescopes, located on 126 different sites, participated in such VLBI sessions (Fig.~\ref{Fig:ICRF3_network_map}). Observations were carried out using the so-called bandwidth synthesis mode, which permits the determination of precise group delay quantities by observing multiple channels spread out across a bandwidth of several hundred MHz, as originally devised by \citet{Rogers1970}. Following acquisition, the data were processed at one of the IVS correlators (in Bonn, Haystack, Kashima, Shanghai, Vienna, or Washington), the VLBA correlator in Soccoro, the Australia Telescope National Facility correlator in Perth, or the DSN processor in Pasadena. Post-processing was accomplished by calibrating the raw phases to make them consistent in all channels and by fringe-fitting these to obtain the group delay quantities which are further fed into geodetic and astrometric software packages for the estimation of source positions (see Sect.~\ref{Sec:analysis} below). Such post-processing was conducted either at the correlators or at the institutions coordinating the relevant VLBI sessions or sets of sessions. Dual-frequency observing at S/X~band has been standard since the early days of geodetic and astrometric VLBI as it permits the calibration of the dispersive delay caused by the ionosphere using a combination of the measurements at the two frequencies. As noted above, a similar scheme was implemented for observing at Ka~band (with simultaneous measurements at X~band), while at K~band observing has remained single-frequency, hence requiring proper modeling of the ionospheric delays for the latter. In general, VLBI sessions are 24-hour long in order to separate parameters for polar motion and nutation and to average out unmodeled geophysical effects which vary on a diurnal basis. Each session generally observes a few tens to a few hundreds of sources depending on the size of the network, the slewing speed of the antennas, the data recording rate, and the objective of the session (i.e., whether it is a survey program). The number of observations collected during a session varies, depending in the first place on the size of the network. Over the years, the amount of data acquired has increased, due to larger networks being used, culminating in 2017 with more than one million observations collected at S/X~band, 0.1~million collected at K~band, and 0.01~million collected at X/Ka~band (see the distribution of observations per year in Fig.~\ref{Fig:Data_distribution}). Characteristics of the data sets at each of the three frequency bands are given in the subsections below.

   \subsection{S/X band (2.3/8.4 GHz)}
\label{Sec:data_SX}
   The data used for ICRF3 in this frequency band include the entire pool of geodetic and astrometric VLBI sessions acquired and made available by the IVS with rare exceptions. The sessions range from August 3, 1979 to March 27, 2018 and come from a variety of programs dedicated to monitor the Earth orientation, establish the terrestrial and celestial reference frames, and maintain and expand these, according to the objectives of the IVS \citep{Schuh2012}. Details on the current observing programs are given in \citet{Nothnagel2017}. Data acquired prior to the establishment of the IVS in 1999 had similar goals and were obtained through ad~hoc arrays organized by cooperations between individual observatories or national agencies and are described in \citet{Ma1998}. Of particular interest for the celestial frame are the Research and Development VLBA~(RDV) sessions conducted jointly with the VLBA six times a year (since~1997) which assemble a network of 15~to 20 stations, allowing for observation of 80--100 sources each time. Such sessions are also essential to image the sources and assess their suitability as fiducial points to materialize the celestial frame (see Sect.~\ref{Sec:ICRF3-defining}). Southern-Hemisphere IVS~sessions are equally important, even if conducted with arrays of only a few stations, as they allow for observations in the far south which otherwise would be impossible with the northern arrays. Unfortunately, despite ongoing efforts \citep{Plank2017}, the paucity of observations in this area of the sky compared to the north remains a deficiency in the data set. Occasional astrometric observations conducted by other VLBI~arrays such as the European VLBI Network and the Australian Long Baseline Array have also been incorporated.

   The generation of ICRF3, furthermore, took advantage of a number of dedicated astrometric sessions conducted with the VLBA at S/X band since the mid 1990s. These include the series of VCS campaigns that took place between 1994 and 2007, already incorporated in ICRF2 \citep[see][]{Fey2015}, along with the campaign that re-observed all VCS sources in 2014--2015, which was initiated specifically for the purpose of ICRF3, as reported in \citet{Gordon2016}. More recently, another 24~such VLBA sessions have been run under the US Naval Observatory share of the VLBA observing time. Those sessions targeted all sources in the VLBI pool meeting one of the following criteria: (i) had not been observed since 2009 (i.e., when ICRF2 was delivered), (ii)~had less than 50 observations, (iii) were observed in three or fewer sessions, or (iv) were among the weakest known optically bright sources listed in \citet{LeBail2016}. The goal here was to further enhance uniformity of the data sets for ICRF3. In all, the VLBA sessions constitute only a small portion of the entire set of S/X~band sessions (200 sessions out of a total of 6206 sessions) but account for 26\% of the data, while more than two-thirds (68\%) of the sources have observations coming exclusively from the VLBA. See \citet{Gordon2017} for further details on the impact of the VLBA observing on the celestial frame.

   \subsection{K band (24 GHz)}
   The data sets in this band are made of 40 VLBA sessions that observed the northern sky (down to mid-southern declinations), supplemented with 16 single-baseline sessions between telescopes in Hartebeesthoek (South Africa) and Hobart (Australia) that observed sources below $-15\degr$ declination (down to the far south). The first VLBA session was conducted on May~15, 2002 and was part of a set of ten such sessions that ultimately led to the first realization of a celestial frame in this frequency band, although not covering the entire sky \citep{Lanyi2010}. Apart from two similar follow-up sessions in 2008, VLBA observing was then interrupted until 2015, after which it was started again (see Fig.~\ref{Fig:Data_distribution}). Since then, another 25~VLBA sessions have been carried out, the bulk of which were run in 2017--2018 under the US Naval Observatory share of the VLBA observing time. The latest session incorporated in this work was conducted on May~5, 2018. Additionally, three archived VLBA sessions dedicated to observing sources in the Galactic plane \citep{Petrov2011} were also included. Observations between Hartebeesthoek and Hobart were initiated at about the same time as the VLBA sessions restarted (first session run on May~4, 2014) to complete the sky coverage in the far south. One southern session also included the Tianma 65 m telescope near Shanghai (China) while another one included the Tidbinbilla 70~m telescope near Canberra (Australia). In all, the VLBA sessions account for 99\% of the data while the Southern-Hemisphere sessions account for~1\% of it, which leads to a majority of sources (66\%) having only VLBA observations, a similar division as that at S/X band.

   \subsection{X/Ka band (8.4/32 GHz)}
   Observations at X/Ka band were initiated in 2005 with the primary goal of building a reference frame for spacecraft navigation, now conducted on the DSN using the Ka~frequency band. The data set includes a total of 168 single-baseline sessions that involved seven telescopes at the three DSN sites in Goldstone (California), Robledo (Spain), and Tidbinbilla (Australia). The first of these sessions took place on July~9, 2005 while the most recent one incorporated was run on January~28, 2018. Occasionally (in about~10\% of the sessions), the European Space Agency (ESA) telescope in Malarg\"ue (Argentina) joined the observations, which was essential to improve the north-south geometry of the network and reduce systematics in the reference frame.

\section{Analysis}
\label{Sec:analysis}

   The principle behind VLBI data analysis is to compare measured quantities with a priori theoretical modeling of the same quantities and to refine underlying models by estimating model parameter corrections that best fit the data. Depending on the objective pursued, such parameters may pertain to the entire data set (e.g., station positions and velocities, source positions,...) or only to individual sessions (e.g., the Earth orientation parameters,...) or even to a small portion of a session (e.g., clock and troposphere parameters which vary on the order of hours). Least-squares methods are generally employed for this estimation. The VLBI modeling, analysis configuration, and software packages used to produce ICRF3 are outlined in the following sections.

   \subsection{Astronomical, geophysical, and instrumental modeling}
   \label{Sec:modeling}
   The measured VLBI quantities (group delay and delay rate) used for ICRF3 were analyzed employing state-of-the-art astronomical and geophysical modeling, generally following the prescriptions of the International Earth Rotation and Reference Systems Service (IERS) \citep{IERS2010}. An extensive review of all the effects to be incorporated in the VLBI model in order to reach the highest accuracy is given in \citet{Sovers1998}. Apart from the effect induced by the galactocentric acceleration of the solar system, only brief information, primarily to identify the models selected, is thus provided here. The interested reader is referred to the \citet{Sovers1998} review for details on the underlying physics. Galactocentric acceleration is addressed specifically in Sect.~\ref{Sec:aberration} as it is the first time this effect is introduced in the modeling used to generate a VLBI celestial reference frame.

   The geometric portion of the VLBI delay, including relativistic effects, was derived consistently with the so-called consensus model \citep{Eubanks1991} which provides 1~ps accuracy. Formulation of the geometric VLBI delay necessitates describing completely the evolution of the dynamic Earth over the period of the observations. This requires specifying the orientation of the Earth's spin axis in inertial space (i.e., precession and nutation), and relative to the Earth's crust (polar motion), and to characterize the daily rotation of the Earth around that axis (UT1). Also to be considered are the various deformations that affect the Earth's crust on which the radiotelescopes are attached. These comprise tectonic plate motions, tidal deformations, and atmospheric pressure loading effects. Modeling of the Earth's spin axis was achieved using the MHB~nutation \citep{Mathews2002} and P03~precession \citep{Capitaine2003,Hilton2006}, further designated as IAU 2000A nutation and IAU 2006 precession after adoption of these models by the IAU. A priori polar motion and UT1 were retrieved from the IERS Rapid Service/Prediction Centre \citep[solution labeled ``finals.data'', see][Sect.~3.5.2]{IERS2018}, to which were added short-period tidal variations, as prescribed by the IERS \citep[see][Chap.~8]{IERS2010}. Initial station (radiotelescope) positions and velocities were taken from the ITRF2014 terrestrial frame \citep{Altamimi2016}, incorporating post-seismic deformation models for sites that were subject to major earthquakes, and further adding deformations due to solid Earth tides, ocean loading, and atmospheric pressure loading. Displacements caused by solid Earth tides were derived following the IERS prescriptions \citep[see][Sect.~7.1.1]{IERS2010}. Ocean loading displacements were obtained from the TPXO.7.2 model \citep{Egbert2002}, supplemented with the FES99 model \citep{Lefevre2002} for the long-period Ssa~tide, while those due to atmospheric pressure loading (both tidal and non-tidal) come from the APLO~model \citep{Petrov2004}. Further displacements caused by the centrifugal effect of polar motion on the solid Earth \citep[see][Sect.~7.1.4]{IERS2010} and the oceans \citep{Desai2002} were also incorporated in the modeling. Calculation of the geometric VLBI delay finally considered a component resulting from the thermal expansion of the antennas which are subject to structural deformations when temperature varies, hence causing a displacement of the position of the reference point of the instruments \citep{Nothnagel2009}.

   Added to the geometric delay were corrections for atmospheric propagation, including the contribution due to high-altitude charged particles (ionosphere) and that due to the neutral component (troposphere). Ionospheric delays are proportional to the inverse of the frequency-squared and were calibrated using the dual-frequency data collected at S/X band and X/Ka band, whereas they were modeled using total electron content maps for K band. Such ionospheric maps are produced daily from global navigation satellite systems and were retrieved from NASA's Crustal Dynamics Data Information System \citep{Noll2010} for the purpose of the K band analysis. Tropospheric delays were derived using the VMF1 mapping function \citep{Boehm2006} to map zenith delay contributions to relevant elevations. It must be noted that observations below $5^{\circ}$ elevation were discarded due to inadequacies in modeling the troposphere at such low elevations. The zenith delays themselves were estimated every 30~minutes at each site using a continuous piecewise linear function during the least-squares parameter adjustment. Additional east-west and north-south tropospheric gradients were also estimated. The interval between such estimates was six hours for the S/X~band and K~band analyses, whereas they were estimated only once for the entire data set at X/Ka~band due to the more limited sky coverage above each site for the latter. Corrections for instrumental delays were further added to the above geometric and atmospheric VLBI delay components. These come from propagation delays in the cables (which have a different length at each telescope), lack of synchronization of the clocks between stations, and clock instabilities. In practice, since it is not possible to calibrate those instrumental delays precisely, they were treated altogether, assuming an overall clock-like behavior at each telescope. A 60-minute continuous piecewise linear function, with quadratic terms when needed, was used to model them accordingly, the parameters of which were estimated during the least-squares parameter adjustment, as in the case of troposphere.

   \subsection{Galactocentric acceleration}
   \label{Sec:aberration}
   The relative motion of the solar system barycenter with respect to the extragalactic reference frame may cause apparent changes in the radio source positions due to aberrational effects. In this motion, only the non-linear part (i.e., the acceleration) is to be considered since the constant part is absorbed into the reported source positions by convention. As noted in \citet{Sovers1998}, the said motion is conveniently decomposed into three components: (i)~the motion of the solar system barycenter with respect to the Galactic center, (ii)~the motion of the Galaxy relative to the Local Group, and (iii)~the motion of the Local Group relative to the extragalactic frame (which may be assumed at rest relative to the cosmic microwave background since the observed radio sources are located at cosmological distances). Of these components, only the first is expected to induce significant acceleration on a timescale of a few decades such as that of our VLBI data set. It is thus solely considered in the rest of this section.

   Based on the rotational property of the Galaxy, the solar system barycenter is known to move around the Galactic center with an orbital period of about 200 million years. Assuming a purely circular rotation, the resulting acceleration vector, which reflects the curvature of the orbit, is directed toward the Galactic center. Classically, this acceleration translates into an aberration effect which takes the form of an apparent overall proper motion for the distant extragalactic radio sources. The amplitude $A_G$ of this proper motion \citep[see, e.g.,][]{Kovalevsky2003} is given by
   \begin{equation}
        A_G={V_0^2\over R_0c},
        \label{Eq:GA}
   \end{equation}
   where $R_0$ is the distance of the barycenter of the solar system to the Galactic center, $V_0$ is the linear speed along its orbit, and $c$~is the speed of light.
   As also pointed out by~\citet{Kovalevsky2003}, the resulting effect varies according to the region of the sky since it depends on the projection of the acceleration vector (which points from the solar system barycenter to the Galactic center) onto the plane of the sky in the source direction. In practice, it is conveniently mapped using Galactic coordinates. For a source at Galactic longitude $l$ and Galactic latitude $b$, the components of the corresponding  proper motion, $\mu_l\cos b$ and $\mu_b$, are then
   \begin{align}
   \mu_l\cos b &= -A_G\sin l, \\
   \mu_b&= -A_G\sin b \cos l.
   \end{align}
   Going further and adopting the IAU-recommended values for the Galactic constants\footnote{Recent estimates of $R_0$ and $V_0$ deviate somewhat from these IAU-recommended values which date back to 1985. See~\citet{Vallee2017}.}, $R_0=8.5$~kpc and $V_0=220$~km/s, a proper motion amplitude $A_G$ of 4~${\mu}$as/yr is obtained from~Eq.~\eqref{Eq:GA}.
   With four decades of accumulated VLBI data, apparent source displacements of up to 150~$\mu$as are thus expected, which is quite significant considering current source position accuracies of a few tens of microarcseconds (see below). Galactic acceleration therefore clearly needs to be considered in the modeling.

   The galactocentric acceleration of the solar system was predicted to induce detectable proper motions for the extragalactic sources soon after the inception of the VLBI astrometric technique \citep{Fanselow1983}. Attempts to detect such proper motions from the VLBI data date back to the 1990s, taking advantage of data sets that were already more than a decade long \citep{Sovers1996,Gwinn1997}. At the same time, plans started to develop to measure those proper motions from future optical space astrometry \citep{Bastian1995,Mignard2002,Kovalevsky2003,Kopeikin2006}. The actual detection of the effect was made by~\citet{Titov2011} who estimated an amplitude of $6.4\pm 1.5$~$\mu$as/yr, in reasonable agreement with the above prediction. This result was derived through a vector spherical harmonics analysis of time series of VLBI source coordinates covering two decades. Several other determinations followed, based on increasing VLBI data span and/or different analysis schemes, including the estimation of the Galactic acceleration amplitude as part of a global VLBI solution~\citep{Xu2012,Titov2013,Titov2018}. In all, the values derived range from 5.2~$\mu$as/yr to 6.4~$\mu$as/yr with an uncertainty of 0.3~$\mu$as/yr for the most recent determinations. Additional estimates can be inferred from measurements of parallax and proper motions of Galactic masers using VLBI phase-referenced techniques~\citep{Reid2009,Brunthaler2011,Honma2012,Reid2014}. These point to somewhat lower values, from 4.8 to 5.4~$\mu$as/yr, with similar uncertainties as the geodetic VLBI determinations. See \citet{MacMillan2019} for an overview of all such determinations either from geodetic VLBI or from maser proper motions. An open question is whether the acceleration vector of the solar system barycenter is offset from the Galactic center, which would happen if the Sun were subject to a specific peculiar motion apart from its circular rotation around the Galactic center. When estimated from geodetic VLBI data, values of this offset range from non-significant (i.e., near $0^{\circ}$) to about $20^{\circ}$, with uncertainties of 5--10$^{\circ}$ \citep{Titov2011,Xu2012,Titov2013,Titov2018}. In all, one cannot be sure whether there is an actual offset or whether some of the above estimates just reflect systematic errors in the data.

   For the present work, we decided to re-determine the amplitude of the Galactic acceleration because we saw no compelling evidence to adopt one or the other of the published values (nor any of the unpublished values known to us) at the time. Additionally, the data sets on which those previous determinations are based are only up to 2016 whereas the data sets for ICRF3 extend to 2018. Because of the high-quality data acquired in the period, we thought these additional two years could make a difference. We also thought that it was most appropriate to use a value determined from a data set that covers the same time span as that for~ICRF3. A dedicated analysis estimating both the amplitude and direction of the acceleration vector of the solar system barycenter was thus performed, prior to constructing ICRF3, based on the almost 40-year long S/X~data set in our hand. While this estimation, ideally, could have been accomplished simultaneously with the determination of the S/X~frame, we decided not to do so because the analysis configuration adopted for ICRF3 treats all sources as global parameters (see Sect.~\ref{Sec:configuration} below) and this scheme is not entirely appropriate for estimating the solar system barycenter acceleration vector. As remarked by \citet{Titov2011}, some sources, mostly observed in the early VLBI sessions, are subject to notable instabilities (due to their having extended and variable structures) which affect significantly the estimation of the acceleration parameters if not filtered out in the analysis process. In the realization of ICRF2, those sources were treated as arc parameters (i.e., a new position was estimated for each session) and denoted as special handling sources \citep{Fey2015}. For our determination, we~adopted a similar approach, meaning that the positions for the 39~sources identified as such in ICRF2 were estimated separately for each session in which they were observed in order to limit the impact of their positional variability on the derived acceleration parameters.

\begin{figure}
   \includegraphics[width=0.87\columnwidth]{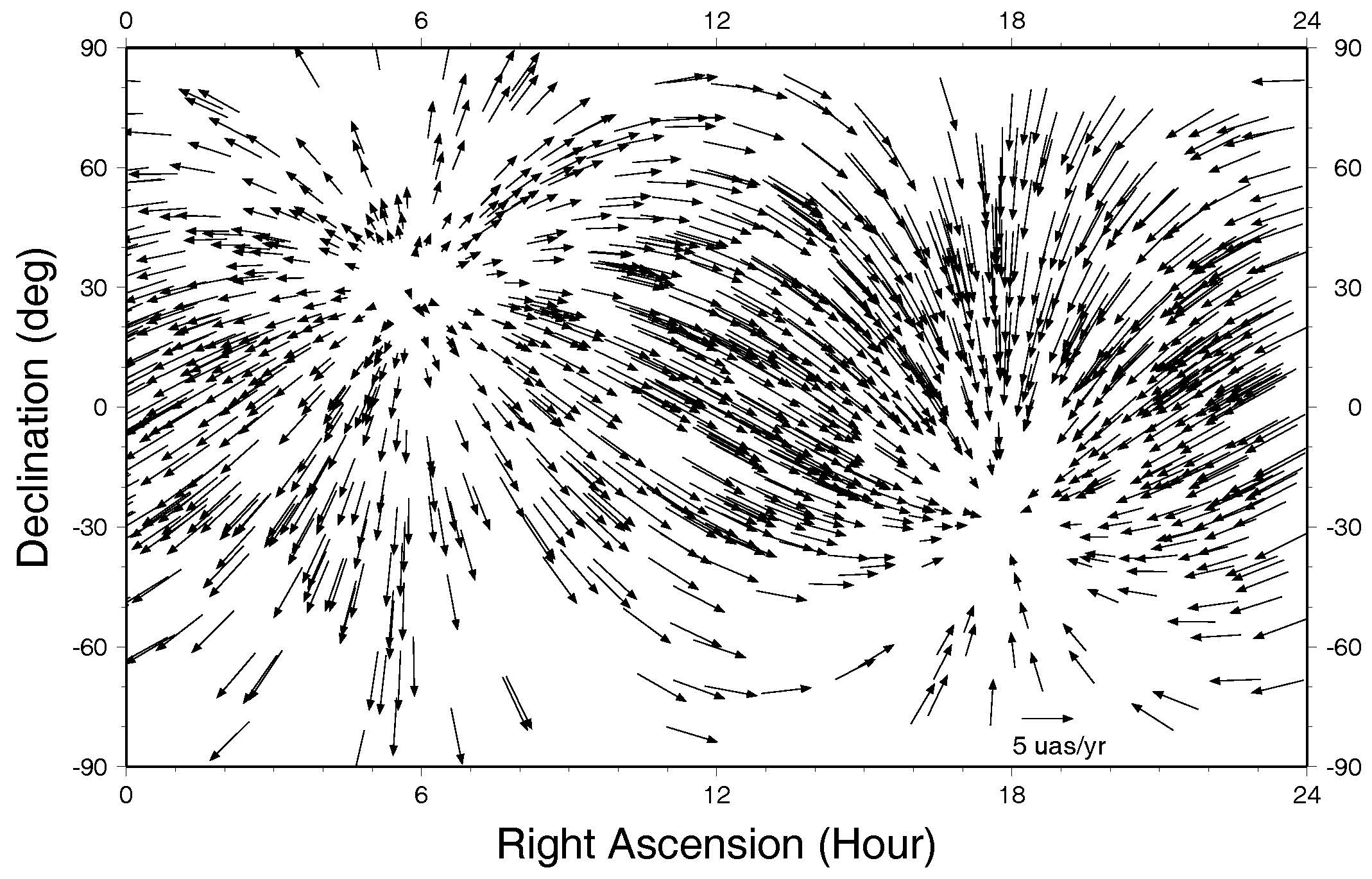}
   \centering
   \caption{Aberration proper motion field (in equatorial coordinates) resulting from a solar system barycenter acceleration vector with amplitude 5.8~$\mu$as/yr, pointing to the Galactic center. The scale is given by the length of the vector bar plotted in the lower right-hand of the figure.}
   \label{Fig:aberration_field}
   \end{figure}

   The dedicated analysis described in the previous paragraph led to a value of $5.83\pm 0.23$~$\mu$as/yr for the amplitude of the solar system barycenter acceleration vector, while the estimated vector was found to point in the direction $\alpha_\mathrm{SSA}=270.2^{\circ}\pm 2.3^{\circ}$, $\delta_\mathrm{SSA}=-20.2^{\circ}\pm 3.6^{\circ}$. This direction is within $10^{\circ}$ of the Galactic center (located at $\alpha_G=266.4^{\circ}$, $\delta_G=-29.0^{\circ}$) and the significance of the offset is less than $2.5\sigma$. Conversely, the amplitude of the acceleration vector is detected at the $25\sigma$ level. Adopting a conservative approach, we decided not to consider any such offset in the construction of ICRF3, in view of its marginal significance, and therefore assumed that the solar system barycenter acceleration vector points to the Galactic center. Concerning the amplitude of the acceleration vector, we adopted the value inferred from our analysis, namely 5.8~$\mu$as/yr. This value was then applied in the modeling for the analysis of all observations considered for ICRF3. The apparent proper motion field induced over the celestial sphere by this correction is depicted in Fig.~\ref{Fig:aberration_field}. Another decision was concerned with the reference epoch of the source coordinates to report in ICRF3 since these are no longer invariant with time due to the application of the acceleration correction. While a perhaps natural choice would have been~2000.0, this date is now more than 20 years back and would have made the acceleration corrections quasi mandatory for use of the frame in present times. We thus decided to leave out that option and adopted instead~2015.0 as reference epoch. Apart from being closer to the present day, this date is also close to the mean epoch of the observations for ICRF3 (2014.5 at S/X~band, 2016.7 at K~band, and 2015.3 at X/Ka~band). Unlike epoch 2000.0, such a choice facilitates the use of the frame in present times where corrections for Galactic acceleration will remain small and may not be necessary unless the highest position accuracy is required. Additionally, it is also very close to the reference epoch of Gaia~DR2, which is J2015.5 \citep{Lindegren2018,Mignard2018}, hence making comparisons between the two frames more straightforward. Finally, it is to be underlined that all our conclusions have been fed into the IVS~working group on Galactic aberration which reported its work in \citet{MacMillan2019}.

   \subsection{Configuration of analysis}
   \label{Sec:configuration}
    Aside from the astronomical, geophysical, and instrumental modeling described above, the remaining elements of the data analysis to configure were concerned with the selection of parameters to estimate, the mechanisms to define the orientation of the terrestrial and celestial frames, and the data weighting scheme. Each of these aspects is discussed in turn below.

    Besides clocks and tropospheric parameters, which must be solved for since they are a source of nuisance and cannot be modeled precisely as discussed in Sect.~\ref{Sec:modeling}, the other parameters that were estimated in the analysis include station positions and velocities, the Earth Orientation Parameters (EOP), and radio source coordinates. Station positions and velocities were estimated globally from the entire data set at each frequency band, resulting in a single position and velocity estimate for each antenna (except for known discontinuities, e.g., due to earthquakes or mechanical movement of the antenna). Radio source positions were treated using a similar scheme, with a single position estimated for each source at each frequency band. No sources were made special cases, unlike in the~ICRF2 analysis where 39~sources with significant position instabilities were estimated session-wise \citep{Fey2015}. We did not repeat this approach because there was no indication in the present case, based on the tests we carried out (see Sect.~\ref{Sec:variations} below), that such sources would degrade the frame if solved globally (which was a concern for~ICRF2). We thus saw no reason to treat them differently than the other sources. Adopting the same approach for all sources also guarantees consistency in the resulting source position uncertainties. Unlike antenna and source parameters, the~EOP were estimated session-wise, with the exception of single-baseline sessions where they were held fixed to their a priori values. Those parameters include offsets and rates for UT1 and the two components of polar motion, and two nutation offsets, all of which were estimated at the midpoint of each session. The EOP define the rotation between the terrestrial and celestial frames and allow for the connection the two frames at any epoch.

   The data analysis was configured in such a way that the terrestrial frame derived from estimating station positions and velocities is consistent with the most recent realization of the International Terrestrial Reference Frame (ITRF), namely ITRF2014 \citep{Altamimi2016}. To this end, loose no-net-translation and no-net-rotation constraints were applied to the positions and velocities of a set of core antennas that do not show any position discontinuities. This scheme ensures that the resulting frame is aligned onto the a priori frame (i.e., ITRF2014) and that their origins coincide. This set of core antennas was comprised of the 38 stations in Table~\ref{Tab:S/X_TRF_constraints} for the S/X band data set, while at K~band it was comprised of all VLBA stations except the MK-VLBA antenna. That station had to be excluded from the constraint because it suffered an earthquake on June~15, 2006 and shows a position discontinuity at that epoch. The X/Ka~band data set was treated differently because it consists mostly of single-baseline sessions where the EOP are kept fixed. In that case, the derived frame is implicitly consistent with the a priori polar motion and UT1 series (which are themselves consistent with ITRF2014).

   \begin{table}
   \caption{Name, two-character code (as in Fig.~\ref{Fig:ICRF3_network_map}), and geographical location of the 38 core antennas used to align the terrestrial reference frame onto ITRF2014 in the S/X band analysis. The nine antennas with the symbol \dag~in superscript were also used for that alignment at K~band.}
   \begin{tabular}{llrrl}
   \hline
   \hline
   \noalign{\smallskip}
\small Code& \small Station name & \small Longitude & \small Latitude & \ \ \ \small Location\\
   \hline
   \noalign{\smallskip}
\ \  \small Kk & \small KOKEE     & \small $-159.67$\ \ \  & \small 22.13\ \     & \ \ \ \small USA\\
\ \  \small Ku & \small KAUAI     & \small $-159.67$\ \ \  & \small 22.13\ \     & \ \ \ \small USA\\
\ \  \small Hc & \small HATCREEK  & \small $-121.47$\ \ \  & \small 40.82\ \     & \ \ \ \small USA\\
\ \  \small Vb & \small VNDNBERG  & \small $-120.62$\ \ \  & \small 34.56\ \     & \ \ \ \small USA\\
\ \  \small Br$^{\dag}$ & \small BR-VLBA   & \small $-119.68$\ \ \  & \small 48.13\ \     & \ \ \ \small USA\\
\ \  \small Oo & \small OVRO\_130 & \small $-118.28$\ \ \  & \small 37.23\ \     & \ \ \ \small USA\\
\ \  \small Ov$^{\dag}$ & \small OV-VLBA   & \small $-118.28$\ \ \  & \small 37.23\ \     & \ \ \ \small USA\\
\ \  \small Kp$^{\dag}$ & \small KP-VLBA   & \small $-111.61$\ \ \  & \small 31.96\ \     & \ \ \ \small USA\\
\ \  \small Pt$^{\dag}$ & \small PIETOWN   & \small $-108.12$\ \ \  & \small 34.30\ \     & \ \ \ \small USA\\
\ \  \small La$^{\dag}$ & \small LA-VLBA   & \small $-106.25$\ \ \  & \small 35.78\ \     & \ \ \ \small USA\\
\ \  \small Fd$^{\dag}$ & \small FD-VLBA   & \small $-103.94$\ \ \  & \small 30.64\ \     & \ \ \ \small USA\\
\ \  \small Nl$^{\dag}$ & \small NL-VLBA   & \small $-91.57$\ \ \   & \small 41.77\ \     & \ \ \ \small USA\\
\ \  \small Ri & \small RICHMOND  & \small $-80.38$\ \ \   & \small 25.61\ \     & \ \ \ \small USA\\
\ \  \small G3 & \small NRAO85\_3 & \small $-79.84$\ \ \   & \small 38.43\ \     & \ \ \ \small USA\\
\ \  \small Gn & \small NRAO20    & \small $-79.83$\ \ \   & \small 38.44\ \     & \ \ \ \small USA\\
\ \  \small Ap & \small ALGOPARK  & \small $-78.07$\ \ \   & \small 45.96\ \     & \ \ \ \small Canada\\
\ \  \small Hn$^{\dag}$ & \small HN-VLBA   & \small $-71.99$\ \ \   & \small 42.93\ \     & \ \ \ \small USA\\
\ \  \small Hs & \small HAYSTACK  & \small $-71.49$\ \ \   & \small 42.62\ \     & \ \ \ \small USA\\
\ \  \small Wf & \small WESTFORD  & \small $-71.49$\ \ \   & \small 42.61\ \     & \ \ \ \small USA\\
\ \  \small St & \small SANTIA12  & \small $-70.67$\ \ \   & \small $-33.15$\ \  & \ \ \ \small Chile\\
\ \  \small Sc$^{\dag}$ & \small SC-VLBA   & \small $-64.58$\ \ \   & \small 17.76\ \     & \ \ \ \small USA\\
\ \  \small Ft & \small FORTLEZA  & \small $-38.43$\ \ \   & \small $-3.88$\ \   & \ \ \ \small Brazil\\
\ \  \small Ys & \small YEBES40M  & \small $-3.09$\ \ \    & \small 40.52\ \     & \ \ \ \small Spain\\
\ \  \small Ny & \small NYALES20  & \small  11.87\ \ \  & \small 78.93\ \     & \ \ \ \small Norway\\
\ \  \small On & \small ONSALA60  & \small  11.93\ \ \  & \small 57.40\ \     & \ \ \ \small Sweden\\
\ \  \small Wz & \small WETTZELL  & \small  12.88\ \ \  & \small 49.15\ \     & \ \ \ \small Germany\\
\ \  \small Nt & \small NOTO      & \small  14.99\ \ \  & \small 36.88\ \     & \ \ \ \small Italy\\
\ \  \small Ma & \small MATERA    & \small  16.70\ \ \  & \small 40.65\ \     & \ \ \ \small Italy\\
\ \  \small Hh & \small HARTRAO   & \small  27.69\ \ \  & \small $-25.89$\ \  & \ \ \ \small South Africa\\
\ \  \small Sv & \small SVETLOE   & \small  29.78\ \ \  & \small 60.53\ \     & \ \ \ \small Russia\\
\ \  \small Yg & \small YARRA12M  & \small 115.35\ \ \  & \small $-29.05$\ \  & \ \ \ \small Australia\\
\ \  \small Sh & \small SESHAN25  & \small 121.20\ \ \  & \small 31.10\ \     & \ \ \ \small China\\
\ \  \small Ke & \small KATH12M   & \small 132.15\ \ \  & \small $-14.38$\ \  & \ \ \ \small Australia\\
\ \  \small Ka & \small KASHIMA   & \small 140.66\ \ \  & \small 35.95\ \     & \ \ \ \small Japan\\
\ \  \small Hb & \small HOBART12  & \small 147.44\ \ \  & \small $-42.80$\ \  & \ \ \ \small Australia\\
\ \  \small Ho & \small HOBART26  & \small 147.44\ \ \  & \small $-42.80$\ \  & \ \ \ \small Australia\\
\ \  \small 45 & \small DSS45     & \small 148.98\ \ \  & \small $-35.40$\ \  & \ \ \ \small Australia\\
\ \  \small Ww & \small WARK12M   & \small 174.83\ \ \  & \small $-36.60$\ \  & \ \ \ \small New Zealand\\
   \hline
   \end{tabular}
   \label{Tab:S/X_TRF_constraints}
   \end{table}

   A similar approach was used to ensure that the celestial reference frames produced at S/X~band, K~band, and X/Ka~band are all aligned and consistent with the previous realization of the ICRF, namely ICRF2 \citep{Fey2015}. This overall alignment was achieved in the first stage by aligning the S/X~band frame onto ICRF2 and in the second stage by aligning the K~band and X/Ka~band frames onto the S/X~band frame. This two-stage approach allowed us to take advantage of the improved S/X~band frame (compared to ICRF2) for the alignment of the K~band and X/Ka~band frames. In practice, the alignment of the S/X~band frame was accomplished by applying a tight (10~$\mu$as/yr) no-net-rotation constraint to the positions of the 295~ICRF2 defining sources. As discussed in Sect.~\ref{Sec:variations} below, alternate analysis solutions, where a fraction of the ICRF2 defining sources (those that show extended structures) were left out from the constraint, were tried, but the impact on the definition of axes was found to be minimal. Therefore, we stuck to the original 295~ICRF2 defining sources for fixing the orientation of the S/X band frame through that constraint. As noted above, the S/X~band frame, once realized, then served as a reference on which to align the K~band and X/Ka~band frames. To this end, a no-net-rotation constraint was applied to the subset of ICRF3 defining sources included in the K~band frame and the same was accomplished for the X/Ka~band frame (see Sect.~\ref{Sec:ICRF3-defining} below for details on the selection of the ICRF3 defining sources). In all, there were 187~usable ICRF3 defining sources at K~band and 174~such sources at X/Ka~band. At K~band, six ICRF3 defining sources included in the frame
   were deemed to be unsuitable for use in the rotation constraints in this band because of too few (< 10) observations. In the case of X/Ka~band, two ICRF3 defining sources included in the frame (0346$+$800 and 0743$-$006) were also left out from the rotation constraints as possible outliers. The sources used to orient the K~band and X/Ka~frames are those labeled as defining sources in Tables~\ref{Tab:ICRF3_K_coordinates} and~\ref{Tab:ICRF3_XKa_coordinates} below with the above restrictions.

   A final aspect of the analysis configuration is the weighting of the individual measurements. Following the usual VLBI practice, the weighting factor $w_i$ assigned to a given observation $i$ was determined as a function of the unit weight $\sigma_0$ as
   \begin{equation}
    w_i={\sigma_0^2\over \sigma_i^2 + \sigma_s^2},
   \end{equation}
   where $\sigma_i$ is the formal uncertainty of that observation, derived on the basis of the signal-to-noise ratio achieved from fringe-fitting and ionosphere calibration process (if applicable), and $\sigma_s$ is a baseline-dependent additive noise calculated for each session. This additive noise was determined through an iterative procedure such that the reduced chi-squared\footnote{The reduced chi-squared $\chi^2_\nu$ is defined as $\chi^2/\nu$ where $\chi^2$ is the sum of the squares of the weighted differences between the observed and calculated quantities and $\nu$ is the degree of freedom which equals the number of observations minus the number of parameters.} of the post-fit residuals on each baseline for each session is about unity. This scheme ensures that the reduced chi-squared is close to unity over each session and furthermore over the entire data set as well.

   \subsection{Analysis software}
   \label{Sec:software}
   An important element of the preparatory work for ICRF3 was that the data analysis was accomplished with several~independent VLBI software packages running concurrently, one of which was also run separately at three institutions. This allowed the working group to check the results derived from these software packages against each other, which was essential to expose any issues, and to gain confidence in the overall data analysis scheme while the work progressed, in particular through the generation of several successive ICRF3 prototypes. Those software packages are the following: CALC-SOLVE \citep[][see appendix]{Caprette1990}, MODEST \citep{Sovers1996}, OCCAM \citep{Titov2004}, QUASAR \citep{Kurdubov2007}, VieVS~\citep{Boehm2018}, and VieVS@GFZ \citep{Nilsson2015}. A detailed description of these software packages is beyond the scope of this paper. In all, they use very similar modeling (as described above). However, they are based on different estimation methods. CALC-SOLVE and VieVS use classical least-squares, while the other software packages employ specialized least-squares or filtering methods -- MODEST is based on the square-root-information filter, OCCAM and QUASAR on the least squares collocation technique, and VieVS@GFZ on Kalman filtering.

   While having such different software packages in hand was of importance for the ICRF3 preparatory work, including numerous tests, the final ICRF3 product was derived from data processing with a unique software package at each frequency band. The S/X~band and K~band data sets were analyzed with CALC-SOLVE while the X/Ka~band data set was analyzed with MODEST. Combination of results from different software packages, for example by combining the individual normal equations, was investigated as an alternate option. However, it requires fine tuning of the input normal equations so that the estimated parameters are defined in a truly identical way in each software package and have consistent a priori settings, which was not possible to achieve within the available time to produce~ICRF3 due to software limitations and other issues. We thus decided to not follow that option and instead to prefer individual determinations.

   The reason why the X/Ka band data set was processed with a different software package than the S/X band and K band data sets was mainly a matter of convenience due to the specificity of the X/Ka band observations which are acquired, correlated, and post-processed through an entirely separate path. The two software packages used in the final data analysis for ICRF3, CALC-SOLVE and MODEST, have been inter-compared, although not recently, and were found to agree at the 1~ps level, as reported in \citet{Ma1998}. Therefore, we are confident that proceeding this way does not limit the consistency of the ICRF3 results.

   \section{Assessment of errors}
   \label{Sec:errors}

   \subsection{Variations in modeling and analysis configuration}
   \label{Sec:variations}
   An important part of the analysis work consisted in assessing the impact of some of the key features adopted for ICRF3 in terms of data selection, modeling or analysis configuration. To this end, a number of alternate analysis solutions were carried out, each of which varying a feature of interest. These alternate solutions were then checked against our original analysis solution, in particular by examining changes in the source positions, which allowed us to quantify the impact of those features on our results, including the level of systematic errors. All such tests were accomplished based on the S/X~band data set and employed the CALC-SOLVE software package to guarantee the consistency of those alternate solutions with our original analysis solution.

\begin{figure}
   \resizebox{\hsize}{!}{\includegraphics[trim=8 7 35 33, clip]{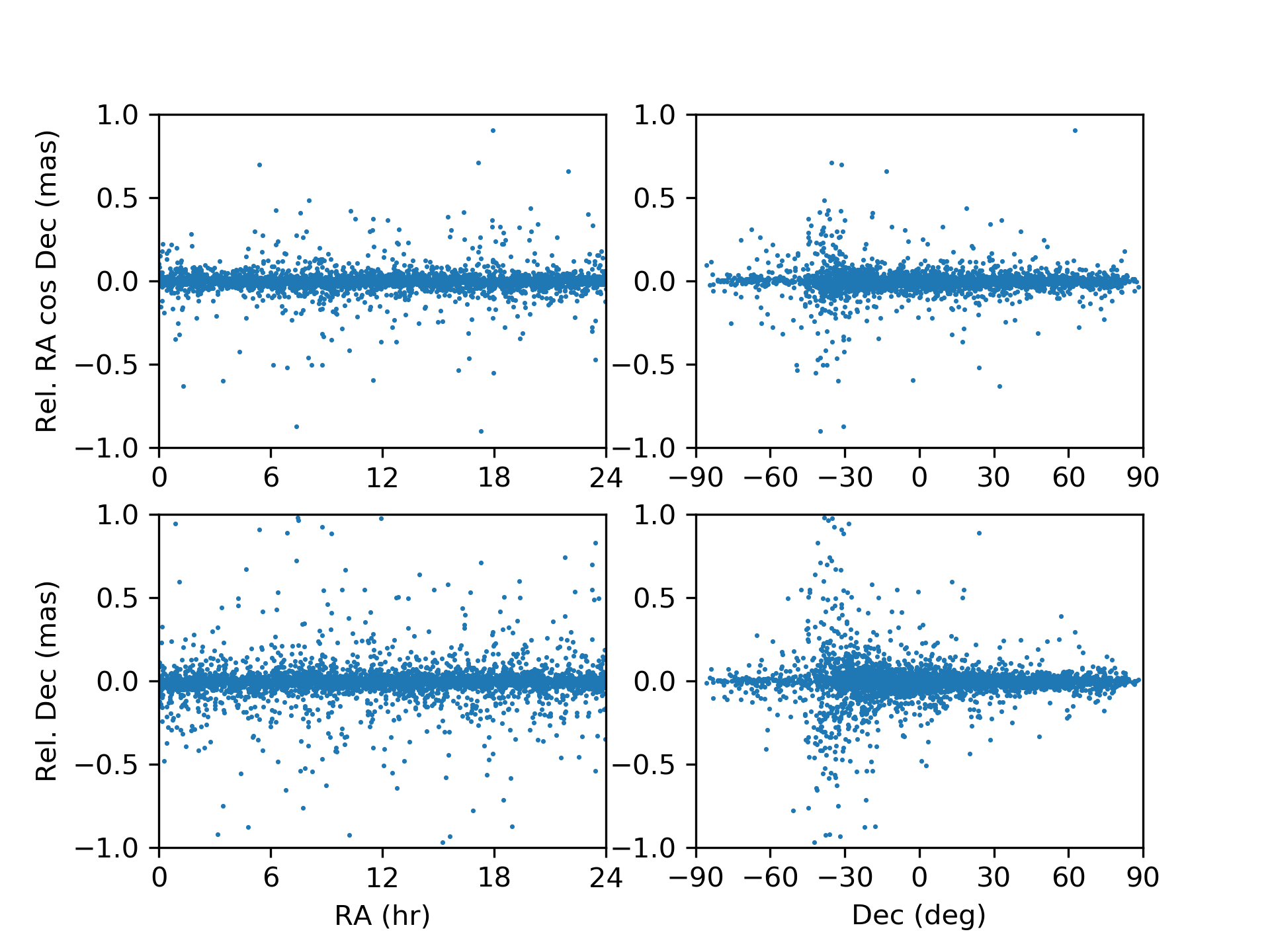}}
   \caption{Variations in the estimated radio source coordinates at S/X band when the observation elevation cutoff angle is changed from $5^{\circ}$ to $7^{\circ}$. The differences are given as $\Delta\alpha\cos\delta$ (upper panels) and $\Delta\delta$ (lower panels) and are plotted as a function of right ascension (left-hand panels) and declination (right-hand panels). Units are milliarcseconds.}
   \label{Fig:cutoff_elevation}
   \end{figure}

   An initial element that was investigated is the elevation cutoff angle for the data to be included into the analysis solutions. Observing at low elevations is necessary to decorrelate estimates of station vertical position from estimates of tropospheric zenith and gradient delays which both depend on the sine of the observation elevation. On the other hand, troposphere modeling errors increase with decreasing elevation because the observed signal passes through increasingly more troposphere. Below some elevation, the expected geometric improvement is overcome by modeling errors, which can thus bias the parameter determination \citep[see, e.g.,][]{Herring1986,Davis1991,MacMillan1994}. Figure~\ref{Fig:cutoff_elevation} compares the estimated radio source coordinates when changing the cutoff elevation angle from $5^{\circ}$ to~$7^{\circ}$.
   As a matter of convenience, the resulting variations (and those for the additional tests presented below) are reported as $\Delta\alpha\cos\delta$ and~$\Delta\delta$.
   While the plots in Fig.~\ref{Fig:cutoff_elevation} show differences up to 1~mas, only a minority of sources are subject to such large differences. The rms scatter is much lower, 6~$\mu$as in $\Delta\alpha\cos\delta$ and 8~$\mu$as in $\Delta\delta$, with reduced chi-squared values of 0.06 and 0.07, respectively, hence indicating that the position variations are statistically not significant. Analysis solutions using even higher cutoff elevation angles ($10^{\circ}$ and $15^{\circ}$) were also carried out to further characterize the impact. These led to rms scatters up to 20~$\mu$as with reduced chi-squared up to 0.4, still below the expected coordinate uncertainties. Results of these tests are reported in Table~\ref{Tab:alternate_solutions}.

   The other elements of the modeling and analysis configuration that were investigated as part of this assessment include (i)~the parametrization of the troposphere, approached by changing the interval between successive estimation of the zenith tropospheric delays from 30~min (the adopted interval) to shorter (20~min) or longer (1~hour and 3~hours) intervals, (ii)~the treatment of the ICRF2 special handling sources, which was tackled by estimating the positions of these sources separately for each session (as for ICRF2) instead of uniquely from the entire data set, and (iii)~the estimation of session-based antenna positions instead of global positions and velocities from all data. The results of these alternate analysis solutions are reported in Table~\ref{Tab:alternate_solutions} in terms of source coordinate variations and reduced chi-squared values in the same way as those regarding the cutoff elevation angle discussed above. In all, it is found that the source coordinate changes do not exceed 20~$\mu$as, while reduced chi-squared values remain lower than 0.4. Based on these tests, it therefore appears that the corresponding choices of parametrization are not a major source of error in the realization of the frame.

   \begin{table}[t]
   \caption{Impact of alternate analysis configurations on the estimated radio source coordinates at S/X band. The differences are given as the rms scatter and reduced chi-squared in right ascension ($\Delta\alpha\cos\delta$) and declination ($\Delta\delta$). Units for the coordinate differences are microarcseconds.}
   \begin{tabular}{l@{\ \ \ \ \ \ }rrc@{\ }crr}
   \hline
   \hline
   \noalign{\smallskip}
   \small Elements of variations& \multicolumn{2}{c}{\small $\Delta\alpha\cos\delta$}&&& \multicolumn{2}{c}{\small $\Delta\delta$}\\
   \cline{2-3}\cline{6-7}
   \noalign{\smallskip}
          & \multicolumn{1}{c}{\small rms} & \small $\chi_\nu^2$\ \ &&& \multicolumn{1}{c}{\small rms} & \small $\chi_\nu^2$ \ \\
          & \small ($\mu$as) &          &&& \small ($\mu$as) &\\
   \noalign{\smallskip}
   \hline
   \noalign{\smallskip}
   \small Elevation angle cutoff\tablefootmark{\ a}\\
   \small \ \ \ \ \ $>7^{\circ}$&  \small 6 \ \ & \small 0.06 &&&  \small 8 \ \ & \small 0.07\\
   \small \ \ \ \ \ $>10^{\circ}$& \small 10 \ \ & \small 0.18 &&& \small 14 \ \ & \small 0.20\\
   \small \ \ \ \ \ $>15^{\circ}$& \small 16 \ \ & \small 0.36 &&& \small 22 \ \ & \small 0.38\\
   \noalign{\smallskip}
   \small Zenith tropospheric delays\tablefootmark{b}\\
   \small \ \ \ \ \ 20 min intervals &  \small 10 \ \ & \small 0.23 &&&  \small 12 \ \ & \small 0.21\\
   \small \ \ \ \ \ 1 hr intervals &  \small 11 \ \ & \small 0.15 &&&  \small 15 \ \ & \small 0.15\\
   \small \ \ \ \ \ 3 hr intervals &  \small 11 \ \ & \small 0.15 &&&  \small 15 \ \ & \small 0.15\\
   \noalign{\smallskip}
   \small Special handling sources\\
   \noalign{\vspace{-2pt}}
   \small coordinates per session&  \small 13 \ \ & \small 0.38 &&&  \small 9 \ \ & \small 0.13\\
   \noalign{\smallskip}
   \small Session-based antenna\\
   \noalign{\vspace{-2pt}}
   \small positions &  \small 8 \ \ & \small 0.11 &&& \small 11 \ \ & \small 0.14\\
   \hline
   \end{tabular}
   \tablefoottext{\scriptsize a}{\footnotesize The reference setting for the elevation angle cutoff is $5^{\circ}$.}\\
   \tablefoottext{\scriptsize b}{\footnotesize The reference setting for the tropospheric delay intervals is 30~min.}
   \label{Tab:alternate_solutions}
   \end{table}

   A final element of the analysis configuration that was tested is the choice of the set of ICRF2 defining sources to be included in the no-net-rotation constraint applied for the alignment of the S/X band frame onto ICRF2. The reason for this test is that a fraction of the ICRF2 defining sources (about~10\%) were found to have extended structures in post-ICRF2 VLBI imaging work. These sources either had not been imaged at the time ICRF2 was built or were subject to structural evolution in the meantime. A notable case is the source 0805$+$406 which shows a double structure with a component separation of about 6~mas, as revealed by VLBI images from the Bordeaux VLBI Image Database (BVID)\footnote{Accessible online at \url{http://bvid.astrophy.u-bordeaux.fr}.} over the period 2010--2016. Its X~band structure index\footnote{The structure index is an indicator of the astrometric suitability of the sources. See \citet{Fey1997} and \citet{Fey2015} for its definition and details on how it is calculated from VLBI source maps.} in BVID has a median value of~5.5, well above the upper limit of~3.0 adopted for the selection of the ICRF2 defining sources. Such a source has very poor astrometric suitability and is improper as a defining source. Additionally, another six ICRF2 defining sources which have a structure index in BVID between 3.5 and~4.0 (0440$+$345, 1038$+$528, 1548$+$056, 1823$+$689, 2106$-$413, and 2326$-$477) were deemed to be not good enough either. In all, if considering all ICRF2 defining sources, 36 of them would not qualify anymore as defining sources according to the original ICRF2 criterion. It should be noted though that the bulk of these (24~sources) come with a structure index value between 3.0 and~3.25, which is just above the upper limit of 3.0 adopted for this selection. In order to test the potential degradation of the orientation of the frame due to these structured sources, we ran four alternate analysis solutions, each leaving out an increased number of ICRF2 defining sources from the constraint, depending on the maximum structure index considered as acceptable (4.0, 3.5, 3.25, 3.0). Since the data, modeling and estimated parameters were the same in the four solutions, the resulting frames then should differ only in orientation due to the slightly different no-net-rotation constraints applied. These four alternate frames were then compared to the frame obtained when all 295~ICRF2 defining sources are included in the constraint. We used for this purpose the formalism developed in Sect.~\ref{Sec:formalism} below. As reported in Table~\ref{Tab:transfer_sources}, the components of the rotations along the three axes, $R_1$, $R_2$, and~$R_3$, as derived from those comparisons, reach 6~$\mu$as at most, which is below the estimated 10~$\mu$as directional stability of ICRF2 \citep{Fey2015} that defines the level to which the two frames should be aligned to be consistent in orientation. We thus concluded that there is no need to worry about these sources, hence our decision to keep all ICRF2 defining sources in the alignment process of ICRF3 onto ICRF2.

   \begin{table}
   \caption{Variations in the orientation of the S/X band frame depending on the subset of ICRF2 defining sources used to align the frame onto ICRF2. Sources in each subset are filtered out based on the maximum structure index (SI). Rotations (in $\mu$as) are measured relative to the~case where all 295~ICRF2 defining sources are considered for the alignment.}
   \begin{tabular}{lrcr@{\small\ }crrr}
   \hline
   \hline
   \noalign{\smallskip}
    & \multicolumn{3}{c}{\small ICRF2 defining sources} && \multicolumn{3}{c}{\small Rotation of the frame}\\
   \cline{2-4}\cline{6-8}
   \noalign{\smallskip}
   \small Max & \small Nb \ \ \ \ & \small Nb & \small \% \ \ \ \ \ && \small $R_1$ \ \ & \small $R_2$ \ \ & \small $R_3$ \ \ \\
   \ \small SI & \small excluded & \small included & \small included && \small ($\mu$as) & \small ($\mu$as) & \small ($\mu$as)\\
   \noalign{\smallskip}
   \hline
   \noalign{\smallskip}
   \small 5.5  & \small  0 \ \ \ \ \ & \small 295 & \small $100$ \ \ \ \ \ && \small - \ \ \ & \small - \ \ \ & \small  - \ \ \ \\
   \small 4.0  & \small  1 \ \ \ \ \ & \small 294 & \small $>99$ \ \ \ \ \ && \small $-1$ \ \ \ & \small $-1$ \ \ \ & \small  4 \ \ \ \\
   \small 3.5  & \small  7 \ \ \ \ \ & \small 288 & \small  98  \ \ \ \ \ && \small $-3$ \ \ \ & \small $-2$ \ \ \ & \small  4 \ \ \ \\
   \small 3.25 & \small 12 \ \ \ \ \ & \small 283 & \small  96  \ \ \ \ \ && \small $-2$ \ \ \ & \small $-2$

    \ \ \ & \small  6 \ \ \ \\
   \small 3.0  & \small 36 \ \ \ \ \ & \small 259 & \small  88  \ \ \ \ \ && \small $-4$ \ \ \ & \small  4 \ \ \ & \small  4 \ \ \ \\
   \hline
   \end{tabular}
   \label{Tab:transfer_sources}
   \end{table}

   \subsection{Accuracy of estimated auxiliary parameters}
   An indirect way to assess the quality of our analysis solutions is to control the accuracy of the auxiliary parameters that were estimated simultaneously with the source coordinates and are thus an integral part of the solutions. The EOP are of particular value in this regard since estimates of these parameters are available for every session, which in some way allows one to check the entire analysis on a statistical basis. In order to evaluate the accuracy of the S/X~band and K~band EOP derived from our analysis solutions, we have compared them with those from the IVS combined series ivs15q2X\footnote{As available online on August~25, 2018 from the IVS products web page at \url{https://ivscc.gsfc.nasa.gov/products-data/products.html}.}. The latter is an official product of the IVS obtained by combining individual EOP series generated by several IVS analysis centers. The comparison was achieved by differencing the EOP estimated for each session with those from the IVS combined series. The largest outliers ($>$5$\sigma$) were then filtered out and a linear slope was fitted to the data and taken out to remove any long term trend between the series prior to computing statistics for the differences. This scheme follows the standard IERS practice for comparing and combining EOP series \citep[see, e.g.,][]{Bizouard2019}. It must be noted that only the post-1994 data were considered in this adjustment in order not to bias the assessment by the earlier less-accurate EOP determinations. After such processing, the weighted rms of the differences and reduced chi-squared were calculated as indicators of the agreement between our series and the ivs15q2X series. The results, including values of the fitted slopes, are given in Table~\ref{Tab:EOP_consistency_SX} for the S/X~band EOP series and in Table~\ref{Tab:EOP_consistency_K} for the K~band EOP series. For completeness, median values of the formal uncertainties for the EOP, as derived from our solutions, and calculated over the same time span, are also reported in these tables.

   \begin{table}
   \caption{Comparison of the EOP estimated at S/X band with those~from the IVS combined series ivs15q2X. Differences are characterized in terms of the weighted rms, reduced chi-squared and slope between those series for each of the five EOP, i.e., the two polar motion components ($x_p$, $y_p$), the daily rotation UT1, and the two nutation offsets~($X$,~$Y$). Median values of the EOP uncertainties resulting from our ICRF3 S/X~band analysis solution are also indicated in the table for comparison.}
   \begin{tabular}{l@{\ \ \ \ }r@{\ \ \ \ }r@{\ \ \ \ }r@{\ \ \ \ }r@{\ \ \ \ }r}
   \hline
   \hline
   \noalign{\smallskip}
   \small \ \ Statistics for EOP & \small $x_p$ \ \ & \small $y_p$ \ \ & \small UT1 & \small $X$\ \ \ & \small $Y$\ \ \ \\
   \small \ \ (3464 data points) & \small ($\mu$as) & \small ($\mu$as) & \small ($\mu$s) \ & \small ($\mu$as) & \small ($\mu$as) \\
   \noalign{\smallskip}
   \hline
   \noalign{\smallskip}
   \small Median uncertainty  & \small 61 & \small 56 & \small 2.6 & \small 55 & \small 56 \\
   \small Difference with IVS series\\
   \small \ \ \ \ \ \ \ \ wrms& \small 76 & \small 79 & \small 6.1 & \small 45 & \small 44 \\
   \small \ \ \ \ \ \ \ \ $\chi_\nu^2$  & \small  2.2 & \small 2.4 & \small 2.6 & \small 1.0 & \small 1.0 \\
   \small \ \ \ \ \ \ \ \ slope\tablefootmark{a} & \small 6.1 & \small 10.3 & \small $-0.50$ & \small 0.3 & \small 0.8 \\
   \small \ \ \ \ \ \ \ \ slope error\tablefootmark{a} & \small $\pm0.2$ & \small $\pm0.2$ & \small $\pm0.04$ & \small $\pm0.1$ & \small $\pm0.1$ \\
   \hline
   \end{tabular}
   \tablefoottext{\scriptsize a}{\footnotesize Units are $\mu$as/yr for ($x_p$, $y_p$), $\mu$s/yr for UT1, and $\mu$as/yr for ($X$, $Y$).}
   \label{Tab:EOP_consistency_SX}
   \end{table}

   By examining the content of Table~\ref{Tab:EOP_consistency_SX}, we first note that the median uncertainties for the five EOP are all about the same, in the range of 40--60 $\mu$as, indicating that they are all determined with similar accuracies from the S/X band analysis solution. Looking at the results of the comparisons, we further note that the nutation offsets are in full agreement with those reported by the IVS (reduced chi-squared values of 1.0 for $X$ and~$Y$). The relative slopes estimated for these parameters are less than 1~$\mu$as/yr, which is not significant at the reported accuracies of 50~$\mu$as. The derived S/X~band frame thus appears to be fully consistent with the currently available IVS nutation series. Differences for polar motion and~UT1 are somewhat larger, with weighted rms of 75--90~$\mu$as and reduced chi-squared values in the range 2.2--2.6. Such increased differences may be due to a slight inconsistency (at the level of 1--2 mm) in fixing the origin and orientation of the terrestrial frame, which may slightly vary depending on the particular set of stations used in the constraints.
   Additionally, the relative slopes for these parameters are also found to have higher values, up to 10~$\mu$as/yr for polar motion component~$y_p$. Again, this may be due to a slight inconsistency in fixing the rotation rate of the terrestrial frame. The consequence is a drift of the series on the long term, although at a limited level (7--12 mm shifts after 40~years). In order to further assess the matter, we have taken an additional step and compared our EOP series to the IERS EOP~14C04 series \citep{Bizouard2019}. The slopes derived for $x_p$, $y_p$, and UT1 from this comparison (in units of $\mu$as/yr for $x_p$ and $y_p$, and $\mu$s/yr for UT1) are 9.1, 9.4, and $-1.64$, to be checked against the results of the comparisons with the ivs15q2X series in Table~\ref{Tab:EOP_consistency_SX}, which are 6.1, 10.3, and $-0.50$. This indicates relative drifts of 3.0, $-0.9$, and $-1.14$ between the IERS EOP~14C04 series and the ivs15q2X series. Calculating the mean of these drifts for the two components of polar motion and UT1, we derive a value of 7~$\mu$as/yr, whereas a value of 8~$\mu$as/yr is inferred when comparing our S/X~band series with the ivs15q2X~series. The inconsistency between our S/X~band series and the IVS combined series, if any, is thus no larger than that between the IERS EOP~14C04 series and the IVS combined series. In any case, those small drifts have no impact on the derived celestial frame since they just reflect the slightly different ways in which the terrestrial frame was fixed and are expected to cancel out when the terrestrial frame and EOP relative rotations are added.

   \begin{table}
   \caption{Comparison of the EOP estimated at K band with those from the IVS combined series ivs15q2X. Differences are characterized in terms of the weighted rms, reduced chi-squared and slope between those series for each of the five EOP, i.e., the two polar motion components ($x_p$, $y_p$), the daily rotation UT1, and the two nutation offsets~($X$,~$Y$). Median values of the EOP uncertainties resulting from our ICRF3 K band analysis solution are also indicated in the table for comparison.}
   \begin{tabular}{l@{\ \ \ }r@{\ \ \ \ }r@{\ \ \ \ }r@{\ \ \ \ }r@{\ \ \ \ }r}
   \hline
   \hline
   \noalign{\smallskip}
   \small \ \ Statistics for EOP & \small $x_p$ \ \ & \small $y_p$ \ \ & \small UT1 & \small $X$\ \ \ & \small $Y$\ \ \ \\
   \small \ \ \ \ (27 data points) & \small ($\mu$as) & \small ($\mu$as) & \small ($\mu$s) \ & \small ($\mu$as) & \small ($\mu$as) \\
   \noalign{\smallskip}
   \hline
   \noalign{\smallskip}
   \small Median uncertainty  & \small 66 & \small 140 & \small 5.5 & \small 43 & \small 42 \\
   \small Difference with IVS series\\
   \small \ \ \ \ \ \ \ \ wrms& \small 153 & \small 246 & \small 14.9 & \small 88 & \small 101 \\
   \small \ \ \ \ \ \ \ \ $\chi_\nu^2$  & \small  4.7 & \small 6.5 & \small 10.2 & \small 3.7 & \small 5.5 \\
   \small \ \ \ \ \ \ \ \ slope\tablefootmark{a} & \small 15.6 & \small 20.1 & \small 0.51 & \small $-1.0$ & \small $-6.8$ \\
   \small \ \ \ \ \ \ \ \ slope error\tablefootmark{a} & \small $\pm8.1$ & \small $\pm11.4$ & \small $\pm1.65$ & \small $\pm3.5$ & \small $\pm3.9$ \\
   \hline
   \end{tabular}
   \tablefoottext{\scriptsize a}{\footnotesize Units are $\mu$as/yr for ($x_p$, $y_p$), $\mu$s/yr for UT1, and $\mu$as/yr for ($X$, $Y$).}
   \label{Tab:EOP_consistency_K}
   \end{table}

   The results of comparing the EOP derived at K band with those from the ivs15q2X~series, which are reported in Table~\ref{Tab:EOP_consistency_K}, are based on a much smaller number of data points (less than~1\% of the number of data points available at S/X band), possibly not statistically meaningful, and thus should be treated with caution. Nevertheless, such results are useful in providing indications on the quality and level of agreement of the estimated EOP at K~band. As indicated by Table~\ref{Tab:EOP_consistency_K}, median uncertainties are in the range between 40 and 80~$\mu$as, which is roughly at the same level as the median uncertainties at S/X band (see Table~\ref{Tab:EOP_consistency_SX}), with the exception of that found for parameter $y_p$ which is much larger~(140~$\mu$as). The latter likely originates in the geometrical configuration of the observing network at K~band (i.e., the VLBA), which does not favor the estimation of that parameter. Comparisons with the ivs15q2X series show differences of 100--250~$\mu$as, which is twice as large as the differences found at S/X~band, reaching a factor of three for parameter $y_p$. Such differences also appear to be fairly significant, with reduced chi-squared values of 4--10, reflecting the presence of systematic errors. In this regard, it is to be pointed out that our K band EOP estimates are fully independent of those in the IVS series since they are derived from different data sets, which is not the case of our S/X~band estimates. For the same reason, interpolation errors may also be larger because there may not be an S/X band session carried out concurrently with each K~band session. As regards polar motion and UT1, differences may also stem from the fixing of the origin and orientation of the terrestrial frame, which may not be attached to ITRF2014 as precisely as at S/X~band, due to the use of only nine stations, all located in the Northern Hemisphere, for this purpose (see Sect.~\ref{Sec:configuration}). On the other hand, the fitted slopes, even if they show higher values than at S/X~band, do not have a significance that exceeds 2$\sigma$. We have thus no indication of significant drifts with respect to ITRF2014. As noted previously, such drifts, in any case, would be absorbed and would have no impact on the derived celestial frame.

   \subsection{Determination of realistic uncertainties}
   The formal uncertainties that come from geodetic and astrometric VLBI analyses are dependent on a number of factors, including the sensitivity of the network, the source flux density, and the number of observations. They get smaller as the sensitivity of the instrumentation (receivers, data acquisition terminals) improves or when using larger antennas, while they deteriorate as the sources become weaker \citep[see, e.g.,][for an investigation of the relationship between source position uncertainty and flux density]{Malkin2016}. Furthermore, the use of the least-squares method to solve for parameters in these analyses implies that the resulting formal uncertainties fall off as the square root of the number of observations. Consequently, such formal uncertainties may become very small, and even reach unrealistic levels, when the number of observations gets large. The number of observations in itself scales with the square of the number of antennas in the array. Larger VLBI arrays are thus more likely to generate formal uncertainties that are too small. Additionally, the proportion of independent VLBI measurements decreases as the size of the array increases. For example, a five-station array (providing ten baselines) delivers ten VLBI delay measurements for a given source scan, among which only five, or 50\%, are fully independent, a percentage that falls down to 22\% for a ten-station array. Not accounting for the induced correlations (in particular related to clocks and troposphere) in the least-squares analysis then results in formal parameter uncertainties that are smaller than they should be in reality considering those correlations. The question of the validity of the formal uncertainties coming out from VLBI analyses has been debated ever since the beginning of VLBI. It was first investigated by \citet{Ryan1993} who concluded that a scaling factor of 1.5 is to be applied to VLBI formal uncertainties to bring them closer to actual uncertainties. Based on this finding, \citet{Ma1998} and \citet{Fey2015} also inflated the formal uncertainties from their analysis solutions by a factor of~1.5 to report ICRF and ICRF2 source coordinate uncertainties. We have not attempted here to redetermine this scaling factor and went along the same lines, in the continuity of the previous ICRF realizations. The scaling factor of 1.5, however, was only applied to the S/X~band and K~band source coordinate uncertainties, the X/Ka~coordinate uncertainties being less likely to be affected by such underestimation because the~X/Ka network consists mostly of single-baseline sessions, as indicated above.

   Apart from this scaling factor, a noise floor was also applied to the source coordinate uncertainties reported in ICRF and ICRF2 so that these do not drop to unrealistic levels when the number of observations for a given source becomes very large. This noise floor was 250~$\mu$as for ICRF and 40~$\mu$as for ICRF2. In practice, those values were added in quadrature to the estimated formal uncertainties, scaled as described above, to derive the final source coordinate uncertainties. The determination of the noise floor in ICRF and ICRF2 was built upon comparisons of catalogs produced from independent analyses and/or independent data sets. In particular, \citet{Fey2015} ran a decimation test in which the ICRF2 data set was divided into two independent subsets and compared the source coordinates estimated from each subset of data to infer the noise level. We have followed a similar approach for ICRF3. To this end, all VLBI sessions were ordered temporally and divided into two subsets selected by even and odd sessions. The source coordinates estimated from the data in each subset were then compared in declination bands of~$10^{\circ}$, limiting the comparison to sources with a minimum of 100~observations and observed in at least two sessions.
   From this comparison, the noise floor in each declination band was derived as the weighted rms of the coordinates differences for the sources in the given declination band divided by the square root of~2. That calculation was achieved separately for right ascension and declination. The reason for treating declination bands separately was to check the overall consistency of the process and to make sure no declination bands stick out, which may occur due to the non-uniform distribution of sources in declination (see Sect.~\ref{Sec:ICRF3-positions} below). Such a scheme was applied to both the S/X~band and K~band data sets but not to the X/Ka~band data set, considering that the amount of data available for the latter was not sufficient to make it meaningful. From our experience in processing data at X/Ka~band, we decided instead to base the noise floor at this frequency band on that at S/X~band.

   \begin{table}
   \caption{Scaling factor and noise floor (in $\mu$as) applied to the estimated formal uncertainties of the source coordinates at each frequency band.}
   \begin{tabular}{c@{\small \ \ \ \ }cccc@{}ccc}
   \hline
   \hline
   \noalign{\smallskip}
    && \multicolumn{2}{c}{\small Right ascension} &&& \multicolumn{2}{c}{\small Declination}\\
   \cline{3-4}\cline{7-8}
   \noalign{\smallskip}
   \small Frequency && \small Scaling & \small Noise floor &&& \small Scaling & \small Noise floor \\
   \small band && \small factor & \small ($\mu$as) &&& \small factor & \small ($\mu$as) \\
   \noalign{\smallskip}
   \hline
   \noalign{\smallskip}
   \small S/X &&  \small 1.5 & \small 30 &&& \small 1.5 & \small 30 \\
   \small K &&    \small 1.5 & \small 30 &&& \small 1.5 & \small 50 \\
   \small X/Ka && \small 1.0 & \small 30 &&& \small 1.0 & \small 30 \\
   \hline
   \end{tabular}
   \label{Tab:noise_level}
   \tablefoot{Inflated coordinate uncertainties are derived with the expressions $(\sigma_{\alpha}\cos\delta)^2 = (s_\alpha\ \sigma_{\alpha\mathrm{,formal}}\cos\delta)^2 + \sigma_{\alpha\cos\delta, 0}^2$ and $\sigma_{\delta}^2 = (s_{\delta}\ \sigma_{\delta,{\rm formal}})^2 + \sigma_{\delta,0}^2$, where $\sigma_{\alpha\mathrm{,formal}}$ and $\sigma_{\delta,{\rm formal}}$ are the formal uncertainties in right ascension and declination (i.e., resulting from the least-squares parameter adjustment in our VLBI analysis solutions) and $\sigma_{\alpha}$ and $\sigma_{\delta}$ are the corresponding inflated uncertainties. The scaling factors are expressed as $s_\alpha$ and $s_\delta$, and the noise floor as $\sigma_{\alpha\cos\delta,0}$ and~$\sigma_{\delta,0}$.
   }
   \end{table}

   The results of the decimation tests at S/X~band show that the noise floor in the various declination bands ranges from 13 to 45~$\mu$as in right ascension and from 12 to 64~$\mu$as in declination (with four declination bands comprising less than 20~sources excluded from that assessment). Taking all sources together (i.e., without separating sources into declination bands), the above numbers average to 26~$\mu$as in right ascension and 30~$\mu$as in declination. A question that arose was whether to treat differently the sources south of $-40^{\circ}$ declination since there appears to be a disparity in the noise floor measured for these sources compared to those further north (44~$\mu$as vs 25~$\mu$as for right ascension and 47~$\mu$as vs 29~$\mu$as for declination). However, we did not do so considering that the assessment at those low declinations relies on far less sources (about 8\% of the total number of usable sources from the decimation test) and that our estimation of the noise floor in the far south only differs by a factor of 1.7 from that further north. Based on these findings, we decided to adopt a noise floor of 30~$\mu$as for both coordinates (right ascension and declination) without a declination dependency. Applying a similar scheme to the K~band data led to values of the noise floor ranging from 18 to 57~$\mu$as in right ascension and from 32 to 113~$\mu$as in declination, with corresponding average values of 33~$\mu$as and 57~$\mu$as, respectively. Still, it must be noted that declinations below~$-30^{\circ}$ could not be assessed with our method because of too few data. Unlike at S/X~band, the decimation tests at K~band indicate that there is a notable difference in the noise floor between right ascension and declination. For this reason, we decided to adopt two different values, 30~$\mu$as for right ascension and 50~$\mu$as for declination. Table~\ref{Tab:noise_level} summarizes the adopted values for the scaling factor and noise floor for the three frequency bands.

   \section{A multi-frequency frame}
   \label{Sec:ICRF3}

   \subsection{Three-frequency VLBI source positions}
   \label{Sec:ICRF3-positions}
   \begin{table*}[t]
   \caption{Statistics about the data and least-squares fits accomplished at the S/X, K, and X/Ka frequency bands. The fits are characterized by the weighted rms of the post-fit delay and delay rate residuals and the corresponding reduced chi-squared values.}
   \begin{center}
   \begin{tabular}{c@{\hskip 11pt}r@{\hskip 7pt}ccc@{\hskip 3pt}c@{\hskip 9pt}rc@{\hskip 10pt}c@{\hskip 10pt}l@{}r@{\hskip 4pt}}
   \hline
   \hline
   \noalign{\smallskip}
          &&& \multicolumn{2}{c}{\small Delay residuals}&&\multicolumn{2}{c}{\small Delay rate residuals}\\
      \cline{4-5}      \cline{7-8}
   \noalign{\smallskip}
   \small Frequency  & \small Number of {\hskip 1pt} && \small {\hskip 0pt} wrms & \small {\hskip 0pt} $\chi_\nu^2$&& \small {\hskip 12pt} wrms & \small $\chi_\nu^2$ && \small {\hskip 34pt} Data span&\\
   \small band       & \small observations&& \small (ps)&&&\small (fs/s){\hskip 1pt} &&&\\
   \noalign{\smallskip}
   \hline
   \noalign{\smallskip}
   \small S/X  & \small 13\,190\,274 {\hskip 1pt} && \small 25.5 & \small 1.05 && \small - \ \ \ \ & \small - && \small
   1979 Aug 03\ \ {\hskip 1pt}--\ \  2018 Mar 27&\\
   \small K    & \small    482\,616 {\hskip 1pt} && \small 17.5 & \small 0.91 && \small 66.6\ \ & \small 0.90 && \small
   2002 May 15\ \ --\ \  2018 May 05&\\
   \small X/Ka & \small     69\,062 {\hskip 1pt} && \small 43.6 & \small 1.00 && \small 102.3\ \ & \small 1.00 && \small
   2005 Jul 09\ \ {\hskip 5.9pt}--\ \ 2018 Jan 28&\\
   \hline
   \end{tabular}
   \end{center}
   \label{Tab:ICRF3_fits}
   \small
   \end{table*}

   \begin{table*}
   \caption{Major features of ICRF3, including the number of sources at each frequency band and statistics about coordinate uncertainties, correlation coefficients between right ascension and declination, and the error ellipse size (semi-major axis). All statistics are given as median values of the said parameters and are provided for the entire S/X, K, and X/Ka band catalogs and for the subset of 600~sources common to the three catalogs.}
   \begin{center}
   \begin{tabular}{c@{\hskip 8pt}r@{\hskip 3pt}rrr@{{\hskip 18pt}}c@{\ \ }c@{\ }c@{\hskip 9pt}r@{\ \ }r@{\ \ \ \ }c@{\ \ }r}
   \hline
   \hline
   \noalign{\smallskip}
          &&& \multicolumn{4}{c}{\small Statistics for all sources}&& \multicolumn{4}{c}{\small Statistics for the common sources}\\
   \noalign{\smallskip}
   \cline{4-7}\cline{9-12}
   \noalign{\smallskip}
   \small Frequency  & \small Number of && \multicolumn{2}{c}{\small Coordinate uncertainty}&\small Correlation &\small {\hskip 1pt}Error ellipse{\hskip 2pt} && \multicolumn{2}{c}{\small Coordinate uncertainty}&\small Correlation & \small {\hskip 1pt}Error ellipse\\
   \small band       & \small sources{ \hskip 4pt} && \small {\hskip 8pt} $\alpha\cos\delta$ & \small $\delta${\hskip 6pt} & \small coeff. &\small ($\mu$as)&& \small {\hskip 14pt} $\alpha\cos\delta$  & \small $\delta${\hskip 14pt} & \small coeff. &\small ($\mu$as) {\hskip 13.3pt} \\
   &&&\small ($\mu$as) {\hskip 1pt} & \small ($\mu$as)&&&& \small ($\mu$as) {\hskip 1pt} & \small ($\mu$as){\hskip 8pt} \\
   \noalign{\smallskip}
   \hline
   \noalign{\smallskip}
   \small S/X  & \small 4536 {\hskip 8pt} && \small  127 {\hskip 3pt} & \small 218 {\hskip 0pt} & \small 0.13& \small 223&& \small 48 {\hskip 3pt} & \small 64 {\hskip 8pt} & \small 0.08&\small 64 {\hskip 16.3pt} \\
   \small K    & \small  824 {\hskip 8pt} && \small  74 {\hskip 3pt} & \small 136 {\hskip 0pt} & \small 0.30& \small 139&& \small 68 {\hskip 3pt} & \small 132 {\hskip 8pt} & \small 0.31&\small 135 {\hskip 16.3pt} \\
   \small X/Ka & \small  678 {\hskip 8pt} && \small  76 {\hskip 3pt} & \small 104 {\hskip 0pt} & \small 0.43& \small 115&& \small 69 {\hskip 3pt} & \small 100 {\hskip 8pt} & \small 0.44&\small 108 {\hskip 16.3pt} \\
   \hline
   \end{tabular}
   \end{center}
   \label{Tab:ICRF3_features}
   \end{table*}

   \begin{figure}
   \centering
   \includegraphics[width=1.01\columnwidth]{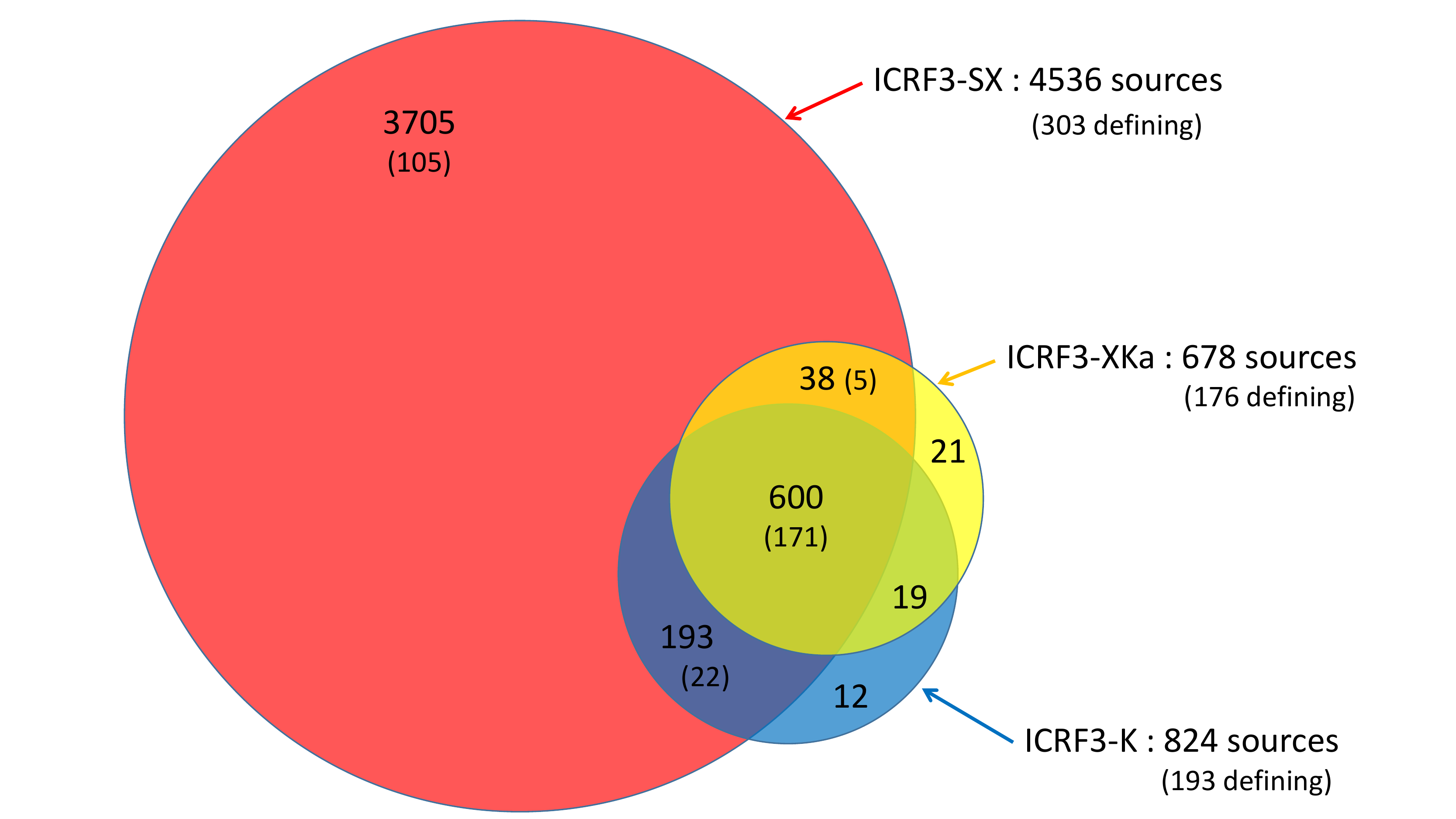}
   \caption{Breakdown of the 4588 sources in ICRF3 according to frequency band. The circle colored red is for S/X band, the one colored blue is for K band, and the one colored yellow is for X/Ka band. The number of sources found in each colored area is printed within that area, with the number of ICRF3 defining sources (see Sect.~\ref{Sec:ICRF3-defining}) given in parentheses.}
   \label{Fig:3-frequency_cross_ID}
   \end{figure}

   The modeling and analysis configuration outlined in Sect.~\ref{Sec:analysis} have been applied to the data sets at S/X~band, K~band, and X/Ka~band described in Sect.~\ref{Sec:data} to produce three separate catalogs in a consistent way and all aligned onto the ICRS. These three catalogs form ICRF3, the first multi-frequency celestial reference frame ever realized. Some statistics about the data and fits performed at the three frequency bands, including the number of observations, their time span, the weighted rms of the post-fit residuals, and the reduced chi-squared are provided in Table~\ref{Tab:ICRF3_fits}, while the major features of ICRF3 (number of sources and median coordinate uncertainties at each frequency band) are summarized in Table~\ref{Tab:ICRF3_features}. The S/X~band catalog comprises 4536 sources, one-third more than in ICRF2, while the K band and X/Ka band catalogs comprise 824~sources and 678 sources, respectively. The diagram in Fig.~\ref{Fig:3-frequency_cross_ID} shows the breakdown of the sources according to frequency band. In all, ICRF3 includes a total 4588 sources, all of which are part of the S/X band catalog, except 52~sources which belong only to the K band and/or X/Ka band catalogs. Also to be noted is that 600 sources are common to the three catalogs. The median coordinate uncertainties in the S/X band catalog are 127~$\mu$as for right ascension\footnote{Here and in the following sections, right ascension always denotes right ascension multiplied by the cosine of the declination.} and 218~$\mu$as for declination, with the two coordinates generally only weakly correlated (median correlation coefficient of 0.13). The median of the error ellipse semi-major axis\footnote{The error ellipse semi-major axis is calculated from the right ascension and declination uncertainties and correlation coefficient between the coordinates. See, e.g., Eq.~(1) of \citet{Mignard2018}.} is 223~$\mu$as. These values represent a factor of~3.4 improvement compared to the ICRF2 figures where such median uncertainties were 396~$\mu$as (for right ascension), 739~$\mu$as (for declination), and 765~$\mu$as (for the error ellipse semi-major axis). This dramatic improvement results from the VLBA campaigns that have been conducted since 2014 to re-observe all the less-observed ICRF2 sources (including the VCS sources), as pointed out in Sect.~\ref{Sec:data_SX}. Compared to the S/X~band catalog, the median uncertainties for the K~band and X/Ka~band catalogs appear to be smaller by a factor of 1.5 to 2 (see the statistics for all sources in Table~\ref{Tab:ICRF3_features}), while correlations between right ascension and declination appear to be stronger (median correlation coefficients of 0.30 and 0.43, respectively). The latter is likely a consequence of the more limited observing configurations at these two frequency bands. The observation that the median uncertainties are lower, however, is a somewhat artificial effect because the bulk of the sources in the S/X~band catalog are VCS-type sources which in terms of positional precision are still not at the level of the most-observed (and most-precise) S/X~band sources, despite the improvement noted above. When calculating median uncertainties solely for the 600~sources common to the three catalogs (see the statistics for the common sources in Table~\ref{Tab:ICRF3_features}), the figures are in fact the other way round. It is the S/X band coordinate uncertainties that turn out to be smaller by a factor of 1.5 to 2 compared to the K~band and X/Ka~band coordinate uncertainties. Notwithstanding this difference, the three catalogs show significantly improved positional precision compared to ICRF2.

   \begin{figure*}
   \includegraphics[width=1.0\textwidth,trim=10 165 30 165, clip]{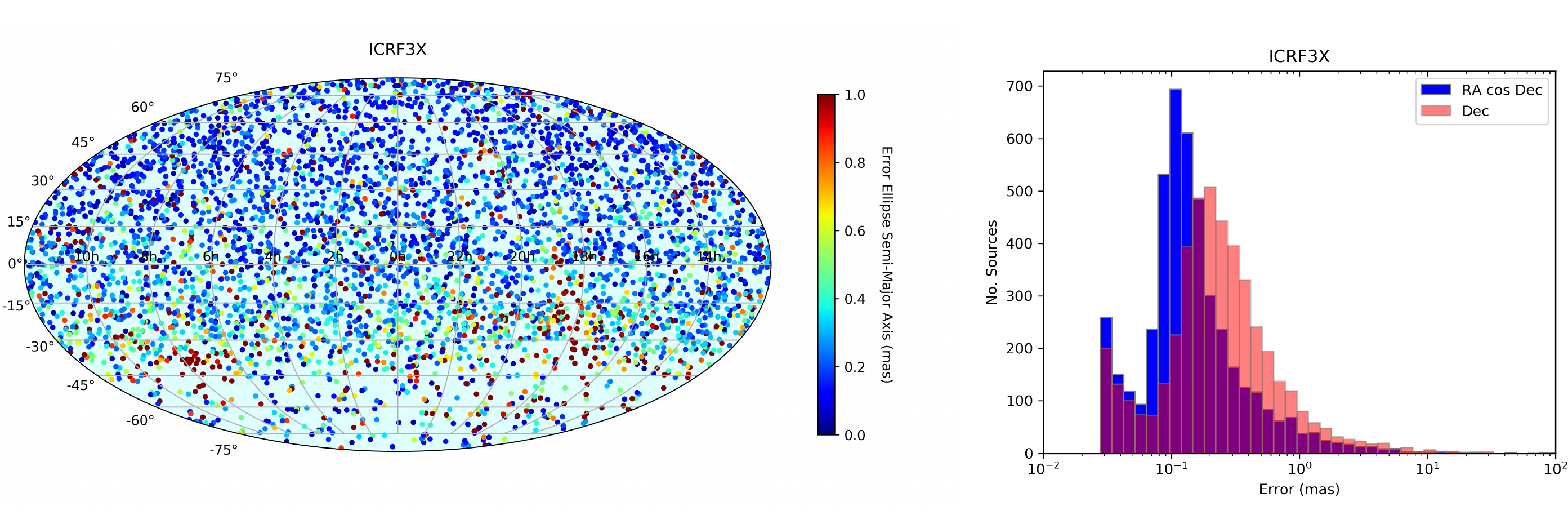}
   \caption{{\it Left:} distribution of the 4536 sources included in the ICRF3 S/X band frame on a Mollweide projection of the celestial sphere. Each source is plotted as a dot color-coded according to its position uncertainty (defined as the semi-major axis of the error ellipse in position).
   {\it Right:} distribution of coordinate uncertainties for the same 4536 sources. Right ascension is shown in blue while declination is shown in salmon. The superimposed portion of the two distributions is shown in purple.}
   \label{Fig:ICRF3_SX}
   \end{figure*}

   \begin{figure*}
   \includegraphics[width=1.0\textwidth,trim=10 165 30 155, clip]{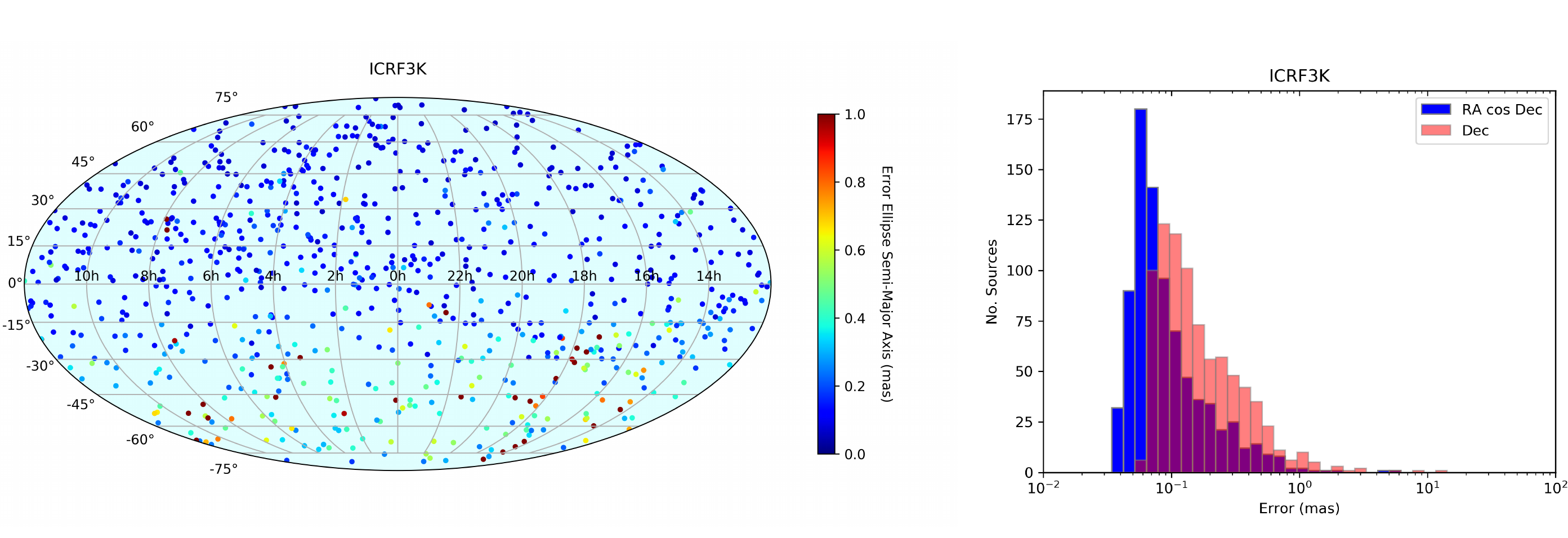}
   \caption{{\it Left:} distribution of the 824 sources included in the ICRF3 K~band frame on a Mollweide projection of the celestial sphere. Each source is plotted as a dot color-coded according to its position uncertainty (defined as the semi-major axis of the error ellipse in position).
   {\it Right:} distribution of coordinate uncertainties for the same 824 sources. Right ascension is shown in blue while declination is shown in salmon. The superimposed portion of the two distributions is shown in purple.
   It must be noted that the scale for the y-axis (number of sources) is different from that in Fig.~\ref{Fig:ICRF3_SX}.}
   \label{Fig:ICRF3_K}
   \end{figure*}

   \begin{figure*}
   \includegraphics[width=1.0\textwidth,trim=10 165 30 155, clip]{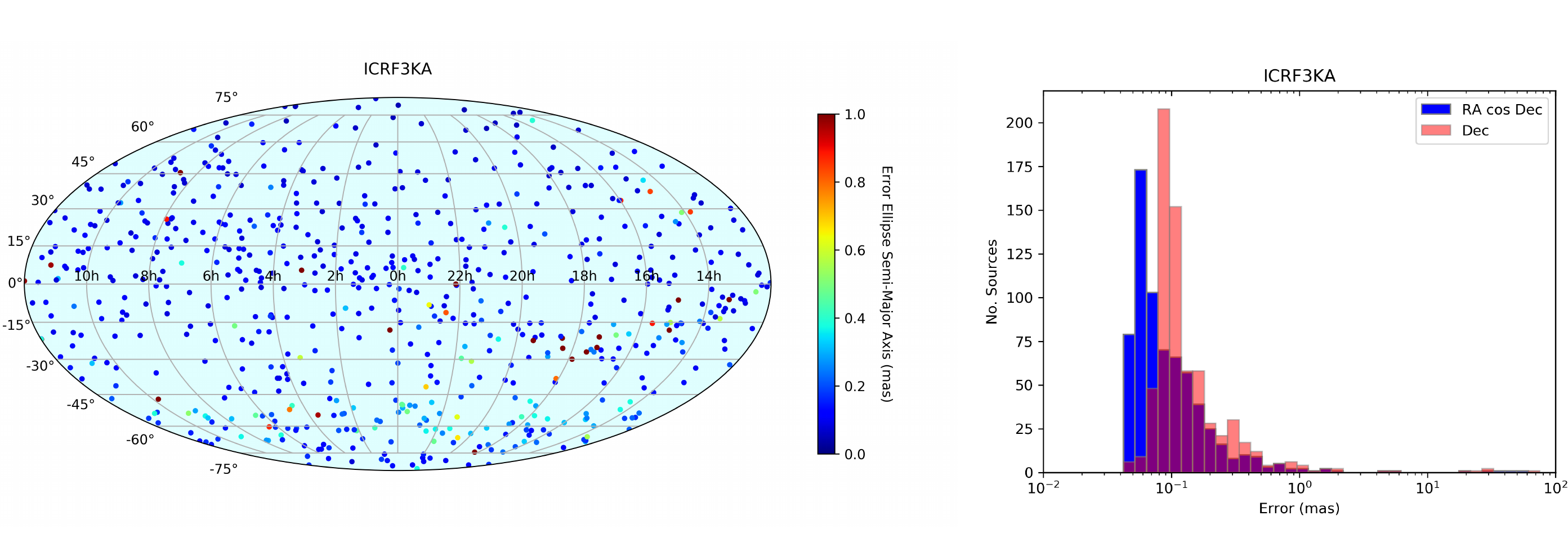}
   \caption{{\it Left:} distribution of the 678 sources included in the ICRF3 X/Ka~band frame on a Mollweide projection of the celestial sphere. Each source is plotted as a dot color-coded according to its position uncertainty (defined as the semi-major axis of the error ellipse in position).
   {\it Right:} distribution of coordinate uncertainties for the same 678 sources. Right ascension is shown in blue while declination is shown in salmon. The~superimposed portion of the two distributions is shown in purple.
   It must be noted that the scale for the y-axis (number of sources) is different from that in~Fig.~\ref{Fig:ICRF3_SX}.}
   \label{Fig:ICRF3_Ka}
   \end{figure*}

   \begin{figure}
   \centering
   \includegraphics[width=0.90\columnwidth, trim=0 0 0 25, clip]{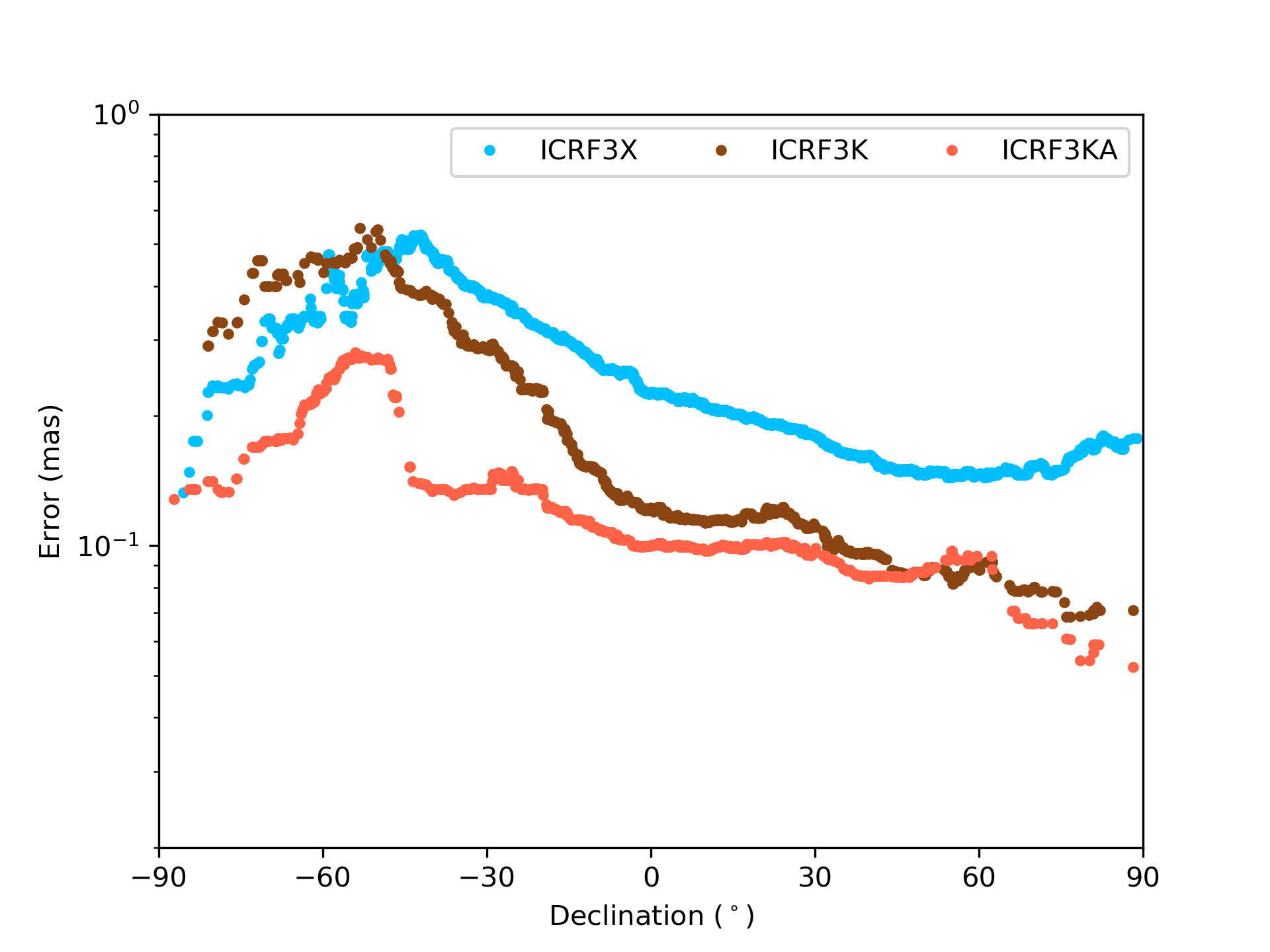}
   \caption{Median source position uncertainty as a function of declination for the S/X band frame (shown in light blue), the K band frame (shown in brown), and the X/Ka band frame (shown in salmon). The quantity represented is the running median for the semi-major axis of the error ellipse in position, determined over declination bins of $15^{\circ}$ in each~frame.}
   \label{Fig:dec_errors}
   \end{figure}

   Figures~\ref{Fig:ICRF3_SX}--\ref{Fig:ICRF3_Ka} show the sky distribution and histogram of position uncertainties for the three catalogs. Not unexpectedly, the sky distribution for the S/X band and K band catalogs (left-hand panels in Figs.~\ref{Fig:ICRF3_SX} and \ref{Fig:ICRF3_K}) has a deficiency of sources south of about $-40^{\circ}$ declination (corresponding to the VLBA southern observing limit). The reason for this deficiency is that the VLBI stations able to observe further south are sparse. At the same time, the sky distribution for the X/Ka~band catalog does not show such a division (see left-hand panel in Fig.~\ref{Fig:ICRF3_Ka}). With only four sites, the VLBI network used at this frequency band is not tailored to densifying the frame (as opposed to the VLBA at the two other frequency bands), hence the more uniform (although not as dense) sky distribution at X/Ka band. The histograms of position uncertainties in the right-hand panels of Figs~\ref{Fig:ICRF3_SX}--\ref{Fig:ICRF3_Ka} show that the distributions of uncertainties for both the K~band and X/Ka~band catalogs peak at about 50--60~$\mu$as in right ascension and 80--90~$\mu$as in declination, while at S/X~band the peaks are near 100~$\mu$as for right ascension and 200~$\mu$as for declination. These numbers are in line with the median uncertainties reported in Table~\ref{Tab:ICRF3_features}. The fact that the uncertainties in declination are a factor of 1.5--2.0 larger than those in right ascension finds its origin in the geometry of the VLBI observing networks which have longer east-west than north-south baselines (see Sect.~\ref{Sec:data}). In Fig.~\ref{Fig:ICRF3_SX}, the histogram of position uncertainties also reveals that the S/X~band distribution of uncertainties has a secondary peak. This peak is just above 30~$\mu$as, at the noise floor, and captures the block of roughly 500~sources that have the most precise positions. It must be noted that no sources with VLBA-only observations are present in this block, meaning that they all have been observed as part of IVS programs (possibly jointly with the VLBA). In all, there are 503~sources in the pool of sources observed by the IVS that have a more precise position than any of the sources observed solely with the~VLBA. For the latter, the position uncertainty (defined as the semi-major axis of the error ellipse in position) is 64~$\mu$as at best and thus does not reach the noise floor. Looking at the other end of the distribution (i.e., where the least precise source positions are found), there are 359~sources with position uncertainty worse than 1~mas (corresponding to 8\% of the S/X~band catalog), including 22 sources (0.5\% of the catalog) that have position uncertainties worse than 10~mas. These numbers contrast with those for ICRF2, where 1428~sources (42\%~of the catalog) had position uncertainties worse than 1~mas, 204~of which (6\%~of the catalog) were found to have position uncertainties worse than 10~mas. The K~band and X/Ka~band frames contain an even smaller portion of sources with less precise positions -- 26 sources at K~band (3\%~of the catalog) and 17~sources at X/Ka~band (2.5\%~of the catalog) have position uncertainties worse than 1~mas, while only a handful of these have position uncertainties worse than 10~mas (one source at K~band and five sources at~X/Ka~band). Based on the color coding of the position uncertainties in Figs.~\ref{Fig:ICRF3_SX}--\ref{Fig:ICRF3_Ka} (left-hand panels), one also sees that the distribution of uncertainties on the sky is not uniform. Overall, the southern sources have generally less precise positions than the northern ones. This is reflected in Fig.~\ref{Fig:dec_errors} which shows how the median position uncertainty varies as a function of declination in the three catalogs. While roughly stable at the highest declinations ($>40^{\circ}$), the S/X~band median position uncertainty degrades regularly when declination goes from $40^{\circ}$ to $-45^{\circ}$, after which it improves again toward $-90^{\circ}$~declination. The K~band and X/Ka~band catalogs show similar properties, with median position uncertainty degrading from $90^{\circ}$ to $-50^{\circ}$ declination and improving further south, from $-50^{\circ}$ to $-90^{\circ}$ declination.

   \subsection{Selection of defining sources}
   \label{Sec:ICRF3-defining}
   Taking advantage of the much extended VLBI data set now available and acknowledging the fact that some of the ICRF2 defining sources were found to be no longer suitable as defining sources (see Sect.~\ref{Sec:variations}), we decided to select a new set of defining sources for ICRF3 based on specific criteria and not considering the set of ICRF2 defining sources as a starting set. For this purpose, three criteria were put forward: (i)~the overall sky distribution of the ICRF3 defining sources, (ii)~the position stability of the individual sources, and (iii)~the compactness of their structures. While the second and third criteria were already considered for ICRF2, the first one (sky distribution) was not a major element of selection. The sole consideration in this regard consisted in splitting the celestial sphere into five declination bands and arranging for a similar number of defining sources to be selected in each band. This resulted in a distribution of the ICRF2~defining sources on the sky not fully optimum although covering the entire sky. Our goal this time was to work toward a set of ICRF3 defining sources that has a uniform distribution on the sky.

   The scheme used to achieve a uniform source distribution consisted in dividing the celestial sphere into equivalent sectors and identifying the most suitable source in each sector based on criteria (ii) and (iii) above. The number of sectors was chosen as a compromise between having more defining sources and having a sufficient number of sources in each sector to identify at least one source with the required properties in terms of position stability and source structure. The choice was made for a total of 324 sectors, splitting right ascensions into 18 equal angular sections, each 1~h 20~min wide, while the z-axis, which joins the southern and northern poles (at $-90^{\circ}$ and $+90^{\circ}$ declination, respectively), was also cut into 18 equal segments. It must be noted that the latter implies that the separation of the declination bands increases toward the poles. All sources within each sector that had observations at S/X~band in at least 20~sessions were then extracted and ranked according to positional stability. For this ranking, we followed the method employed for ICRF2~\citep[see][]{Fey2015} and defined the source position stability as
   \begin{equation}
    s=\sqrt{\mathrm{wrms_{\alpha\cos\delta}^2}\ \ \chi_{\nu,\hspace{1.5pt}\alpha\cos\delta}^2 + \mathrm{wrms_{\delta}^2}\ \ \chi_{\nu,\hspace{1.5pt}\delta}^2}\ ,
    \label{Eq:ranking}
   \end{equation}
   where $\mathrm{wrms_{\alpha\cos\delta}}$ and $\mathrm{wrms_{\delta}}$ are the weighted rms of the coordinate variations about the weighted mean coordinates for right ascension and declination, respectively, and $\chi_{\nu,\hspace{1.5pt}\alpha\cos\delta}^2$ and~$\chi_{\nu,\hspace{1.5pt}\delta}^2$ are the reduced chi-squared of the fit to the mean coordinates. The coordinate time series used in the calculation were obtained from four different analysis solutions, one where the positions of the ICRF2 defining sources were solved globally (i.e., estimated once from the entire data set), while the positions for the rest of the sources were estimated session-wise, and three others where the positions for one-third of the ICRF2 defining sources, selected in turn, were estimated session-wise, while the positions for all other sources were solved globally. In this scheme, no-net-rotation constraints were applied either to the full set of ICRF2 defining sources (in the first case) or only to two-thirds of it (corresponding to the portion of defining sources that was solved globally in the three other solutions), in order to fix the orientation of the frame. Following this ranking, the VLBI morphology of all extracted sources was closely examined to assess their compactness and hence suitability as defining sources. This was achieved through visual inspection of the multi-epoch BVID images available, supplemented with those from the Radio Reference Frame Image Database\footnote{Accessible online at \url{https://www.usno.navy.mil/USNO/astrometry/vlbi-products/rrfid}.}, looking in particular for structural variations, and by checking source structure indices and their variability with time (as available from~BVID). Purpose-made VLBI images of sources in the Southern Hemisphere were also used to supplement the material from those two databases. Based on this assessment, the sources within each sector were separated into three categories, those that show minimal structure, usually qualified as point-like or quasi point-like (category~A), those that show moderately extended structures (category~B), and those that show extended or very extended structures (category~C). From that categorization, the selection of the defining sources was then performed by choosing the top-ranked source from category A within each sector or in the case that there is no such source the top-ranked source from category~B.

   \begin{figure}
   \centering
   \includegraphics[width=1.0\columnwidth, trim=10 95 10 100, clip]{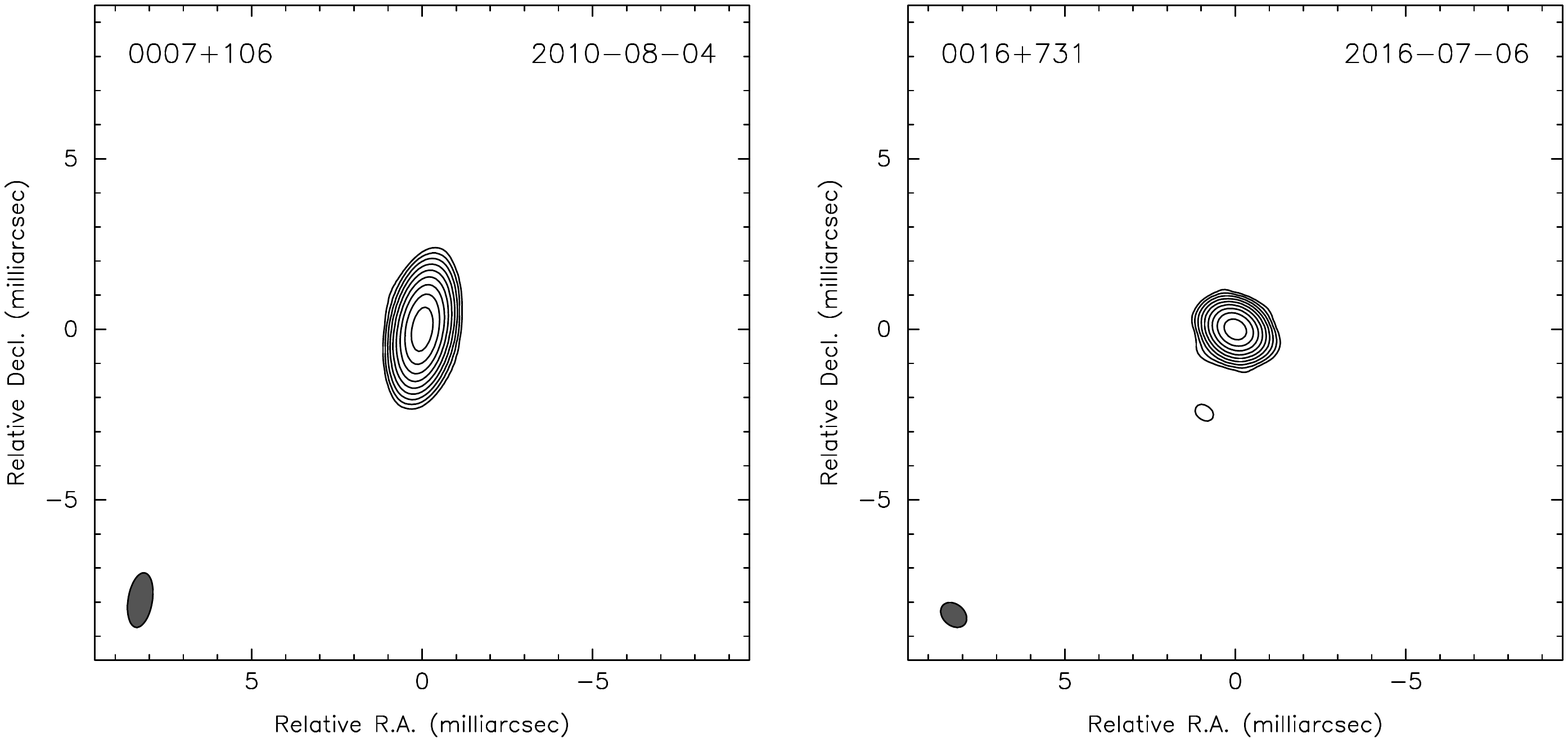}
   \caption{VLBI maps at X band of two ICRF3 defining sources with structural morphology in category A. Contour levels are drawn at $\pm0.25$, 0.5, 1, 2, 4, 8, 16, 32, and 64$\%$ of the peak brightness. The labels in each plot specify the source name and the epoch of the observations. These maps (retrieved from BVID) were made from data from RDV sessions.}
   \label{Fig:ICRF3_def_catA}
   \end{figure}

   \begin{figure}
   \centering
   \includegraphics[width=1.0\columnwidth, trim=10 95 10 100, clip]{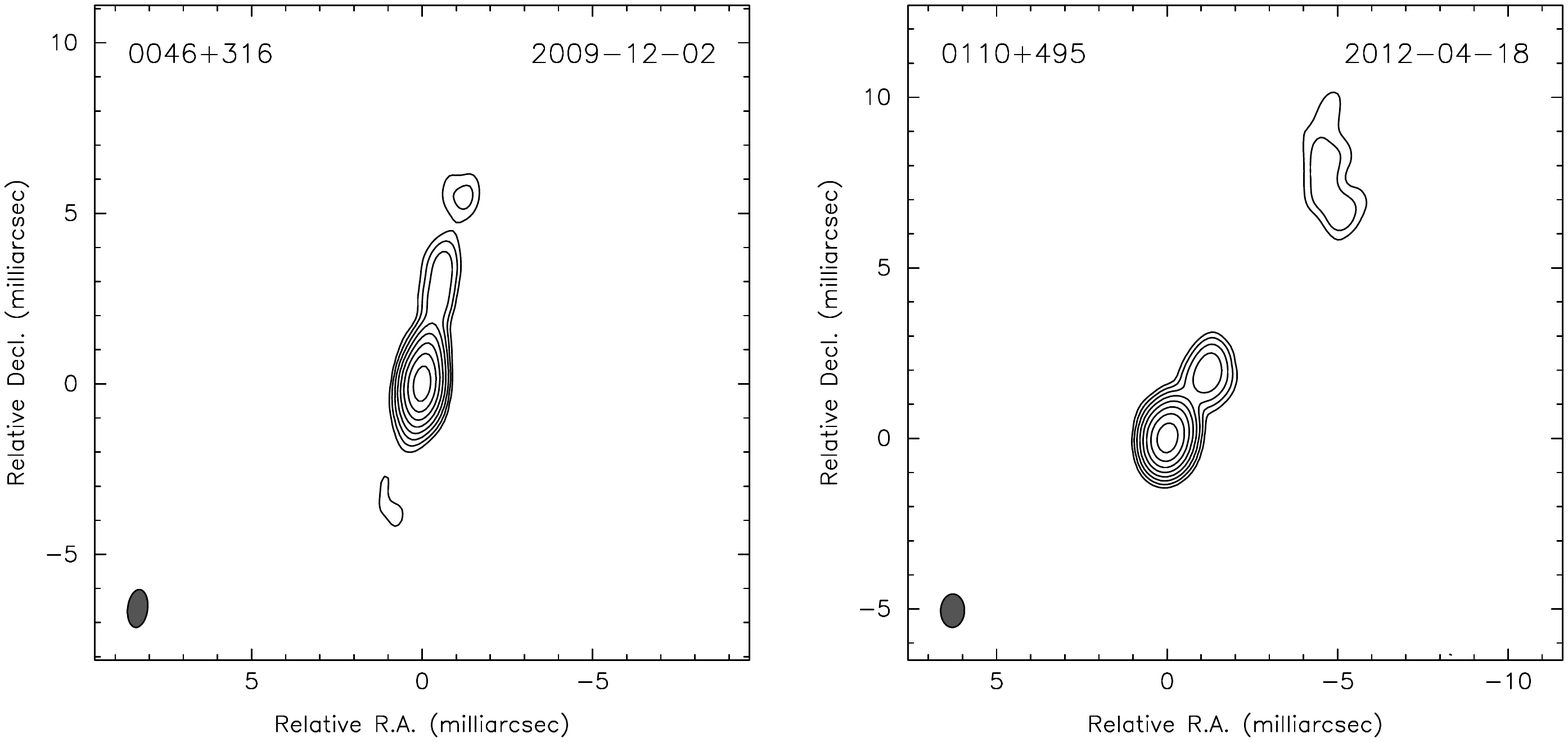}
   \caption{VLBI maps at X band of two ICRF3 defining sources with structural morphology in category B. Contour levels are drawn at $\pm0.5$, 1, 2, 4, 8, 16, 32, and 64$\%$ of the peak brightness. The labels in each plot specify the source name and the epoch of the observations. These maps (retrieved from BVID) were made from data from RDV sessions.}
   \label{Fig:ICRF3_def_catB}
   \end{figure}

   \begin{figure}
   \centering
   \includegraphics[width=1.0\columnwidth, trim=10 95 10 100, clip]{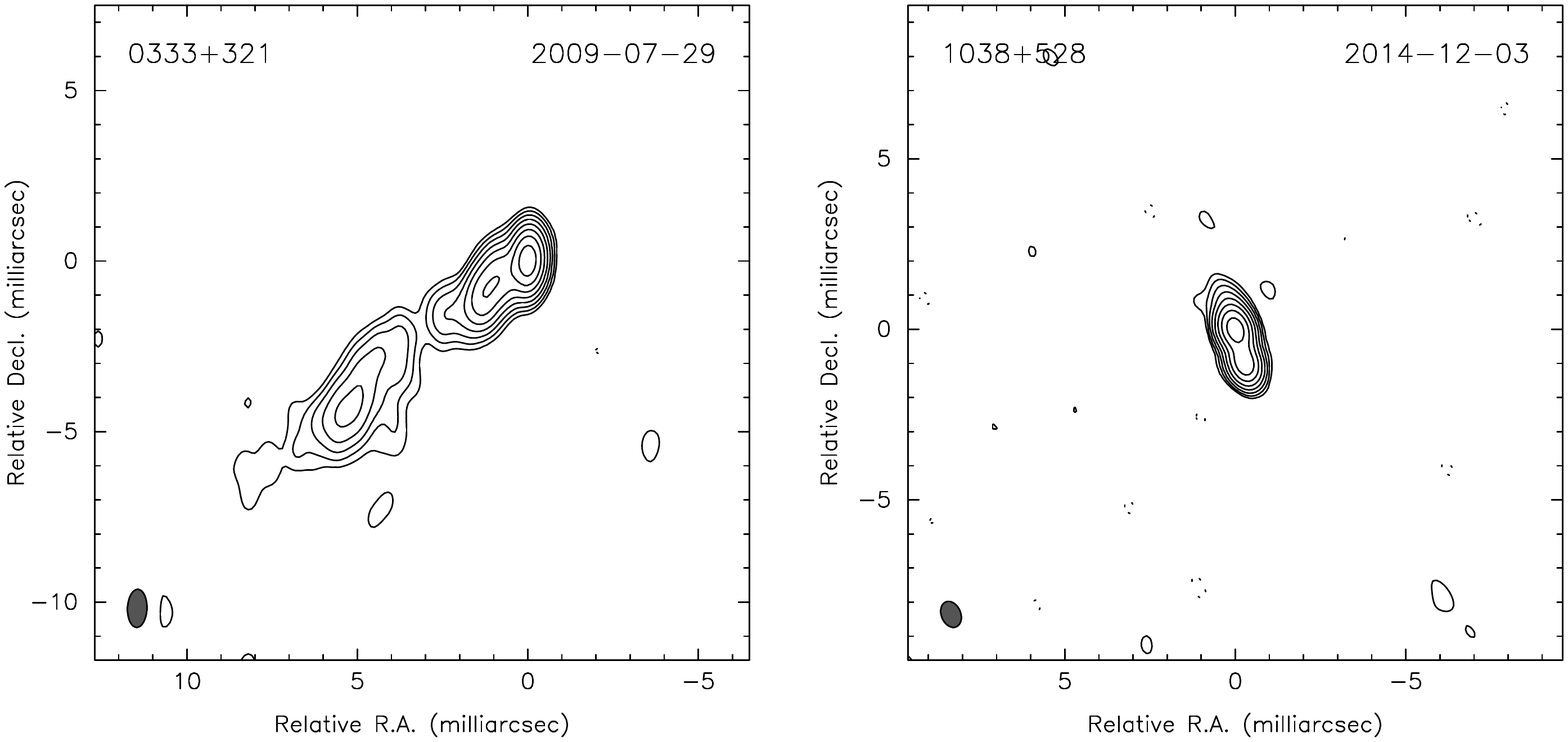}
   \caption{VLBI maps at X band of two ICRF3 sources with structural morphology in category C. Contour levels are drawn at $\pm0.5$, 1, 2, 4, 8, 16, 32, and 64$\%$ of the peak brightness. The labels in each plot specify the source name and the epoch of the observations. It should be emphasized that the source pictured in the right-hand panel has a double morphology with two closely-spaced components. These maps (retrieved from BVID) were made from data from RDV sessions.}
   \label{Fig:ICRF3_def_catC}
   \end{figure}

   \begin{figure}[t]
   \centering
   \includegraphics[width=1.00\columnwidth, trim=0 10 30 0, clip]{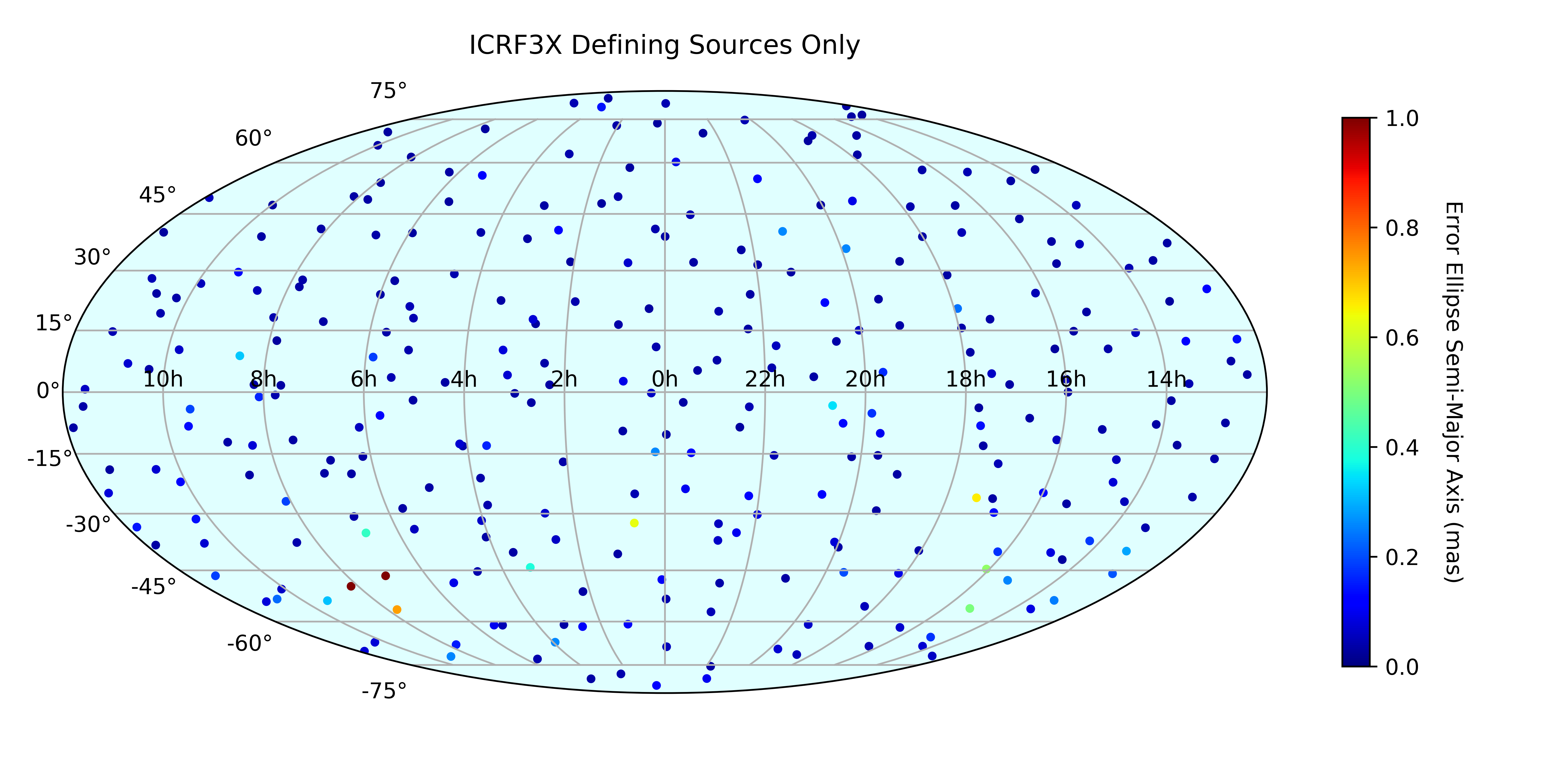}
   \caption{Sky distribution of the 303 ICRF3 defining sources. Each source is plotted as a dot color-coded according to its position uncertainty in the S/X band frame (where the position uncertainty is defined as the semi-major axis of the error ellipse in position).}
   \label{Fig:ICRF3_SX_def}
   \end{figure}

   \begin{figure}
   \centering
   \includegraphics[width=0.85\columnwidth, trim=0 0 0 20, clip]{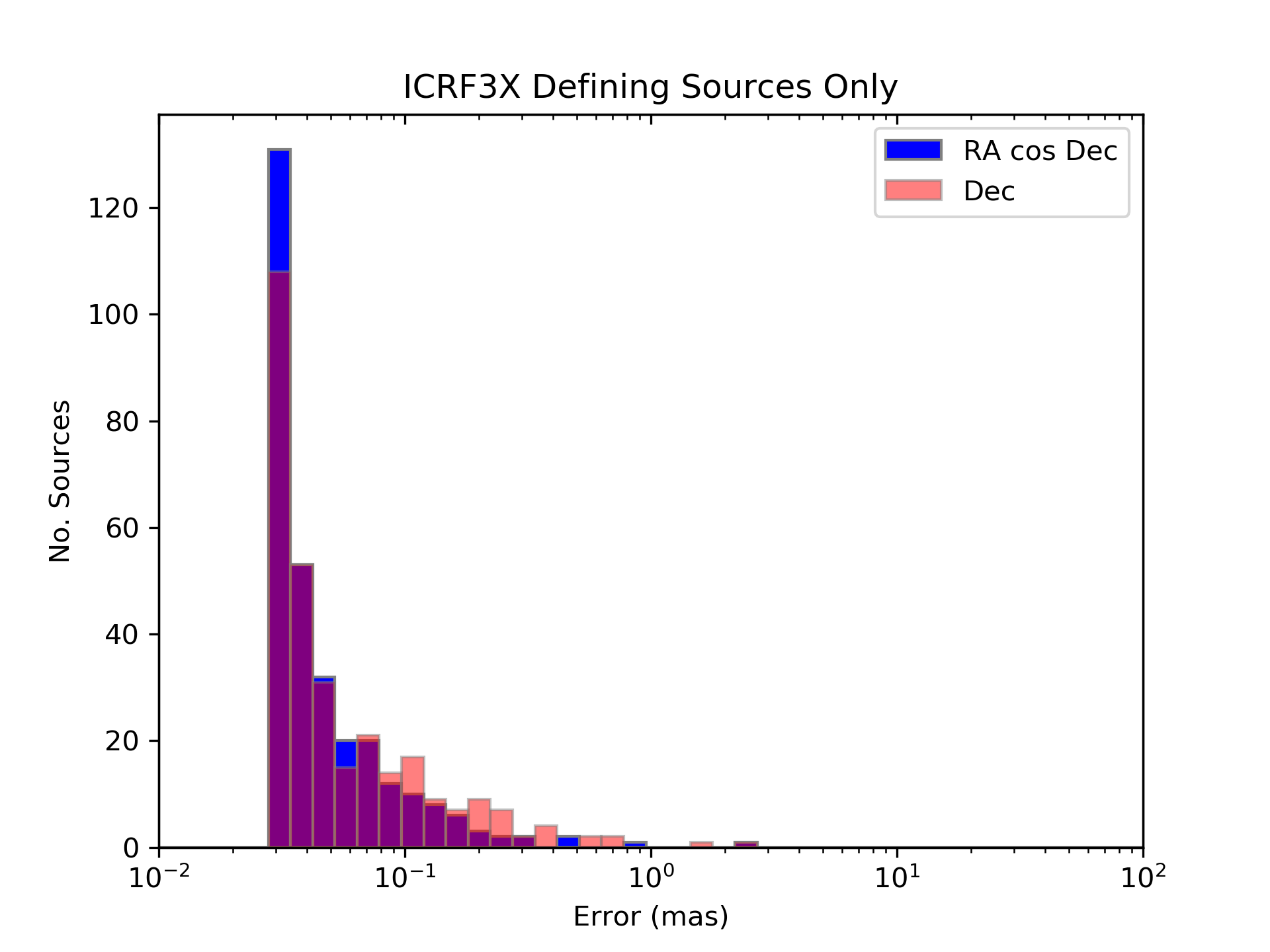}
   \caption{Distribution of coordinate uncertainties of the 303~ICRF3 defining sources at S/X band. Right ascension is shown in blue while declination is shown in salmon. The superimposed portion of the two distributions is shown in purple.}
   \label{Fig:ICRF3_SX_def_errors}
   \end{figure}

   Following the above scheme, an initial pool of 702 sources having observations in at least 20 sessions in the S/X band catalog was identified as a subset of potential defining sources. These sources cover 322 sectors, thus leaving two sectors empty, while each of the other (non-empty) sectors includes between one and eight sources. Moving further with the source categorization, 216~sectors were found to contain at least one source from category~A, while 62~sectors had only category~B (and~C) sources and 19~sectors had only category~C sources. In the remaining 25~sectors, source structure could not be assessed because of the lack of images. The choice was then made to leave out the 19~sectors with only category~C sources since these have poor astrometric quality due to their having extended VLBI structures and cannot be deemed suitable as defining sources. On the other hand, the top-ranked sources in the 25~sectors where no images were available were kept in since there was no reason to assume that the structure of those sources would be inadequate. In all, this leaves a total of 303 defining sources where 216 of these (72\%) have either good or excellent astrometric suitability (i.e., are category~A sources) and 62~others (20\%) have acceptable (if not ideal) quality (corresponding to category~B sources), the astrometric suitability of the 25~remaining sources (8\%) being unknown. See Table~\ref{Tab:empty_sectors} below for the identification of the 21~sectors where no suitable defining source was found and Tables~\ref{Tab:no_images} and~\ref{Tab:category_B_sectors} for the identification of the defining sources for which structure was either not assessed or found to fall into category~B.
   Figures~\ref{Fig:ICRF3_def_catA} and~\ref{Fig:ICRF3_def_catB} provide examples of images at X~band for category~A and category~B defining sources. For comparison purposes, examples of images at X~band of category~C sources (excluded from the algorithm of selection for the defining sources) are shown in Fig.~\ref{Fig:ICRF3_def_catC}. Of interest is that among the 303~defining sources, 246~were ranked first (within their sector) in terms of source position stability, while 40~were ranked second, 12~were ranked third, four were ranked fourth, and one was ranked fifth, hence indicating good overall consistency between position stability and source compactness. In terms of observing characteristics, the vast majority of the selected sources fall into the pool of sources observed by the IVS -- only 21 of them have observations that were conducted solely with the VLBA.

   \begin{table*}[t]
   \caption{Number of ICRF3 defining sources at each frequency band and statistics about coordinate uncertainties, correlation coefficients between right ascension and declination, and the error ellipse size (semi-major axis) for these sources. All statistics are given as median values of the said parameters and are provided
   for all defining sources in each catalog and for the 171 defining sources common to the three catalogs.}
   \begin{center}
   \begin{tabular}{c@{\hskip 8pt}r@{\hskip 3pt}rrr@{{\hskip 18pt}}c@{\ \ }c@{\ }c@{\hskip 9pt}r@{\ \ }r@{\ \ \ \ }c@{\ \ }r}
   \hline
   \hline
   \noalign{\smallskip}
            &&& \multicolumn{4}{c}{\small Statistics for all defining sources}&& \multicolumn{4}{c}{\small Statistics for the common defining sources}\\
   \noalign{\smallskip}
   \cline{4-7}\cline{9-12}
   \noalign{\smallskip}
   \small Frequency  & \small Number of && \multicolumn{2}{c}{\small Coordinate uncertainty}&\small Correlation &\small {\hskip 1pt}Error ellipse{\hskip 2pt} && \multicolumn{2}{c}{\small Coordinate uncertainty}&\small Correlation & \small {\hskip 1pt}Error ellipse\\
   \small band       & \small sources{ \hskip 4pt} && \small {\hskip 8pt} $\alpha\cos\delta$ & \small $\delta${\hskip 6pt} & \small coeff. &\small ($\mu$as)&& \small {\hskip 14pt} $\alpha\cos\delta$  & \small $\delta${\hskip 14pt} & \small coeff. &\small ($\mu$as) {\hskip 13.3pt} \\
   &&&\small ($\mu$as) {\hskip 1pt} & \small ($\mu$as)&&&& \small ($\mu$as) {\hskip 1pt} & \small ($\mu$as){\hskip 8pt} \\
   \noalign{\smallskip}
   \hline
   \noalign{\smallskip}
   \small S/X  & \small  303 {\hskip 8pt} && \small   36 {\hskip 3pt} & \small  41 {\hskip 0pt} & \small 0.07& \small {\hskip 2pt} 41&& \small 33 {\hskip 3pt} & \small  35 {\hskip 8pt} & \small 0.06&\small  35 {\hskip 16.3pt} \\
   \small K    & \small  193 {\hskip 8pt} && \small   63 {\hskip 3pt} & \small 120 {\hskip 0pt} & \small 0.28& \small 122&& \small 60 {\hskip 3pt} & \small 113 {\hskip 8pt} & \small 0.27&\small 115 {\hskip 16.3pt} \\
   \small X/Ka & \small  176 {\hskip 8pt} && \small   62 {\hskip 3pt} & \small  92 {\hskip 0pt} & \small 0.40& \small {\hskip 2pt} 98&& \small 62 {\hskip 3pt} & \small  92 {\hskip 8pt} & \small 0.40&\small  97 {\hskip 16.3pt} \\
   \hline
   \end{tabular}
   \end{center}
   \label{Tab:ICRF3_def_errors}
   \end{table*}

   As expected, the sky distribution of the defining sources thus selected ends up being fairly uniform due to the scheme~adopted for this selection (see Fig.~\ref{Fig:ICRF3_SX_def}). Looking at the distribution of coordinate uncertainties in Fig.~\ref{Fig:ICRF3_SX_def_errors}, it is striking that the defining sources, in their vast majority, show very precise positions despite this quantity not being used as a selection criterion. It is also notable that the histograms of uncertainties in right ascension and  declination are superimposed, a situation that differs from that observed when considering the entire S/X~band catalog where the peaks for the uncertainties in right ascension and declination are shifted relative to one another (see Fig.~\ref{Fig:ICRF3_SX}). These specificities are further reflected by the values of the corresponding median uncertainties which are close to the noise floor and similar for right ascension and declination (36~$\mu$as and 41~$\mu$as, respectively), as reported in Table~\ref{Tab:ICRF3_def_errors}. Correlation coefficients between right ascension and declination have a median value that is even smaller than when computed for the entire catalog (0.07 vs 0.13), therefore indicating very weak correlation between the two coordinates in general.
   At K~band and X/Ka~band, there are no such striking differences between the defining sources and the rest of the catalog sources. While median uncertainties appear to be slightly better for the defining sources (by about 10--15\%), the magnitude of these uncertainties remains worse by a factor of two to three when compared to the S/X~band median uncertainties (see Table~\ref{Tab:ICRF3_def_errors}). Median correlation coefficients do not show any significant differences either, whether computed for all catalog sources (Table~\ref{Tab:ICRF3_features}) or solely for the subset of defining sources (Table~\ref{Tab:ICRF3_def_errors}). The absence of apparent specificities for the defining sources at K~band or X/Ka~band is not unexpected since the algorithm for selecting the defining sources was only tailored to the properties of the sources and configuration of the observations at S/X~band. This finding may also reflect the fact that the sources tend to be more compact at higher frequencies \citep{Charlot2010}, thus mitigating the differences observed at S/X band (in terms of structural properties) between the defining and non-defining sources, or simply biases in the selection of the sources targeted for observation at K~band and X/Ka~band.

\subsection{Practical use of the frame}

   The coordinates of the 4588 sources comprised in ICRF3 along with their uncertainties are given in Tables~\ref{Tab:ICRF3_SX_coordinates}--\ref{Tab:ICRF3_XKa_coordinates}. Table~\ref{Tab:ICRF3_SX_coordinates} is for the S/X~band frame, Table~\ref{Tab:ICRF3_K_coordinates} is for the K~band frame, and Table~\ref{Tab:ICRF3_XKa_coordinates} is for the X/Ka~band frame. Besides source coordinates, the three tables also include proper information to identify each source (ICRF designation and IERS name) and details about the VLBI sessions (first and last sessions in which a source was observed, mean epoch of the sessions, number of sessions), the observations (number of VLBI delays and delay rates used to estimate the source position), and the characteristics of the errors (correlation coefficient between right ascension and declination). The ICRF3 defining sources are also identified in these tables, as are those sources that were observed solely with the VLBA. The total number of sources in this condition (i.e., with VLBA-only data) is 3084 for the S/X~band frame and 544 for the K~band frame. For the sources that have been imaged, indicators about their astrometric suitability (i.e., source structure indices and compactness) are available from the BVID database.
   As supplemental information, the basic optical characteristics of most ICRF3 sources may be found in the Optical Characteristics of Astrometric Radio Sources catalog \citep{Malkin2018}.

\begin{sidewaystable*}
\small
\caption{Coordinates at epoch 2015.0 for the 4536 sources included in the S/X band frame. The coordinates reported in the table may be propagated at other epochs using Eqs.~\eqref{Eq:GA_alpha} and~\eqref{Eq:GA_delta}.}
\begin{tabular}{c@{\hskip 8pt}l@{}c@{\hskip 5pt}l@{\hskip 5pt}r@{\hskip 10pt}r@{\hskip 16pt}rr@{}c@{\ \ \ \ \ }r@{\hskip 9pt}ccc@{}c@{\hskip 10pt}rr@{\hskip 17pt}r}
\hline
\hline
\noalign{\smallskip}
\multicolumn{2}{c}{Source identification}
&&Cat.\tablefootmark{c}
&\ \hfill Right ascension\hfill\
&\ \hfill Declination\hfill\
&\multicolumn{2}{l}{Coordinate uncertainty}
&&Correl.\tablefootmark{d}
&\multicolumn{3}{c}{Epoch of sessions\tablefootmark{\ e}}
&&\multicolumn{3}{c}{Observations\tablefootmark{\ f}}
\\
\cline{1-2}\cline{7-8}\cline{11-13}\cline{15-17}
\noalign{\smallskip}
ICRF designation\tablefootmark{\ a} &
IERS name\tablefootmark{\ b} &
&&
\ \hfill (h\ \ m\ \ s)\hfill\  &
\ \hfill ($\degr$\ \ \ $\arcmin$\ \ \ $\arcsec$)\hfill\  &
\ \hfill (s)\hfill\ &
\ \hfill ($\arcsec$)\hfill\ &&
&
Mean &
First &
Last &
&
$N_{\rm ses}$ &
$N_{\rm del}$ &
$N_{\rm rat}$
\\
\noalign{\smallskip}
\hline
\noalign{\smallskip}
ICRF J000020.3$-$322101 & {\hskip 4pt}2357$-$326 && {\hskip 6pt}V & 00 00 20.39997606 & $-32$ 21 01.2337415 & 0.00000804 & 0.0002624 && $-$0.0602\ \ \ & 56559.8 & 52306.7 & 57776.0 &&    4 &   237 &    0\\
ICRF J000027.0$+$030715 & {\hskip 4pt}2357$+$028 &&               & 00 00 27.02251377 &    03 07 15.6463606 & 0.00005931 & 0.0003421 && $-$0.0119\ \ \ & 57974.7 & 57974.7 & 57974.7 &&    1 &    28 &    0\\
ICRF J000053.0$+$405401 & {\hskip 4pt}2358$+$406 && {\hskip 6pt}V & 00 00 53.08106320 &    40 54 01.8096518 & 0.00001504 & 0.0002670 && $-$0.1654\ \ \ & 56460.2 & 50242.8 & 57809.9 &&    3 &   152 &    0\\
ICRF J000105.3$-$155107 & {\hskip 4pt}2358$-$161 && {\hskip 6pt}V & 00 01 05.32873479 & $-15$ 51 07.0752302 & 0.00000702 & 0.0002261 && $-$0.2106\ \ \ & 56338.4 & 50632.3 & 58137.6 &&    4 &   316 &    0\\
ICRF J000107.0$+$605122 & {\hskip 4pt}2358$+$605 && {\hskip 6pt}V & 00 01 07.09981547 &    60 51 22.7980875 & 0.00003378 & 0.0001948 &&    0.1619\ \ \ & 57160.2 & 52306.7 & 57836.8 &&    3 &   172 &    0\\
ICRF J000108.6$+$191433 & {\hskip 4pt}2358$+$189 &&               & 00 01 08.62156616 &    19 14 33.8017136 & 0.00000260 & 0.0000472 && $-$0.0314\ \ \ & 55771.9 & 50085.5 & 58205.8 &&  168 &  3584 &    0\\
\hline
\label{Tab:ICRF3_SX_coordinates}
\end{tabular}
\vspace{-11pt}
\tablefoot{The content printed here corresponds to the entries for the first six sources. The table in its entirety is available in electronic form at the CDS
via anonymous ftp to cdsarc.u-strasbg.fr (130.79.128.5) or via \url{http://cdsweb.u-strasbg.fr/cgi-bin/qcat?J/A+A/}.
The meaning of the footnotes is the following:}\\
\tablefoottext{a}{The ICRF designations were derived from the source coordinates with the format ICRF JHHMMSS.s$+$DDMMSS or ICRF JHHMMSS.s$-$DDMMSS, according to the original ICRF prescriptions.}\\
\tablefoottext{b}{The complete format for the IERS names, constructed from previous B1950.0 coordinates, includes acronym and epoch and is IERS BHHMM$+$DDd or IERS BHHMM$-$DDd.}\\
\tablefoottext{c}{The ICRF3 defining sources are identified with a ``D'' in this column. Additionally, the sources observed solely with the VLBA are marked with a ``V''.}\\
\tablefoottext{d}{The value given in this column indicates the correlation coefficient between the estimated right ascension and declination coordinates.}\\
\tablefoottext{e}{The values given in these three columns are expressed as Modified Julian Date (MJD), i.e., JD $-$ 2\,400\,000.5.}\\
\tablefoottext{f}{The three columns indicate the number of sessions in which the source was observed ($N_{\rm ses}$) and the number of delay and delay rate observations used to estimate its coordinates ($N_{\rm del}$ and $N_{\rm rat}$).}
\\
\smallskip
\smallskip
\small
\caption{Coordinates at epoch 2015.0 for the 824 sources included in the K band frame. The coordinates reported in the table may be propagated at other epochs using Eqs.~\eqref{Eq:GA_alpha} and \eqref{Eq:GA_delta}.}
\begin{tabular}{c@{\hskip 8pt}l@{}c@{\hskip 5pt}l@{\hskip 5pt}r@{\hskip 10pt}r@{\hskip 16pt}rr@{}c@{\ \ \ \ \ }r@{\hskip 9pt}ccc@{}c@{\hskip 10pt}rrr}
\hline
\hline
\noalign{\smallskip}
\multicolumn{2}{c}{Source identification}
&&Cat.\tablefootmark{c}
&\ \hfill Right ascension\hfill\
&\ \hfill Declination\hfill\
&\multicolumn{2}{c}{Coordinate uncertainty}
&&Correl.\tablefootmark{d}
&\multicolumn{3}{c}{Epoch of sessions\tablefootmark{\ e}}
&&\multicolumn{3}{c}{Observations\tablefootmark{\ f}}
\\
\cline{1-2}\cline{7-8}\cline{11-13}\cline{15-17}
\noalign{\smallskip}
ICRF designation\tablefootmark{\ a} &
IERS name\tablefootmark{\ b} &
&&
\ \hfill (h\ \ m\ \ s)\hfill\  &
\ \hfill ($\degr$\ \ \ $\arcmin$\ \ \ $\arcsec$)\hfill\  &
\ \hfill (s)\hfill\ &
\ \hfill ($\arcsec$)\hfill\ &&
&
Mean &
First &
Last &
&
$N_{\rm ses}$ &
$N_{\rm del}$ &
$N_{\rm rat}$
\\
\noalign{\smallskip}
\hline
\noalign{\smallskip}
ICRF J000435.6$-$473619 & {\hskip 4pt}0002$-$478 && D{\hskip 6pt}  & 00 04 35.65553663 & $-$47 36 19.6034797 & 0.00004466 & 0.0010827 &&    0.6589\ \ \ & 57961.2 & 57599.3 & 58067.8 &&    7 &    13 &    13\\
ICRF J000504.3$+$542824 & {\hskip 4pt}0002$+$541 && {\hskip 6pt}V  & 00 05 04.36334104 &    54 28 24.9244365 & 0.00000840 & 0.0000905 &&    0.0477\ \ \ & 57561.2 & 53898.6 & 58181.5 &&    9 &   732 &   732\\
ICRF J000557.1$+$382015 & {\hskip 4pt}0003$+$380 && {\hskip 6pt}V  & 00 05 57.17539180 &    38 20 15.1489842 & 0.00000450 & 0.0000737 && $-$0.1988\ \ \ & 57909.1 & 57372.1 & 58195.8 &&   10 &   724 &   724\\
ICRF J000613.8$-$062335 & {\hskip 4pt}0003$-$066 &&                & 00 06 13.89288203 & $-$06 23 35.3357441 & 0.00000321 & 0.0001061 && $-$0.4903\ \ \ & 57819.0 & 56782.4 & 58181.5 &&    9 &   658 &   658\\
ICRF J000903.9$+$062821 & {\hskip 4pt}0006$+$061 && {\hskip 6pt}V  & 00 09 03.93185280 &    06 28 21.2397824 & 0.00000772 & 0.0002257 && $-$0.3528\ \ \ & 57837.4 & 52782.5 & 58195.8 &&    7 &   316 &   316\\
ICRF J001031.0$+$105829 & {\hskip 4pt}0007$+$106 && DV & 00 10 31.00590225 &    10 58 29.5043805 & 0.00000325 & 0.0000920 && $-$0.3330\ \ \ & 55132.2 & 52782.5 & 58244.2 &&   12 &  1043 &  1043\\
\hline
\label{Tab:ICRF3_K_coordinates}
\end{tabular}
\vspace{-11pt}
\tablefoot{The content printed here corresponds to the entries for the first six sources. The table in its entirety is available in electronic form at the CDS
via anonymous ftp to cdsarc.u-strasbg.fr (130.79.128.5) or via \url{http://cdsweb.u-strasbg.fr/cgi-bin/qcat?J/A+A/}.
See Table~\ref{Tab:ICRF3_SX_coordinates} for the meaning of the footnotes.}
\\
\smallskip
\smallskip
\smallskip
\small
\caption{Coordinates at epoch 2015.0 for the 678 sources included in the X/Ka band frame. The coordinates reported in the table may be propagated at other epochs using Eqs.~\eqref{Eq:GA_alpha} and~\eqref{Eq:GA_delta}.}
\begin{tabular}{c@{\hskip 8pt}l@{}c@{\hskip 5pt}l@{\hskip 5pt}r@{\hskip 10pt}r@{\hskip 16pt}rr@{}c@{\ \ \ \ \ }r@{\hskip 9pt}ccc@{}c@{\hskip 10pt}r@{\hskip 16pt}r@{\hskip 16pt}r}
\hline
\hline
\noalign{\smallskip}
\multicolumn{2}{c}{Source identification}
&&Cat.\tablefootmark{c}
&\ \hfill Right ascension\hfill\
&\ \hfill Declination\hfill\
&\multicolumn{2}{c}{Coordinate uncertainty}
&&Correl.\tablefootmark{d}
&\multicolumn{3}{c}{Epoch of sessions\tablefootmark{\ e}}
&&\multicolumn{3}{c}{Observations\tablefootmark{\ f}}
\\
\cline{1-2}\cline{7-8}\cline{11-13}\cline{15-17}
\noalign{\smallskip}
ICRF designation\tablefootmark{\ a} &
IERS name\tablefootmark{\ b} &
&&
\ \hfill (h\ \ m\ \ s)\hfill\  &
\ \hfill ($\degr$\ \ \ $\arcmin$\ \ \ $\arcsec$)\hfill\  &
\ \hfill (s)\hfill\ &
\ \hfill ($\arcsec$)\hfill\ &&
&
Mean &
First &
Last &
&
$N_{\rm ses}$ &
$N_{\rm del}$ &
$N_{\rm rat}$
\\
\noalign{\smallskip}
\hline
\noalign{\smallskip}
ICRF J000435.6$-$473619 & {\hskip 4pt}0002$-$478 && D{\hskip 6pt} & 00 04 35.65546616 & $-$47 36 19.6047565 & 0.00001340 & 0.0001584 && $-$0.4333\ \ \ & 57086.5 & 56297.6 & 58027.3 &&   23 &    30 &    30\\
ICRF J000504.3$+$542824 & {\hskip 4pt}0002$+$541 &&               & 00 05 04.36338455 &    54 28 24.9241634 & 0.00001033 & 0.0001306 &&    0.1878\ \ \ & 57364.8 & 55304.5 & 58146.7 &&   27 &    54 &    54\\
ICRF J000557.1$+$382015 & {\hskip 4pt}0003$+$380 &&               & 00 05 57.17541201 &    38 20 15.1487480 & 0.00000509 & 0.0000814 && $-$0.3088\ \ \ & 56844.3 & 53561.7 & 58146.9 &&   88 &   126 &   126\\
ICRF J000613.8$-$062335 & {\hskip 4pt}0003$-$066 &&               & 00 06 13.89287585 & $-$06 23 35.3351018 & 0.00000627 & 0.0001293 && $-$0.5731\ \ \ & 56545.8 & 53561.6 & 58146.8 &&   92 &   137 &   137\\
ICRF J000903.9$+$062821 & {\hskip 4pt}0006$+$061 &&               & 00 09 03.93183737 &    06 28 21.2399247 & 0.00000619 & 0.0001190 && $-$0.6048\ \ \ & 56785.3 & 53651.5 & 58146.8 &&   77 &   110 &   110\\
ICRF J001031.0$+$105829 & {\hskip 4pt}0007$+$106 && D{\hskip 6pt} & 00 10 31.00590425 &    10 58 29.5042938 & 0.00000375 & 0.0000883 && $-$0.4731\ \ \ & 56132.3 & 53694.3 & 58146.9 &&  113 &   179 &   179\\
\hline
\label{Tab:ICRF3_XKa_coordinates}
\end{tabular}
\vspace{-11pt}
\tablefoot{The content printed here corresponds to the entries for the first six sources. The table in its entirety is available in electronic form at the CDS
via anonymous ftp to cdsarc.u-strasbg.fr (130.79.128.5) or via \url{http://cdsweb.u-strasbg.fr/cgi-bin/qcat?J/A+A/}.
See Table~\ref{Tab:ICRF3_SX_coordinates} for the meaning of the footnotes.}
\end{sidewaystable*}

   The ICRF3 source coordinates reported in Tables~\ref{Tab:ICRF3_SX_coordinates}--\ref{Tab:ICRF3_XKa_coordinates} are provided for epoch 2015.0. As explained above, these coordinates should be propagated for observations at epochs away from that epoch using a Galactic acceleration amplitude of 5.8~$\mu$as/yr. In practice, this may be accomplished using the formulas
   \begin{align}
   \alpha_t &= \alpha_{t_0} + \Delta\mu_{\alpha} \left(t-t_0\right), \label{Eq:GA_alpha}\\
   \delta_t &= \delta_{t_0} + \Delta\mu_{\delta} \left(t-t_0\right),
   \label{Eq:GA_delta}
   \end{align}
   where ($\alpha_{t_0}$, $\delta_{t_0}$) are the ICRF3 source coordinates (i.e., the coordinates at epoch $t_0=2015.0$), while ($\alpha_t$, $\delta_t$) are the coordinates at epoch $t$. The expressions for the components of the proper motion induced by Galactic acceleration ($\Delta\mu_{\alpha}$, $\Delta\mu_{\delta}$) may be found, for instance, in \citet{MacMillan2019} and are given by
   \begin{align}
   \Delta\mu_{\alpha}\cos\delta &= -A_1\sin\alpha + A_2\cos\alpha, \\
   \Delta\mu_{\delta} &=-A_1\cos\alpha\sin\delta - A_2\sin\alpha\sin\delta + A_3\cos\delta,
   \end{align}
   where the $A_i$ parameters are the barycentric components of the Galactic acceleration vector, scaled by $1/c$ as in Eq.~\eqref{Eq:GA}. These may be expressed as $A_G (\cos\delta_G\cos\alpha_G, \cos\delta_G\sin\alpha_G, \sin\delta_G)$, where ($\alpha_G,\delta_G$) are the equatorial coordinates of the vector direction ($\alpha_G=266.4\degr$, $\delta_G=-29.0\degr$) and $A_G$ is the amplitude of this vector, as defined above ($A_G=5.8$~$\mu$as/yr). It is to be pointed out that propagation of the ICRF3 coordinates at other epochs using the above equations is required only for the most accurate needs. For observations within ten years of the reference epoch of the frame (2015.0), such adjustments may not be necessary unless a positional accuracy better than 100~$\mu$as is desired.

   Apart from such considerations on propagation of the source coordinates, the practical use of ICRF3 also implies that one has to select positions in one of the S/X, K, or X/Ka band catalogs for sources that have positions available at more than one frequency band. In this respect, we recommend that the S/X band positions be used at first since these are on average more accurate than the K band or X/Ka~band positions (see Tables~\ref{Tab:ICRF3_features} and~\ref{Tab:ICRF3_def_errors}). However, when positions at K~band or X/Ka~band are specifically desired, these should be favored over the S/X~band positions. This is the case in particular when ICRF sources are used as calibrators to observe water masers at K~band \citep[e.g.,][]{Immer2013} or as fiducial references for spacecraft navigation at X/Ka band as deep space flights move to Ka~band links \citep{Morabito2017}.

\section{Analysis of ICRF3}
\label{Sec:discussion}

   Aside from internal assessments, comparisons of ICRF3 with independent realizations of the extragalactic frame are essential to control its quality, its alignment onto the ICRS, and potential deformations of the frame. The two independent frames used for this purpose are the predecessor of ICRF3, ICRF2 \citep{Fey2015}, and the recently released Gaia-CRF2 frame in the optical domain \citep{Mignard2018}. Also important is the assessment of the consistency of the individual source positions at the three different ICRF3 frequencies and with optical ones in Gaia-CRF2 since the measured positions may depend at some level on the observed frequency band due to possible physical offsets of the emission in the different bands. All such checks are reported below, including a description of the approach devised for these comparisons, prior to the presentation of the results.

\subsection{Modeling rotations and deformations between catalogs}
\label{Sec:formalism}
   The scheme that we used for comparing catalogs follows that developed by \citet{Mignard2012} and is based on vector spherical harmonics decomposition. In this scheme, the coordinate differences between catalogs are modeled by a transformation that takes the global rotation between the catalogs into account as well as the low-degree deformations. Mathematically, the corresponding coordinate transformation includes three rotations, three ``glide'' parameters that characterize the dipolar deformation of the coordinate field \citep[see][]{Mignard2012}, and ten quadrupole terms representing degree-2 deformations. Adopting the notation of \citet{Titov2013}, the coordinate differences between the catalogs may be expressed as
   \begin{align}
   \Delta\alpha\cos\delta =&\ \ \ \ \, R_1\cos\alpha\sin\delta + R_2\sin\alpha\sin\delta -R_3\cos\delta \nonumber\\
                                &-D_1\sin\alpha+D_2\cos\alpha \nonumber\\
                                &+M_{20}\sin{2\delta} \nonumber\\
                                &+\left(E_{21}^{{\rm Re}}\sin\alpha + E_{21}^{{\rm Im}}\cos\alpha\right)\sin\delta \nonumber\\
                                &-\left(M_{21}^{{\rm Re}}\cos\alpha - M_{21}^{{\rm Im}}\sin\alpha\right)\cos{2\delta} \nonumber\\
                                &-2\left(E_{22}^{{\rm Re}}\sin{2\alpha} + E_{22}^{{\rm Im}}\cos{2\alpha}\right)\cos\delta \nonumber\\
                                &-\left(M_{22}^{{\rm Re}}\cos{2\alpha} - M_{22}^{{\rm Im}}\sin{2\alpha}\right)\sin{2\delta},
                                \label{Eq:VSH_alpha}\\
   \Delta\delta                =&-R_1\sin\alpha + R_2\cos\alpha \nonumber\\
                                &-D_1\cos\alpha\sin\delta - D_2\sin\alpha\sin\delta + D_3\cos\delta \nonumber\\
                                &+E_{20}\sin{2\delta} \nonumber\\
                                &-\left(E_{21}^{{\rm Re}}\cos\alpha - E_{21}^{{\rm Im}}\sin\alpha\right)\cos{2\delta} \nonumber\\
                                &-\left(M_{21}^{{\rm Re}}\sin\alpha + M_{21}^{{\rm Im}}\cos\alpha\right)\sin\delta \nonumber\\
                                &-\left(E_{22}^{{\rm Re}}\cos{2\alpha} - E_{22}^{{\rm Im}}\sin{2\alpha}\right)\sin{2\delta} \nonumber\\
                                &+2\left(M_{22}^{{\rm Re}}\sin{2\alpha} + M_{22}^{{\rm Im}}\cos{2\alpha}\right)\cos\delta,
   \label{Eq:VSH_delta}
   \end{align}
   where the $R_i$ parameters denote the rotations, the $D_i$ parameters denote the glide terms, and the $E_{2i}$ and $M_{2i}$ parameters denote the poloidal (or electric-type) and toroidal (or magnetic-type) quadrupole terms (each of which with a real and an imaginary part). For every pair of catalogs that we have compared (see the following subsections), the 16~parameters of the transformation were obtained by a least-squares fit to the coordinate differences for the common sources. In this fit, the coordinate differences were weighted by the inverse of the sum of the squared coordinate uncertainties in the two catalogs. In order not to bias the determination, the fit was performed after excluding the outliers. As a criterion, a source was considered as an outlier if the angular separation between the measured positions in the two catalogs normalized by its formal uncertainty (a quantity hereafter referred to as normalized separation) is larger than~5. Additionally, all sources with an angular separation larger than 5~mas or with an error ellipse in position with a semi-major axis larger than 5~mas in either catalog were also discarded.

   \subsection{Comparison with ICRF2}
\begin{figure}
   \centering
   \includegraphics[width=0.80\columnwidth, trim=0 15 0 36, clip]{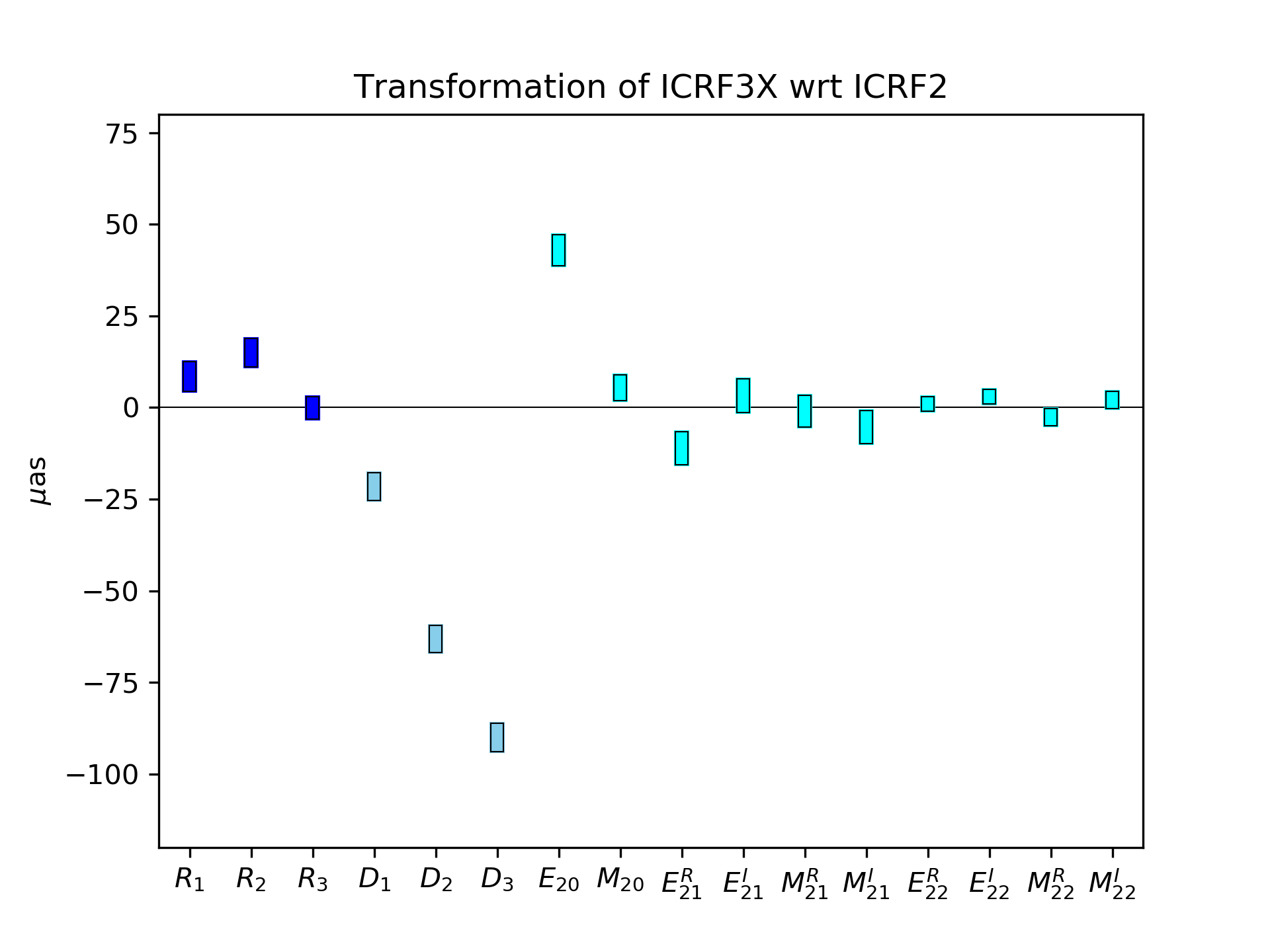}
   \caption{Bar chart showing the values of the 16 parameters of the transformation between the ICRF3 S/X band frame and ICRF2. The $R_i$~parameters are for the rotations, the $D_i$ parameters are for the glide terms, and the $E_{2i}$ and $M_{2i}$ parameters are for the quadrupole terms. See Eqs.~\eqref{Eq:VSH_alpha} and~\eqref{Eq:VSH_delta} for further details on the transformation.}
   \label{Fig:comp_ICRF3X-ICRF2}
   \end{figure}

   The coordinate transformation described in the previous section was first applied to comparing the ICRF3 S/X band frame with its predecessor, ICRF2, also constructed at S/X~band. Since the purpose of the analysis is to assess the deformations between the two frames and not only to check their alignment, the comparison was carried out based on all sources common to the two catalogs and not just the defining sources. After elimination of the outliers, there were a total of 2918 such sources. To facilitate the evaluation of the results, the 16~parameters derived from the fit (i.e., three rotations, three glide parameters, and ten quadrupole terms) are plotted in the form of a bar chart in Fig.~\ref{Fig:comp_ICRF3X-ICRF2}. As expected, the rotations, which reach 15~$\mu$as at most, are small. These cannot be exactly zero because the no-net-rotation constraints imposed to align ICRF3 onto ICRF2 were applied only to the ICRF2 defining sources (see Sect.~\ref{Sec:configuration}). Additionally, the meaning of that alignment in the (new) context of inclusion of Galactic acceleration in the modeling remains somewhat uncertain since ICRF2, unlike ICRF3, did not have a reference epoch. Looking at the other parameters, it is striking that the glide terms stand out, whereas all quadrupole terms but one are not significant. The $D_2$ and $D_3$ glide terms show values of $-63\pm 4$~$\mu$as and $-90\pm 4$~$\mu$as, respectively, well above the expected deformations between the two frames. The only significant quadrupole parameter is the $E_{20}$ term which shows a value of $43\pm 4$~$\mu$as. All the other quadrupole terms are found to be no larger than $10$~$\mu$as.

\begin{figure}
   \centering
   \includegraphics[width=0.80\columnwidth, trim=0 15 0 37, clip]{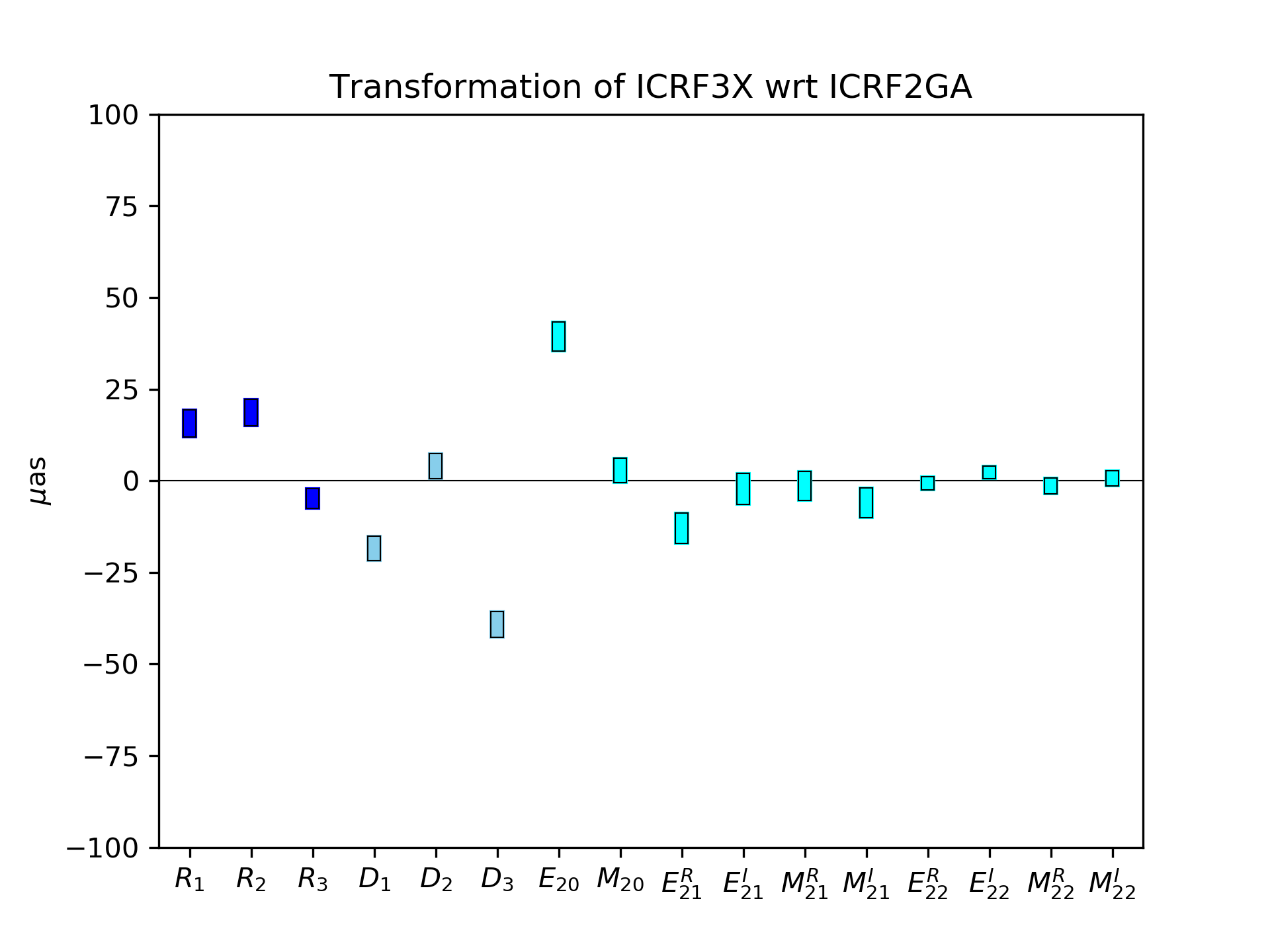}
   \caption{Bar chart showing the values of the 16 parameters of the transformation between the ICRF3 S/X band frame and the reproduced ICRF2, the latter incorporating Galactic acceleration in the modeling. The $R_i$~parameters are for the rotations, the $D_i$ parameters are for the glide terms, and the $E_{2i}$ and $M_{2i}$ parameters are for the quadrupole terms. See Eqs.~\eqref{Eq:VSH_alpha} and~\eqref{Eq:VSH_delta} for further details on the transformation.}
   \label{Fig:comp_ICRF3X-ICRF2_GA}
   \end{figure}

   In order to try to get insights into such systematics, we produced several variants of the ICRF3 S/X band frame by changing the reference epoch of the catalog or alternately by not considering Galactic acceleration in the modeling. Interestingly, the $D_2$ and $D_3$ glide terms for these variants were found to vary by several tens of microarcseconds in the comparison to ICRF2, in line with the level of the systematics observed for those terms. Such findings are not unexpected since Galactic acceleration manifests itself as a dipolar deformation in the source coordinates \citep[e.g.,][]{Titov2013}. Moving further, and noting that \citet{Mignard2016} mentioned this phenomenon as a possibility for explaining the observed glide between ICRF2 and the Gaia Data Release 1 (Gaia DR1) auxiliary quasar solution, we decided to reproduce an equivalent of ICRF2 by considering only the stretch of data used for ICRF2 (i.e., including only the VLBI sessions up to March 2009 in the solution) and to make a variant that adds Galactic acceleration in the modeling, as implemented for ICRF3. To guarantee the maximum consistency, those two analyses were conducted by employing the same software package as that used for ICRF2, namely CALC-SOLVE \citep[see][]{Fey2015}. Looking at the results, we first observed that our ``reproduced'' ICRF2 shows similar deformations as the original ICRF2 when compared to ICRF3, hence ruling out the possibility that ICRF2 was in error. Most importantly, the ICRF2 variant that incorporates Galactic acceleration modeling was found to have much reduced glide terms compared to the original or reproduced ICRF2. This is illustrated by the bar chart in Fig.~\ref{Fig:comp_ICRF3X-ICRF2_GA} which shows that the $D_2$ term has now vanished while the $D_3$ term has been cut by more than half (down to a value of $-39\pm 4$~$\mu$as) in this variant, hence indicating that the deformations between the two frames, in large part, stem from Galactic acceleration not accounted for in ICRF2. This is somehow not surprising since the data set for ICRF2 already covered 30~years, enough for Galactic acceleration effects to emerge, even though the accuracy of the frame was lower than that of~ICRF3.

   The transformation parameters for the above comparisons are reported in full in Table~\ref{Tab:comp_ICRF3X-all} below. Also included in that table are the values of the parameters derived for the additional comparisons that we have accomplished (i.e., relative to the Gaia-CRF2 frame and for the K~band and X/Ka~band frames), the details of which are presented in the following subsections.

\subsection{Comparison with Gaia-CRF2}

   The ICRF3 S/X band frame was compared in a second stage with the Gaia-CRF2 frame, a fully independent frame constructed in the optical domain. For this comparison, a pre-requisite was to identify the sample of sources common to the two frames. While \citet{Mignard2018} made available such a sample, we found it necessary to make an update since the original cross-identification of the sources was based on a prototype version of ICRF3 which differs from ICRF3, the final frame not being available by the time of publication. For our determination, we used the same criteria as those originally employed to identify the sources in common with the ICRF3 prototype \citep{Lindegren2018}. Accordingly, a radius of 0.1\arcsec was used for the positional matching, supplemented by Gaia-specific conditions whose purpose was to reduce the risk of contamination of the quasar sample by Galactic stars. These conditions ensured (i)~that a minimum number of eight field-of-view transits was used for each source, (ii)~that the source astrometric parameters were obtained exclusively from five-parameter solutions (i.e., including position, parallax, and proper motion), and (iii)~that the normalized parallaxes and proper motions are less than 5 \citep[see][Eq.~14]{Lindegren2018}. Applying this scheme, a total of 3373~common sources were identified from the positional matching. The additional Gaia-specific conditions imposed led to discarding 390 of these (90~from condition~(i), 275~from condition (ii), and 25~from condition~(iii)), hence leaving a total of 2983 common sources. This is about 6\% larger than the number of such sources found by \citet{Mignard2018}~when comparing Gaia-CRF2 to the ICRF3 prototype (2983 vs 2820~common sources).

   \begin{figure}[t]
   \centering
   \includegraphics[width=0.80\columnwidth, trim=0 15 0 36, clip]{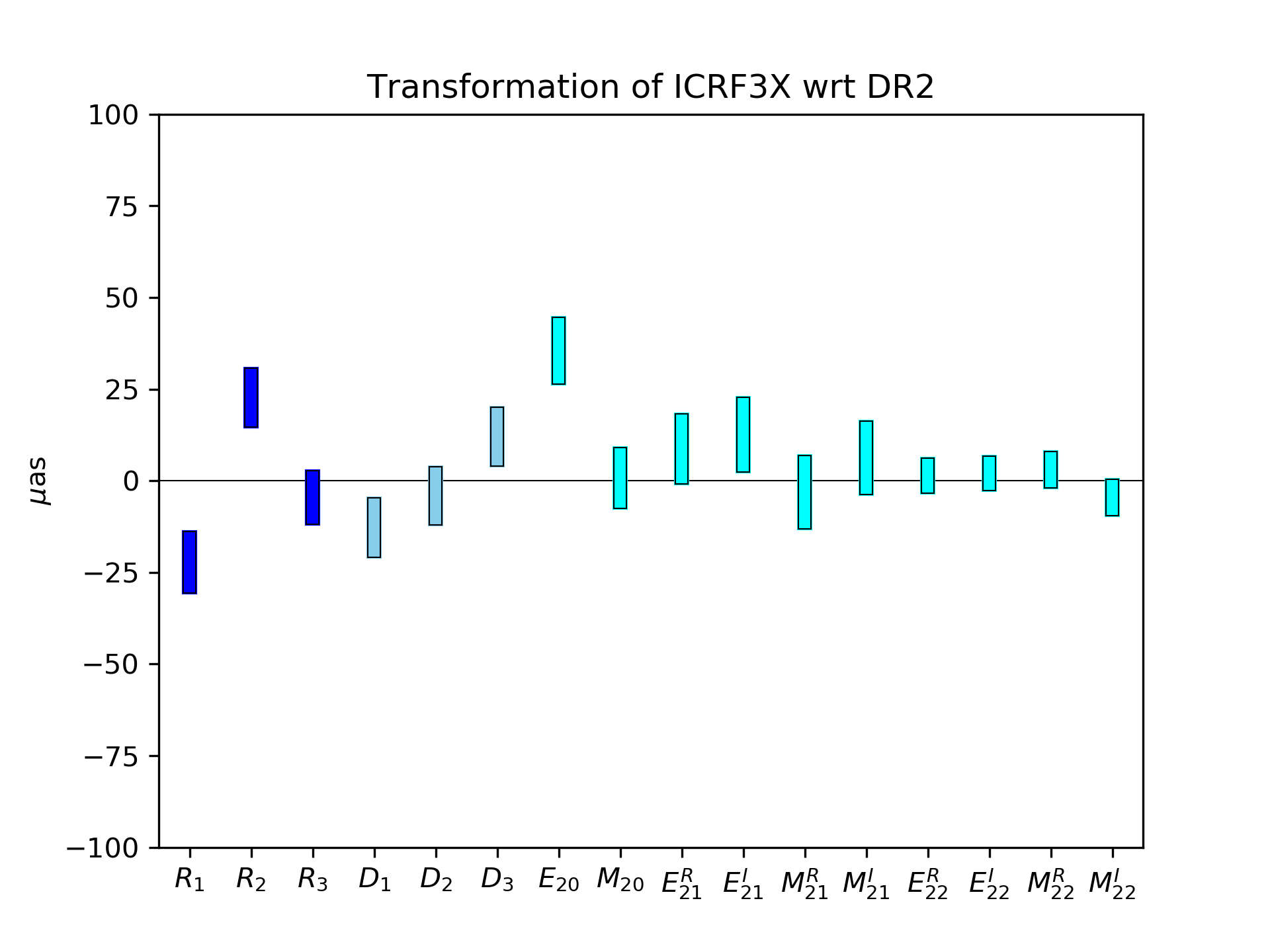}
   \caption{Bar chart showing the values of the 16 parameters of the transformation between the ICRF3 S/X band frame and the Gaia-CRF2 frame. The $R_i$~parameters are for the rotations, the $D_i$ parameters are for the glide terms, and the $E_{2i}$ and $M_{2i}$ parameters are for the quadrupole terms. See Eqs.~\eqref{Eq:VSH_alpha} and~\eqref{Eq:VSH_delta} for further details on the transformation.}
   \label{Fig:comp_ICRF3X-Gaia-CRF2}
   \end{figure}

   The transformation parameters between the two frames were determined by assuming that the epochs of the two frames are the same. Though not identical, the ICRF3 and Gaia-CRF2 epochs are indeed very close, 2015.0 for ICRF3 and 2015.5 for Gaia-CRF2 (mean epoch of the observations). As a result, the 5.8~$\mu$as/yr correction for Galactic acceleration would induce changes in the ICRF3 positions by 3~$\mu$as at most, equivalent to one tenth of the ICRF3 noise floor, a value not regarded as significant. After elimination of the outliers, a total of 2612~sources (out of the initial 2983 common sources) was left for the comparison, corresponding to 87.5\% of the original sample. This means that 12.5\% of the common sources show significantly discrepant radio and optical positions according to the criteria set above, a percentage comparable to that reported by \citet{Mignard2018}.
   The results of the fit for the 16~parameters of the transformation are plotted as a bar chart in Fig.~\ref{Fig:comp_ICRF3X-Gaia-CRF2}, in the same way as previously.
   As further information, the corresponding parameter values are reported in Table~\ref{Tab:comp_ICRF3X-all}. A visual inspection of the bar chart in Fig.~\ref{Fig:comp_ICRF3X-Gaia-CRF2} reveals in the first place that the two frames are very close. In particular, there is no such a dipolar deformation as that seen for the comparison to ICRF2. The rotation parameters are within 25~$\mu$as and have only marginal significance (at most $2.8\sigma$, see the error values in Table~\ref{Tab:comp_ICRF3X-all}), which means that the two frames are reasonably well aligned. The dipolar terms are less than 15~$\mu$as and are non-significant, as are also all of the quadrupole terms but one. The only possibly significant such term is $E_{20}$ (as for the comparison with ICRF2), which has a value of $35\pm 9$~$\mu$as (corresponding to $3.9\sigma$). This term translates into a zonal deformation of the said value, peaking at $\pm 45\degr$~declination (with opposite values in the north and in the south) and vanishing at 0 and $\pm 90\degr$ declination. Such a deformation, if real and inherent to ICRF3, might come from the asymmetry of the VLBI networks which include many more east-west baselines than north-south baselines, as reflected by Fig.~\ref{Fig:ICRF3_network_map}.

   Based on the above comparison, the ICRF3 S/X band frame and Gaia-CRF2 frame were found to be consistent at the 30~$\mu$as level, in agreement with the ICRF3 noise floor, which is also 30~$\mu$as. This contrasts with the comparison to ICRF2 discussed in the previous subsection which revealed relative deformations up to about 100~$\mu$as between the two frames. Considering that ICRF3 and Gaia-CRF2 are fully independent frames, it is likely then that the observed deformations are inherent to ICRF2.

   \subsection{Intercomparison of the three individual catalogs}
   \label{Sec:intercomparison}
   In addition to the comparisons relative to the ICRF2 and Gaia-CRF2 frames, we also performed internal comparisons between the three independent catalogs that form ICRF3. For this purpose, the S/X~band catalog was used as a reference against which the K~band and X/Ka~band catalogs were compared. Employing the same scheme as before for the outlier elimination, a total of 768~sources was left for the comparison with the S/X~band catalog at K~band, corresponding to 97\% of the sources common to the two catalogs. At X/Ka~band, the percentage of sources remaining after such filtering was lower, with only 87\% of the common sources (i.e., 556~sources) left for the comparison.

   \begin{figure}[t]
   \centering
   \includegraphics[width=0.80\columnwidth, trim=0 15 0 36, clip]{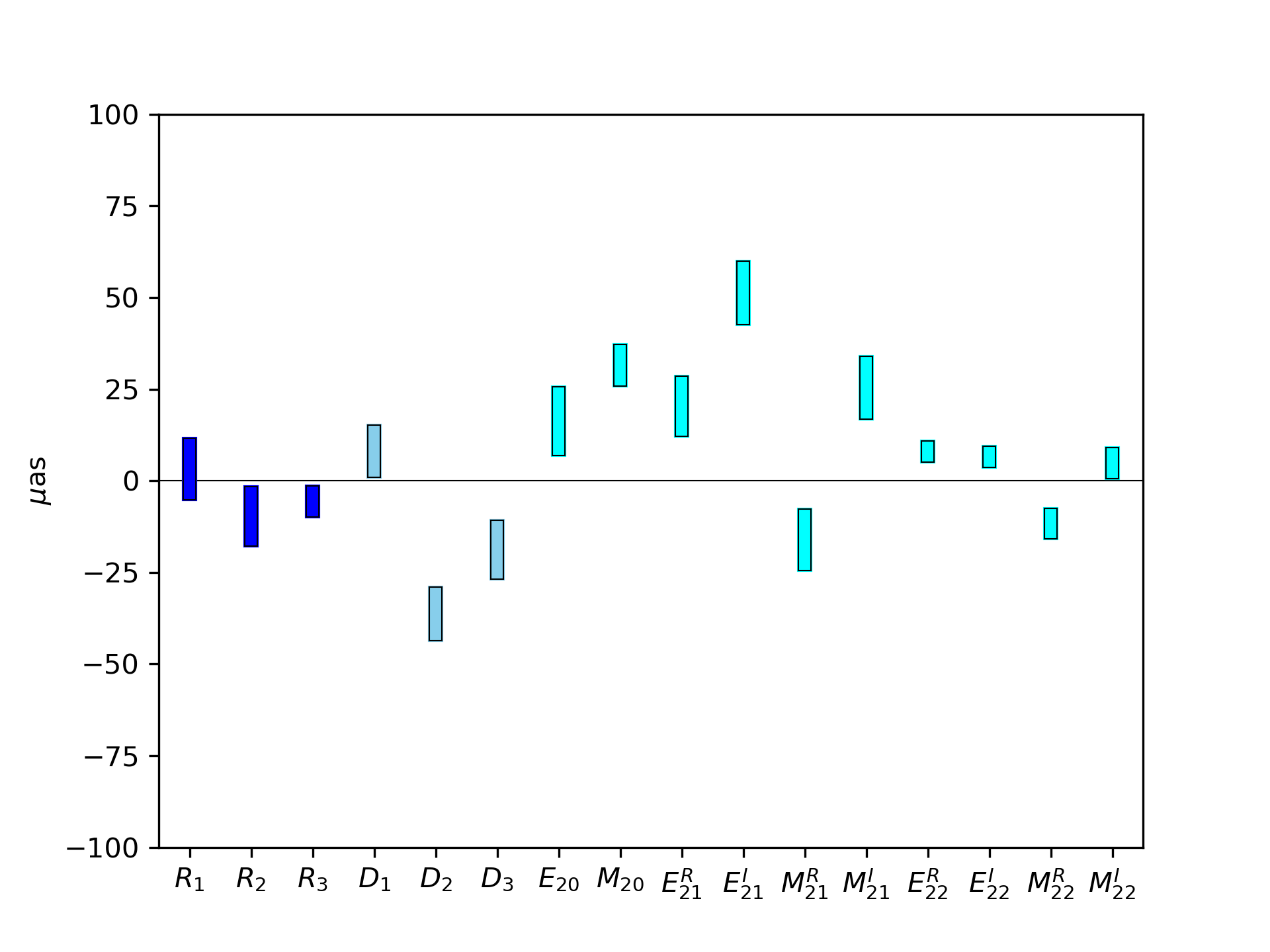}
   \caption{Bar chart showing the values of the 16 parameters of the transformation between the ICRF3 S/X band and K band frames. The $R_i$~parameters are for the rotations, the $D_i$ parameters are for the glide terms, and the $E_{2i}$ and $M_{2i}$ parameters are for the quadrupole terms. See Eqs.~\eqref{Eq:VSH_alpha} and~\eqref{Eq:VSH_delta} for further details on the transformation.}
   \label{Fig:comp_ICRF3X-ICRF3K}
   \end{figure}

\begin{figure}[t]
   \centering
   \includegraphics[width=0.80\columnwidth, trim=0 15 0 36, clip]{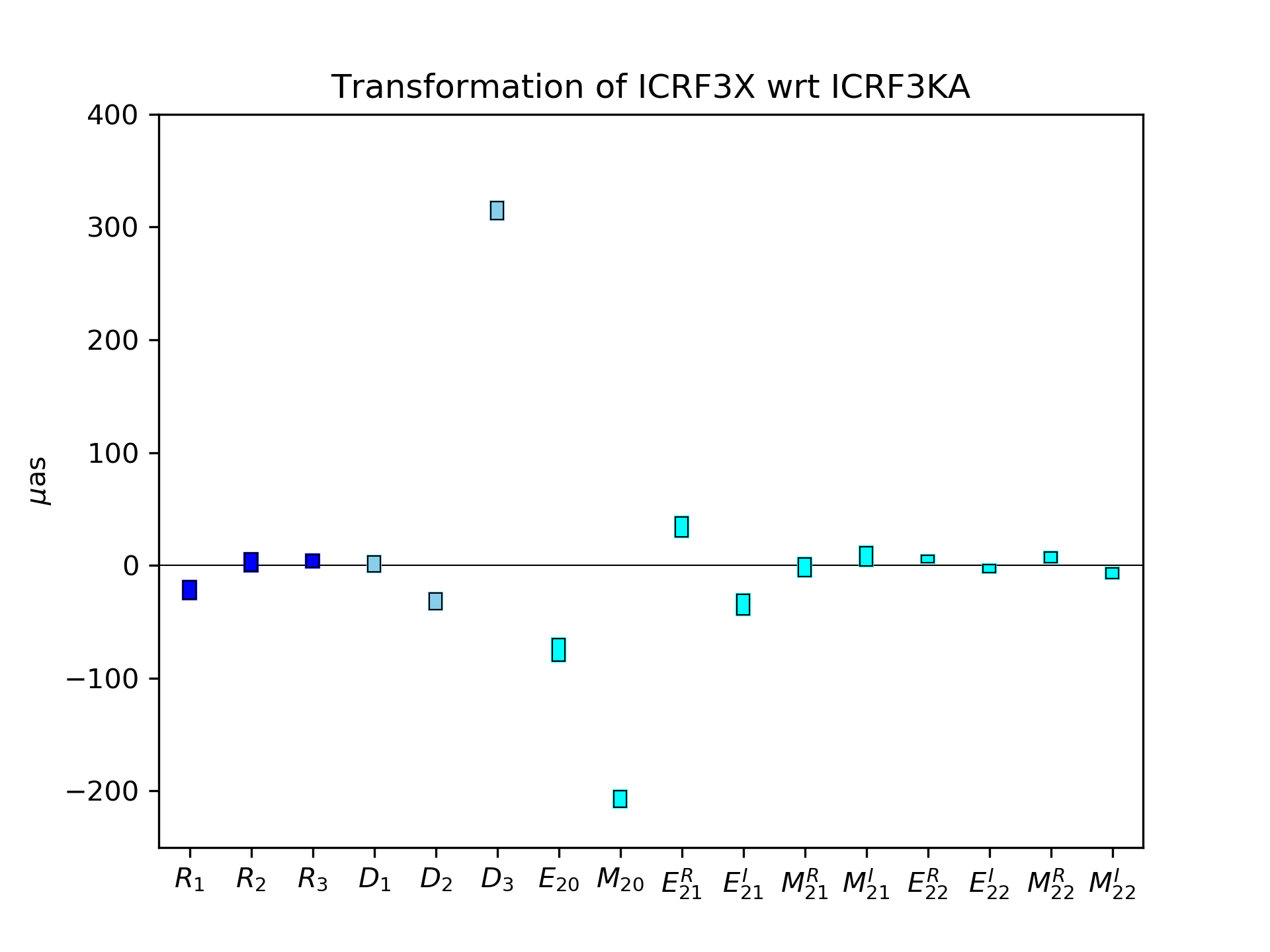}
   \caption{Bar chart showing the values of the 16 parameters of the transformation between the ICRF3 S/X band and X/Ka band frames. The $R_i$~parameters are for the rotations, the $D_i$ parameters are for the glide terms, and the $E_{2i}$ and $M_{2i}$ parameters are for the quadrupole terms. It should be emphasized that the scale is different from that in Figs.~\ref{Fig:comp_ICRF3X-ICRF2}--\ref{Fig:comp_ICRF3X-ICRF3K}. See Eqs.~\eqref{Eq:VSH_alpha} and~\eqref{Eq:VSH_delta} for further details on the transformation.}
   \label{Fig:comp_ICRF3X-ICRF3XKa}
   \end{figure}

   The bar chart in Fig.~\ref{Fig:comp_ICRF3X-ICRF3K} shows the level of the 16~transformation parameters derived from fitting coordinate differences between the K~band and S/X~band catalogs, while that in Fig.~\ref{Fig:comp_ICRF3X-ICRF3XKa} shows the equivalent for the X/Ka band catalog. Looking at the rotations, these charts confirm that the K band and X/Ka~band catalogs are properly aligned onto the S/X band frame, a property to be expected as a result of the analysis configuration used to build those catalogs (see Sect.~\ref{Sec:configuration}). With a single exception, all rotation values are lower than 10~$\mu$as and have a significance at the level of one sigma or less (see transformation parameters in Table~\ref{Tab:comp_ICRF3X-all}). As already remarked when addressing the alignment to ICRF2, such rotations cannot be exactly zero because the no-net-rotation constraints imposed to align the K~band and X/Ka~band frames onto the S/X~band frame were applied solely to the ICRF3 defining sources.
   \begin{table}
   \caption{Parameters of the transformations between ICRF3 (S/X~band frame) and the ICRF2 and Gaia-CRF2 frames, including rotation, glide (dipole terms), and quadrupole terms (all in microarcseconds). Transformation parameters are also given for the catalogs at the two other radio frequencies that form ICRF3 (K band and X/Ka band). The number of common sources, outliers, and used sources is provided in each~case.}
   \small
   \begin{tabular}{@{\ \ }l@{\ \ }r@{\ \ \ }r@{\ \ }r@{\ \ }c@{\ \ }r@{\ \ \ }r@{\ \ }}
   \hline
   \hline
   \noalign{\smallskip}
     & \multicolumn{3}{c}{Reference frame} && \multicolumn{2}{c}{ICRF3 catalog}\\
   \cline{2-4}\cline{6-7}
   \noalign{\smallskip}
   Parameters & ICRF2 & ICRF2$^{{\rm GA}}$ & Gaia-CRF2 && K band & X/Ka band\\
   \noalign{\smallskip}
   \hline
   \noalign{\smallskip}
    Nb sources\\
    \ \ \ \ \ common   & 3414 & 3414\ \ \  & 2983\ \ \ \ \ && 793\ \ & 638\ \ \ \ \\
    \ \ \ \ \ outliers &  496 &  382\ \ \  &  371\ \ \ \ \ &&  25\ \ &  82\ \ \ \ \\
    \ \ \ \ \ used & 2918 & 3032\ \ \  & 2612\ \ \ \ \ && 768\ \ & 556\ \ \ \ \\
   \noalign{\smallskip}
   \multicolumn{2}{l}{Rotation}\\
   \ \ \ \ \ $R_1$ & $  8\pm 4$ & $ 16\pm 4$\ \  & $-22\pm 8\ \ \ $ && $  3\pm 8$ & $ -22\pm 8\ \ $\\
   \ \ \ \ \ $R_2$ & $ 15\pm 4$ & $ 19\pm 4$\ \  & $ 23\pm 8\ \ \ $ && $-10\pm 8$ & $   3\pm 8\ \ $\\
   \ \ \ \ \ $R_3$ & $  0\pm 3$ & $ -5\pm 3$\ \  & $ -5\pm 7\ \ \ $ && $ -6\pm 4$ & $   4\pm 5\ \ $\\
   \noalign{\smallskip}
   \multicolumn{2}{l}{Glide (dipole)}\\
   \ \ \ \ \ $D_1$ & $-22\pm 4$ & $-18\pm 3$\ \  & $-13\pm 8\ \ \ $ && $  8\pm 7$ & $   1\pm 7\ \ $\\
   \ \ \ \ \ $D_2$ & $-63\pm 4$ & $  4\pm 3$\ \  & $ -4\pm 8\ \ \ $ && $-36\pm 7$ & $ -32\pm 7\ \ $\\
   \ \ \ \ \ $D_3$ & $-90\pm 4$ & $-39\pm 4$\ \  & $ 12\pm 8\ \ \ $ && $-19\pm 8$ & $ 314\pm 8\ \ $\\
   \noalign{\smallskip}
   \multicolumn{2}{l}{Quadrupole}\\
   \ \ \ \ \ $E_{20}$ & $ 43\pm 4$ & $ 39\pm 4$\ \  & $ 35\pm 9\ \ \ $ && $ 16\pm 9$ & $ -75\pm10$    \\
   \noalign{\vspace{0pt}}
   \ \ \ \ \ $M_{20}$ & $  5\pm 4$ & $  3\pm 3$\ \  & $  1\pm 8\ \ \ $ && $ 31\pm 6$ & $-207\pm 7\ \ $\\
   \noalign{\vspace{1pt}}
   \ \ \ \ \ $E_{21}^{\rm Re}$ & $-11\pm 5$ & $-13\pm 4$\ \  & $  9\pm10$     \ && $ 20\pm 8$ & $  34\pm 9\ \ $\\
   \noalign{\vspace{1pt}}
   \ \ \ \ \ $E_{21}^{\rm Im}$ & $  3\pm 5$ & $ -2\pm 4$\ \  & $ 13\pm10$     \ && $ 51\pm 9$ & $ -35\pm 9\ \ $\\
   \noalign{\vspace{1pt}}
   \ \ \ \ \ $M_{21}^{\rm Re}$ & $ -1\pm 4$ & $ -1\pm 4$\ \  & $ -3\pm10$     \ && $-16\pm 8$ & $  -2\pm 8\ \ $\\
   \noalign{\vspace{1pt}}
   \ \ \ \ \ $M_{21}^{\rm Im}$ & $ -5\pm 4$ & $ -6\pm 4$\ \  & $  6\pm10$     \ && $ 25\pm 9$ & $   8\pm 9\ \ $\\
   \noalign{\vspace{1pt}}
   \ \ \ \ \ $E_{22}^{\rm Re}$ & $  1\pm 2$ & $ -1\pm 2$\ \  & $  1\pm 5\ \ \ $ && $  8\pm 3$ & $   6\pm 3\ \ $\\
   \noalign{\vspace{1pt}}
   \ \ \ \ \ $E_{22}^{\rm Im}$ & $  3\pm 2$ & $  2\pm 2$\ \  & $  2\pm 5\ \ \ $ && $  7\pm 3$ & $  -3\pm 4\ \ $\\
   \noalign{\vspace{1pt}}
   \ \ \ \ \ $M_{22}^{\rm Re}$ & $ -3\pm 2$ & $ -1\pm 2$\ \  & $  3\pm 5\ \ \ $ && $-12\pm 4$ & $   7\pm 5\ \ $\\
   \noalign{\vspace{1pt}}
   \ \ \ \ \ $M_{22}^{\rm Im}$ & $  2\pm 2$ & $  1\pm 2$\ \  & $ -5\pm 5\ \ \ $ && $  5\pm 4$ & $  -7\pm 5\ \ $\\
   \noalign{\smallskip}
   \hline
   \end{tabular}
   \label{Tab:comp_ICRF3X-all}
   \tablefoot{ICRF2$^{{\rm GA}}$ denotes the equivalent of ICRF2 that was reproduced after incorporating Galactic acceleration in the modeling. }
   \end{table}
   \begin{figure*}[t]
   \centering
   \includegraphics[width=1.01\textwidth, trim=0 190 0 190, clip]{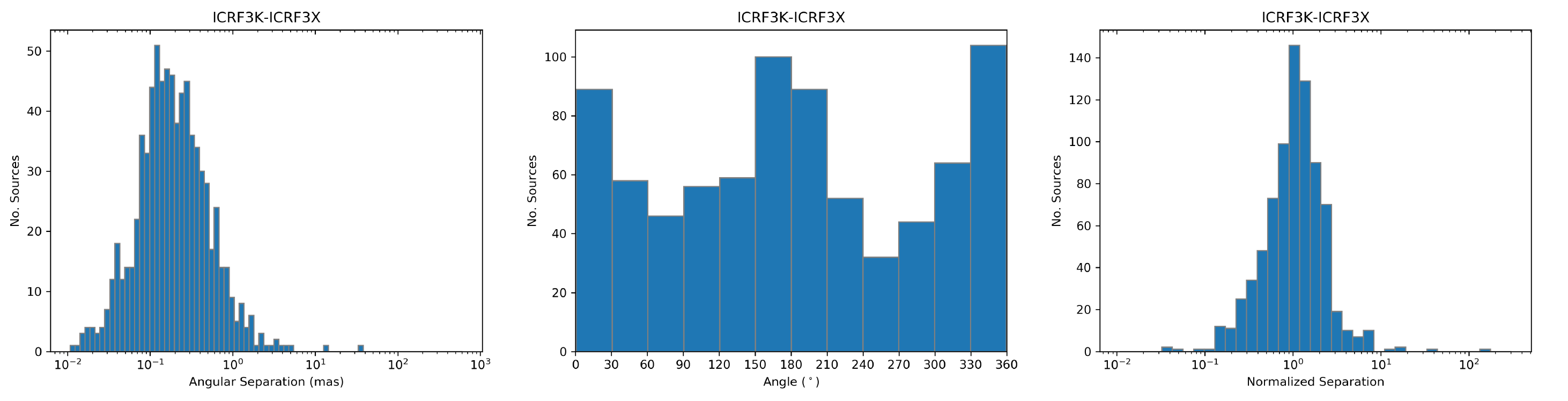}
   \caption{Comparison of the ICRF3 positions at S/X band and K band for the 793 sources common to the two catalogs. Position differences were derived after applying the transformation in Table~\ref{Tab:comp_ICRF3X-all} to the K~band source coordinates. The histogram in the left-hand panel shows the distribution of the angular separation between the two sets of positions, that in the middle panel shows the distribution of the direction of the offset vector joining those positions (counted counter clockwise), and that in the right-hand panel the distribution of the normalized separation.}
   \label{Fig:X-K_offsets}
   \end{figure*}
   \begin{figure*}
   \centering
   \includegraphics[width=1.01\textwidth, trim=0 190 0 190, clip]{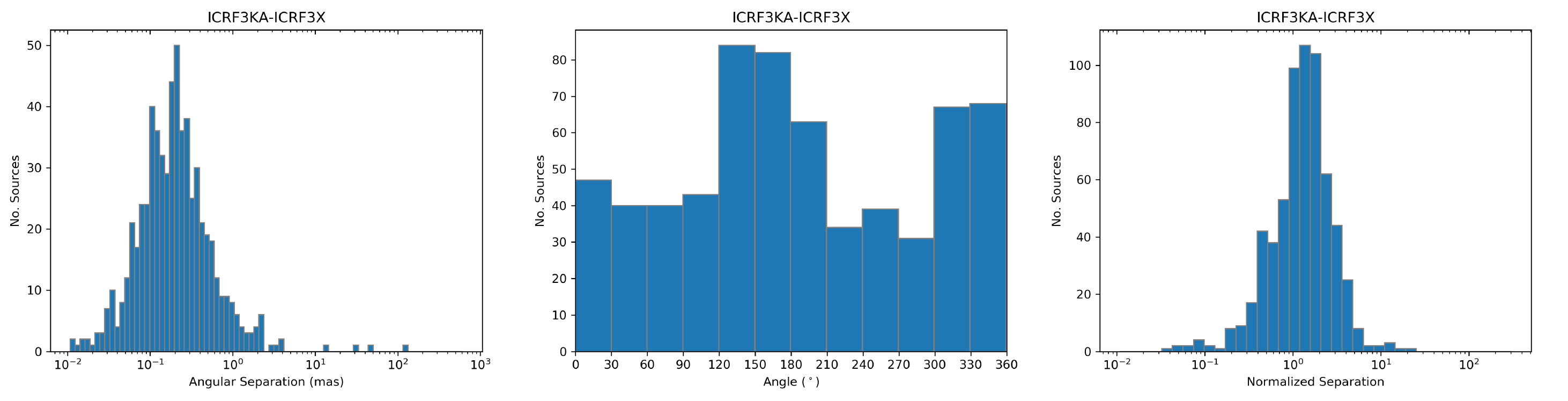}
   \caption{Comparison of the ICRF3 positions at S/X band and X/Ka band for the 638 sources common to the two catalogs. Position differences were derived after applying the transformation in Table~\ref{Tab:comp_ICRF3X-all} to the X/Ka band source coordinates. The histogram in the left-hand panel shows the distribution of the angular separation between the two sets of positions, that in the middle panel shows the distribution of the direction of the offset vector joining those positions (counted counter clockwise), and that in the right-hand panel the distribution of the normalized separation.}
   \label{Fig:X-Ka_offsets}
   \end{figure*}
   The only rotation parameter departing from zero in these comparisons is the rotation $R_1$ for the X/Ka~band catalog comparison, which has a value of $-22\pm 8$~$\mu$as, hence with a marginal significance of $2.7\sigma$. Unlike rotation parameters, the deformation parameters reveal a different picture at K~band and X/Ka~band. At K~band, all dipole and quadrupole terms are within 50~$\mu$as. The largest such terms are the dipole term $D_2$ ($-36\pm 7$~$\mu$as) and the quadrupole terms $M_{20}$~($31\pm 6$~$\mu$as) and $E_{21}^{{\rm Im}}$ ($51\pm 9$~$\mu$as), each of which showing roughly a $5\sigma$ significance (see Table~\ref{Tab:comp_ICRF3X-all}). Considering the weakness of the observing in the far south, such moderate deformations of the K~band frame are not unexpected. It is also worth noting that this level of deformation is consistent with the noise floor of the frame (30~$\mu$as in right ascension and 50~$\mu$as in declination), as reported in Table~\ref{Tab:noise_level}. In contrast, the bar chart for the X/Ka band catalog (Fig.~\ref{Fig:comp_ICRF3X-ICRF3XKa}) reveals much stronger deformations. In particular, there are three terms that stick out: the glide term $D_3$ ($314\pm 8$~$\mu$as) and the quadrupole terms $E_{20}$ ($-75\pm 10$~$\mu$as) and $M_{20}$~($-207\pm 7$~$\mu$as). As noted previously, the X/Ka band network is comprised of only four sites and thus has an inherent limited observing geometry, which is a possible reason for the deformations seen. Such deformations may also explain the larger number of outliers and the slight misalignment of the frame (reflected by the non-zero value of the $R_1$ rotation parameter) noted above. In the future, a specific effort should be placed on reducing these systematics by strengthening the observing geometry at this frequency band.


   \subsection{Consistency of source positions at the three frequencies}
   \label{Sec:comp_3-frequency-positions}

   While the previous sections deal with rotations and deformations between catalogs, it is also of interest to compare the individual source positions measured at the three ICRF3 frequencies. As indicated by the source breakdown in Fig.~\ref{Fig:3-frequency_cross_ID}, the S/X~band catalog has 793 sources in common with the K~band catalog and 638~sources in common with the X/Ka~band catalog, which provides adequate material for the desired comparisons. For this purpose, the S/X~band positions were adopted as a reference against which the K~band and X/Ka~band positions were compared. Prior to the comparison of individual source coordinates, the 16-parameter transformations in Table~\ref{Tab:comp_ICRF3X-all} were applied to the K~band and X/Ka band catalogs in order to free the comparisons from catalog deformations. Aside from source coordinate differences, we examined also the angular separation and direction of the offset vector joining the measured positions at K~band (or X/Ka~band) and S/X~band, along with the corresponding normalized separation. The results of these comparisons are presented graphically in Figs.~\ref{Fig:X-K_offsets} and~\ref{Fig:X-Ka_offsets} for those three quantities.

   Looking at the distribution of angular separations (left-hand panels in Figs.~\ref{Fig:X-K_offsets} and~\ref{Fig:X-Ka_offsets}), both histograms are found to peak at about 0.15--0.20~mas, which is consistent with the median ellipse error of the catalogs (see common sources in Table~\ref{Tab:ICRF3_features}), though slightly larger. Examining further the K~band distribution, the first quartile (25\% of data) is at 0.10~mas, the second quartile (50\% of data) is at 0.18~mas, while the third quartile (75\% of data) is at 0.36~mas. Values for the X/Ka~band distribution are very similar, 0.10~mas for the first quartile, 0.19~mas for the second quartile, and 0.34~mas for the third quartile. As regards offset vector directions, the histograms in Figs.~\ref{Fig:X-K_offsets} and~\ref{Fig:X-Ka_offsets} (middle panels) are indicative of non-uniform distributions. In both cases, the observed directions show an excess at $0^\circ$ and~$180^\circ$, that is, along the declination axis. In a separate check, we also compared directly the K~band and X/Ka~band positions (plots not shown) and found a similar excess, hence ruling out that it comes purely from the S/X~band data. Such an excess is not expected from the source physics since any jet-like features that could induce offsets between positions measured at different frequencies should have a random orientation due to the VLBI jets having no preferred direction in the sky. It is thus most likely that the observed excess arises from remaining catalog systematics, especially in declination, not fully removed by the applied transformations.

   \begin{table}
   \caption{List of the 46 sources for which the normalized separation~$N_L$ between the S/X and K~band positions is above 3. Position differences are characterized by the right ascension and declination offsets and the length and direction of the offset vector joining those positions.}
   \small
      \begin{tabular}{@{\ \ }c@{\hskip 11.0pt}r@{\hskip 5.5pt}r@{\hskip 0.5pt}c@{\hskip 12.0pt}c@{\hskip 0.5pt}r@{\hskip 8.5pt}r@{\hskip 4.5pt}r@{\hskip 1.0pt}}
   \hline
   \hline
   \noalign{\smallskip}
     & \multicolumn{2}{c}{Coordinate offsets}&&&\multicolumn{3}{c}{Offset vector}\\
   \cline{2-3}\cline{6-8}
   \noalign{\smallskip}
   Source & $\Delta\cos\delta${\hskip 3.0pt} & $\Delta\delta${\hskip 14pt} &&& Length\tablefootmark{\dag}{\hskip 0.0pt}{\hskip 0.5pt} & Direction\tablefootmark{\ddag}{\hskip 0.0pt} & $N_L${\hskip 2.5pt}\ \\
   name   & ($\mu$as){\hskip 6.0pt}          & ($\mu$as){\hskip 10pt}       &&& ($\mu$as){\hskip 7.5pt} & ($\degr$){\hskip 16.5pt} & \\
   \noalign{\smallskip}
   \hline
   \noalign{\smallskip}
   0003$-$066 &  $-86\pm 57$ & $-423\pm 111$             &&& $431\pm 109$              & $  191\pm {\hskip 4.5pt}8${\hskip 6.0pt} & $   4.0${\hskip 3pt}\ \\
   0014$+$813 & $-140\pm 76$ & $-438\pm {\hskip 4.5pt}95$&&& $460\pm {\hskip 4.5pt}94$ & $  198\pm              10${\hskip 6.0pt} & $   4.9${\hskip 3pt}\ \\
   0112$-$017 & $-601\pm 86$ & $ -62\pm 149$             &&& $604\pm {\hskip 4.5pt}88$ & $  264\pm              14${\hskip 6.0pt} & $   6.9${\hskip 3pt}\ \\
   0146$+$056 & $ 589\pm 77$ & $-338\pm 135$             &&& $679\pm {\hskip 4.5pt}95$ & $  120\pm              10${\hskip 6.0pt} & $   7.2${\hskip 3pt}\ \\
   0212$+$735 & $ 330\pm 62$ & $-320\pm {\hskip 4.5pt}73$&&& $460\pm {\hskip 4.5pt}68$ & $  134\pm {\hskip 4.5pt}8${\hskip 6.0pt} & $   6.8${\hskip 3pt}\ \\
   0229$+$131 & $-308\pm 64$ & $  57\pm 107$             &&& $313\pm {\hskip 4.5pt}66$ & $  281\pm              19${\hskip 6.0pt} & $   4.7${\hskip 3pt}\ \\
   \hline
   \end{tabular}
   \label{Tab:K-outliers}
   \tablefoot{The content here is printed only for the first six sources. The table in its entirety is available in electronic form from the CDS at \url{http://cdsweb.u-strasbg.fr/cgi-bin/qcat?J/A+A/}.}
   \tablefoottext{\dag}{The vector length is also denoted as angular separation in the text.}\\
   \tablefoottext{\ddag}{The vector direction is counted counter clockwise from north to east.}
   \end{table}

   \begin{table}
   \caption{List of the 70 sources for which the normalized separation~$N_L$ between the S/X and X/Ka~band positions is above~3. Position differences are characterized by the right ascension and declination offsets and the length and direction of the offset vector joining those positions.}
   \small
      \begin{tabular}{@{\ \ }c@{\hskip 9.0pt}r@{\hskip 5.0pt}r@{\hskip 0.5pt}c@{\hskip 10.5pt}c@{\hskip 0.5pt}r@{\hskip 8.5pt}r@{\hskip 4.5pt}r@{\hskip 1.0pt}}
   \hline
   \hline
   \noalign{\smallskip}
     & \multicolumn{2}{c}{Coordinate offsets}&&&\multicolumn{3}{c}{Offset vector}\\
   \cline{2-3}\cline{6-8}
   \noalign{\smallskip}
   Source & $\Delta\cos\delta${\hskip 5.5pt} & $\Delta\delta${\hskip 14pt} &&& Length\tablefootmark{\dag}{\hskip 0.5pt} & Direction\tablefootmark{\ddag}{\hskip 0.0pt} & $N_L${\hskip 2.5pt}\ \\
   name   & ($\mu$as){\hskip 8.5pt}          & ($\mu$as){\hskip 10pt}       &&& ($\mu$as){\hskip 7.5pt} & ($\degr$){\hskip 16.5pt} & \\
   \noalign{\smallskip}
   \hline
   \noalign{\smallskip}
   0003$-$066 & $-109\pm {\hskip 4.5pt}98$ &  $544\pm 133$               &&& $555\pm 132$              & $  349\pm 10${\hskip 6.0pt} & $   4.2${\hskip 3pt}\ \\
   0038$-$020 & $-147\pm 114$              &  $674\pm 191$               &&& $690\pm 188$              & $  348\pm 10${\hskip 6.0pt} & $   3.7${\hskip 3pt}\ \\
   0059$+$581 & $ 182\pm {\hskip 4.5pt}58$ & $-163\pm {\hskip 4.5pt}69$  &&& $244\pm {\hskip 4.5pt}63$ & $  132\pm 15${\hskip 6.0pt} & $   3.9${\hskip 3pt}\ \\
   0112$-$017 & $-593\pm {\hskip 4.5pt}96$ &  $116\pm 131$               &&& $604\pm {\hskip 4.5pt}98$ & $  281\pm 12${\hskip 6.0pt} & $   6.2${\hskip 3pt}\ \\
   0119$+$115 & $ -23\pm {\hskip 4.5pt}62$ & $-324\pm {\hskip 4.5pt}93$  &&& $325\pm {\hskip 4.5pt}93$ & $  184\pm 11${\hskip 6.0pt} & $   3.5${\hskip 3pt}\ \\
   0122$-$003 & $-101\pm 126$              &  $823\pm 205$               &&& $829\pm 204$              & $  353\pm {\hskip 4.5pt}9${\hskip 6.0pt} & $   4.1${\hskip 3pt}\ \\
   \hline
   \end{tabular}
   \label{Tab:Ka-outliers}
   \tablefoot{The content here is printed only for the first six sources. The table in its entirety is available in electronic form from the CDS at \url{http://cdsweb.u-strasbg.fr/cgi-bin/qcat?J/A+A/}.}
   \tablefoottext{\dag}{The vector length is also denoted as angular separation in the text.}\\
   \tablefoottext{\ddag}{The vector direction is counted counter clockwise from north to east.}
   \end{table}

   Looking at the right-hand panels in Figs.~\ref{Fig:X-K_offsets} and~\ref{Fig:X-Ka_offsets}, the distribution of normalized separations is found to peak near~1 in the case of the K~band comparison and around~1.5 in the case of the X/Ka~band comparison. The median value of the distribution is 1.09 for the former and is 1.32 for the latter. This is an indirect indication that source position uncertainties in all are reasonably well estimated, though possibly somewhat underestimated at X/Ka~band. A further noticeable feature in these histograms is that a visible portion of the sources show normalized separations above~3. There are 46~sources in that condition (i.e., 6\% of the set of common sources), including five ICRF3 defining sources (0403$-$132, 1245$-$457, 1448$+$762, 1745$+$624, and 2229$+$695), in the case of the K~band comparison (see Table~\ref{Tab:K-outliers}) and 70~such sources (i.e., 11\%~of the set of common sources), including 18 ICRF3 defining sources, in the case of the X/Ka~band comparison (see Table~\ref{Tab:Ka-outliers}). We suspect that the higher percentage found for the latter is due to the uncertainty in the angular separations being underestimated, as reflected by the median normalized separation of 1.32 noted above. Indeed, if assuming a 30\% underestimation, hence considering only sources with normalized separations above~3.9, the number of sources that stand out falls down to~32 (i.e., 5\%~of the set of common sources), including three ICRF3 defining sources (0607$-$157, 0648$-$165, and 0743$-$006). In any case, that percentage of outliers (i.e., about 5\% for both the K~band and X/Ka~band comparisons) is much higher than anticipated for a Rayleigh distribution (where only~1.1\%, i.e., 7--8 sources, would be expected) and it only reflects the existence of systematics in the positions of these sources. It is of interest that a number of sources in those two tables are known to have extended VLBI structures, for example 0430$+$052 (3C120),  0923$+$392 (4C39.25), or 2251$+$158 (3C454.3). Additionally, about half of the sources that are found to deviate in the K~band comparison (24~sources) happen to deviate also in the X/Ka~band comparison. Taking a step further, we examined the X~band structure indices (from BVID) for these sources. For the 39~deviating sources in Table~\ref{Tab:K-outliers} (K~band comparison) for which structure was assessed, the structure indices range from 1.6 to~4.9, with a mean value of~3.3, hence reflecting a predominance of significantly-structured sources. The structure indices for the 61~deviating sources in Table~\ref{Tab:Ka-outliers} (X/Ka~band comparison) for which structure was assessed cover the same range, from 1.6 to 4.9, but have a mean value that is somewhat lower (3.0). On the other hand, if reducing the sample to the 32~sources with normalized separations above~3.9 (see the discussion about underestimation of the uncertainties above), the mean value of the structure index goes up to 3.3. The X/Ka~band comparison therefore also provides indication that the deviating sources have significant structures. In all, those extended morphologies may well explain the observed position inconsistencies between S/X and K or X/Ka~band. The further investigation and interpretation of these offsets, however, is beyond the scope of this paper.

   \subsection{Consistency of radio and optical source positions}
   \label{Sec:comp_radio-optical-positions}

\begin{figure*}
   \centering
   \includegraphics[width=1.01\textwidth, trim=0 190 0 190, clip]{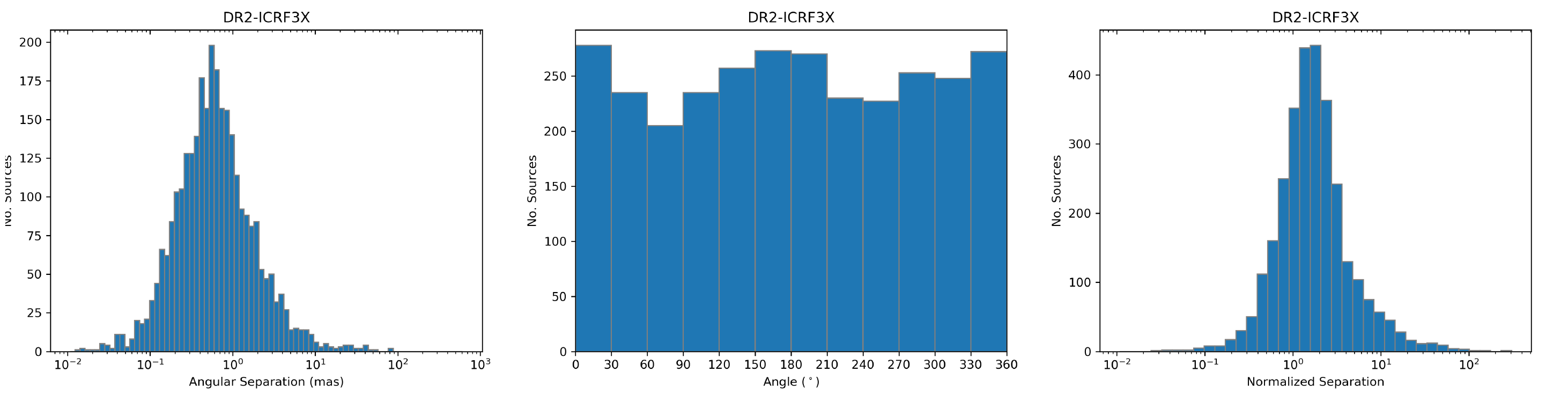}
   \caption{Comparison of the ICRF3 positions at S/X band with the Gaia-CRF2 positions for the 2983 sources common to the two frames. Position differences were derived after applying the transformation in Table~\ref{Tab:comp_ICRF3X-all} to the Gaia-CRF2 source coordinates. The histogram in the left-hand panel shows the distribution of the angular separation between the two sets of positions, that in the middle panel shows the distribution of the direction of the offset vector joining those positions (counted counter clockwise), and that in the right-hand panel the distribution of the normalized separation.}
   \label{Fig:X-DR2_offsets}
   \end{figure*}

   The individual ICRF3 source positions may also be compared to the Gaia-CRF2 optical positions. For this characterization, we used the S/X band catalog since it contains the largest number of sources in common (2983 sources, among which 250 defining sources) and adopted the same scheme as above for the comparison, meaning that the transformation in Table~\ref{Tab:comp_ICRF3X-all} was applied to the Gaia-CRF2 frame prior to comparing the radio and optical positions. As before, the separations, offset vector directions, and normalized separations between the two sets of positions were computed, and the distribution of these is shown in Fig.~\ref{Fig:X-DR2_offsets}.

   Looking at the distribution of angular separations (left-hand panel in Fig.~\ref{Fig:X-DR2_offsets}), the histogram is found to peak at about 0.6~mas, with the first quartile at 0.30~mas, the second quartile at 0.58~mas, and the third quartile at 1.13~mas. These values are roughly a factor of three larger than those obtained when comparing the S/X~band catalog to the K~band or X/Ka~band catalog. Such a difference might be explained, at least in part, by the common sources in this comparison having generally larger position uncertainties than those involved in the previous comparisons. This is true both for the Gaia-CRF2 positions, where the median value of the ellipse error semi-major axis for the 2983~common sources is 0.26~mas, and the S/X~band positions, where this quantity is 0.19~mas. In contrast, the corresponding quantities are 0.14~mas and 0.09~mas in the K~band to S/X~band comparison (with 793~common sources) and 0.11~mas and 0.07~mas in the X/Ka~band to S/X~band comparison (with 638 common sources). The reason why the S/X~band median position uncertainty is larger when comparing to the Gaia-CRF2 catalog than when comparing to the K~band or X/Ka~band catalog is that the common sources fall predominantly into the VLBA-only category in that comparison, unlike those involved in the K~band and X/Ka~band comparisons where other sources (i.e., observed also by the~IVS) predominate. Looking now at the middle panel in Fig.~\ref{Fig:X-DR2_offsets}, which plots the direction of the offset vectors, the distribution is found to be much more uniform than when comparing to the K~band and X/Ka~band catalogs (see Figs.~\ref{Fig:X-K_offsets} and~\ref{Fig:X-Ka_offsets}), though there is still a small excess along the declination axis. The presence of this excess indicates that the S/X~band catalog is not entirely free from declination systematics since there is no reason why such systematics should come from the Gaia-CRF2 catalog where declination plays no particular role due to the frame being built from space. The effect, however, is less pronounced than that observed in the K~band and X/Ka~band catalogs.

   \begin{table}
   \caption{List of the 653 sources for which the normalized separation $N_L$ between the S/X and Gaia-CRF2 positions is above 3. Position differences are characterized by the right ascension and declination offsets and the length and direction of the offset vector joining those positions.}
   \small
   \begin{tabular}{@{\ }c@{\hskip 6.0pt}r@{\hskip 3.0pt}r@{\hskip 0.5pt}c@{\hskip 7.5pt}c@{\hskip 0.5pt}r@{\hskip 5.5pt}r@{\hskip 0.5pt}r@{\hskip 0.0pt}}
   \hline
   \hline
   \noalign{\smallskip}
     & \multicolumn{2}{c}{Coordinate offsets}&&&\multicolumn{3}{c}{Offset vector}\\
   \cline{2-3}\cline{6-8}
   \noalign{\smallskip}
   Source & $\Delta\cos\delta${\hskip 7pt} & $\Delta\delta${\hskip 14pt} &&& Length\tablefootmark{\dag}{\hskip 2.5pt} & Direction\tablefootmark{\ddag}{\hskip 0.0pt} & $N_L${\hskip 4.0pt}\ \\
   name   & ($\mu$as){\hskip 10pt}          & ($\mu$as){\hskip 10pt}       &&& ($\mu$as){\hskip 10.5pt} & ($\degr$){\hskip 16.5pt} & \\
   \noalign{\smallskip}
   \hline
   \noalign{\smallskip}
   2357$-$326 & $-6265\pm              177$ & $ -618\pm              289$  &&& $6295\pm              178$ & $264\pm {\hskip 4.5pt}3${\hskip 6.0pt} & $35.3${\hskip 3pt}\ \\
   2359$-$221 & $-1702\pm              518$ & $ 1114\pm              567$  &&& $2034\pm              533$ & $303\pm              16${\hskip 6.0pt} & $ 3.8${\hskip 3pt}\ \\
   0000$-$199 & $-1053\pm              339$ & $ 1584\pm              359$  &&& $1902\pm              353$ & $326\pm              10${\hskip 6.0pt} & $ 5.4${\hskip 3pt}\ \\
   0001$-$120 & $ 1108\pm              319$ & $   50\pm              328$  &&& $1109\pm              319$ & $ 87\pm              17${\hskip 6.0pt} & $ 3.5${\hskip 3pt}\ \\
   0003$+$380 & $ 6425\pm              233$ & $-5409\pm              185$  &&& $8399\pm              214$ & $130\pm {\hskip 4.5pt}1${\hskip 6.0pt} & $39.2${\hskip 3pt}\ \\
   0003$-$066 & $  -47\pm {\hskip 4.5pt}99$ & $  227\pm {\hskip 4.5pt}70$  &&& $ 232\pm {\hskip 4.5pt}71$ & $348\pm              24${\hskip 6.0pt} & $ 3.3${\hskip 3pt}\ \\
   \hline
   \end{tabular}
   \label{Tab:DR2-outliers}
   \tablefoot{The content here is printed only for the first six sources. The table in its entirety is available in electronic form from the CDS at \url{http://cdsweb.u-strasbg.fr/cgi-bin/qcat?J/A+A/}.}
   \tablefoottext{\dag}{The vector length is also denoted as angular separation in the text.}\\
   \tablefoottext{\ddag}{The vector direction is counted counter clockwise from north to east.}
   \end{table}

   Of particular interest is the examination of the distribution of normalized separations, which is shown in the right-hand panel of Fig.~\ref{Fig:X-DR2_offsets}. The histogram peaks between 1 and 2, with the median value of the distribution at 1.64, hence indicating possibly an overall underestimation of the position uncertainties (which for the VLBI part would be in contrast with the findings in the previous subsection) or the presence of real physical offsets between the measured optical and radio positions. Another noticeable feature in that distribution is that a significant portion of the sources show large normalized separations (see the higher end of the histogram). Looking at this feature in detail, a total of 653~sources are found to have a normalized separation larger than~3, including 59~defining sources. Interestingly, the percentage of sources that stand out is similar when considering the entire set of common sources or only the defining sources (about 22\% for the former and 24\% for the latter). Furthermore, there are 144~sources among these that show normalized separations larger than~10 (corresponding to 4.8\% of the common sources), including four defining sources (1.6\% of the defining sources in common). All 653~sources in this condition (i.e., with a normalized separation larger than 3) are listed in Table~\ref{Tab:DR2-outliers}, together with the measured coordinate offsets, offset vectors, and normalized separations. The existence of such notable VLBI-Gaia position offsets for a significant portion of the sources was first revealed by \citet{Mignard2016}, \citet{Kovalev2017}, and \citet{Petrov2017a} based on the Gaia DR1 catalog \citep{Lindegren2016} and interpreted as the manifestation of optical jets on scales 1--100~mas \citep{Petrov2017b}. The release of the Gaia~DR2 catalog \citep{Lindegren2018} confirmed those findings with a higher level of significance \citep{Mignard2018,Petrov2019,Plavin2019}. Following the lines of the previous section, we examined the structure indices of the sources with significant positional offsets when available. Out of the 653 sources in Table~\ref{Tab:DR2-outliers}, 267 were found to have an X~band structure index in BVID. The mean value of these structure indices is~3.4, with a dichotomy between the 50~defining sources (mean structure index of~2.7) and the 217~other sources (mean structure index of~3.5), hence indicating that sources with significant structure predominate. The ICRF3 data thus also suggest that the large Gaia-VLBI offsets observed for a notable fraction of the sources are likely a manifestation of source structure.

   \section{Adoption of ICRF3 by the IAU}
   \label{Sec:adoption}

   As stated in the mandate of the working group, ICRF3 was presented at the 30th IAU General Assembly held in Vienna (Austria) on August~20--31, 2018. The resolution adopted by the General Assembly, referenced as Resolution B2\footnote{See Resolution B2 at \url{https://iau.org/static/resolutions/IAU2018_ResolB2_English.pdf}.}, resolved that the fundamental ICRS realization shall be ICRF3 from January~1, 2019, that the organizations responsible for astrometric and geodetic VLBI observing programs (e.g., IVS) shall take appropriate measures to both maintain and improve ICRF3, and that the organizations responsible for defining high-accuracy reference frames at other wavelengths, together with the IERS, shall take appropriate measures to align those reference frames onto ICRF3 with the highest possible accuracy. As a result of this adoption, ICRF3 replaced ICRF2 as the fundamental celestial reference frame to use for all applications on January~1, 2019.

   \section{Evolution of the ICRF}
   \label{Sec:evolution}
   The material presented in the previous sections already provides directions at which to target prospective VLBI observing along with modeling improvements and refinements of the analysis configuration in order to enhance the ICRF in the future. Each of these aspects is discussed in turn in the following paragraphs.

   On the observing side, a special effort should be made in the first place to strengthen the celestial frame in the far south. As noted above, the S/X band frame shows a deficit of sources at the most southern declinations (Fig.~\ref{Fig:ICRF3_SX}). While the number of sources is similar in the south and in the north for declinations between $-35^{\degr}$ and $+35^{\degr}$ (1461 vs 1460 sources), the number of sources further south ($<-35^{\degr}$ declination) is lower by a factor of 2.5 than that further north ($>35^{\degr}$ declination) (460 vs~1155~sources). The same applies to the K~band frame, although the deficit of sources in the south is only by a factor of 1.5 in this case. Additionally, it is found that the source position accuracy deteriorates with declination (Fig.~\ref{Fig:dec_errors}). For such reasons, it is desirable that a proper observing strategy be devised for these two frequency bands to both increase the source density and augment the source position accuracy in the south. Efforts in this direction have already been engaged and IVS sessions in the south are now strengthened \citep{deWitt2019}. At X/Ka~band, the frame is more uniform but suffers from systematics. The focus of the observing in the future should then be placed on reducing these systematics.

   A second direction for prospective VLBI observing lies in the further increase in the source position accuracy for the bulk of the sources in the S/X band frame. Figure~\ref{Fig:ICRF3_SX} shows that there are two peaks in the distribution of position uncertainties, a primary peak at 100--200~$\mu$as capturing the majority of the sources and a secondary peak just above the noise floor (30~$\mu$as) that captures the roughly 500~sources that have the most precise positions. As noted above, the primary peak (already existing in ICRF2) was brought down from 350--700~$\mu$as in ICRF2 to the present error level thanks to a number of VLBA campaigns conducted since 2014 (see Sect.~\ref{Sec:data_SX}). Acquiring more data on the corresponding such 3000 sources will bring this peak further down and even closer to the secondary peak at the noise floor. A factor of 3--4 (the same as that gained from ICRF2 to ICRF3) is still to be gained to merge the two peaks. In this respect, continuation of the VLBA campaigns will be decisive to achieve this goal. In this process, observation of the optically bright ICRF3 sources (i.e., detected by Gaia) will be also of special interest to strengthen the alignment between the radio and optical frames.

   Above all, it will be essential to monitor closely the defining sources. For this purpose, it would be desirable that these be incorporated into the regular geodetic VLBI observing programs, such as those carried out by the IVS, in order to control their astrometric stability. Additionally, it will be important that these sources be imaged on a regular basis to track potential source structure changes. This especially applies to the 25~ICRF3 defining sources for which we had no VLBI images in hand -- the list of which is given in Table~\ref{Tab:no_images} -- so that the brightness distribution of those sources, and hence their suitability as defining sources, may be assessed. A specific effort should also be made toward observing the~ICRF3 defining sources at K~band and X/Ka~band since not all of them are included the respective catalogs and the amount of data for those that do remains too limited in some cases. Having such a plan is necessary to strengthen the alignment between the three catalogs in future realizations of the~ICRF. Beyond that, prospective VLBI observing should also aim at filling in the 21 empty sectors, which means to search and identify a suitable defining source for every such sector. These sectors are identified in Table~\ref{Tab:empty_sectors} based on the corresponding right ascension and declination range. The table also lists the sources originally selected in these sectors but then discarded due to their having too extended (category~C) structures (see Sect.~\ref{Sec:ICRF3-defining}). In the longer term, one should also seek to replace the 62 defining sources that show moderately extended (i.e. category~B) structures by new ones which are more point-like. Such sources are listed in Table~\ref{Tab:category_B_sectors} along with the identification of the corresponding sectors where new sources are to~be~found. For the goal of finding new sources, either to fill in the empty sectors in Table~\ref{Tab:empty_sectors} or to replace the category~B sources in Table~\ref{Tab:category_B_sectors}, the VLBA, again, should be important since it has the capability to observe weaker sources than geodetic VLBI networks.

   \begin{table}
   \caption{ICRF3 defining sources for which VLBI structure was not assessed prior to their selection as defining sources.}
   \small
   \begin{tabular}{ccccc}
   \hline
   \hline
   \noalign{\smallskip}
   \multicolumn{5}{c}{Source names}\\
   \noalign{\smallskip}
   \hline
   \noalign{\smallskip}
   0009$-$148 & 0044$-$846 & 0227$+$403 & 0642$-$349 & 0742$-$562\\
   0802$-$010 & 0804$-$267 & 0841$-$607 & 0926$-$039 & 0930$-$080\\
   1016$-$311 & 1036$-$529 & 1101$-$536 & 1245$-$457 & 1312$-$533\\
   1325$-$558 & 1412$-$368 & 1511$-$558 & 1556$-$245 & 1606$-$398\\
   1753$+$204 & 2037$+$216 & 2111$+$400 & 2121$+$547 & 2220$-$351\\
   \noalign{\smallskip}
   \hline
   \end{tabular}
   \label{Tab:no_images}
   \end{table}

   \begin{table}
   \caption{Identification of the 21 sectors over the celestial sphere where no suitable defining source was found, either because there is no ICRF3 source in that sector or because the source available (indicated in parentheses) has poor (category C) structure. Each sector is numbered and defined by a range in right ascension and a range in declination.}
   \small
   \begin{tabular}{r@{\hskip 14pt}r@{\hskip 14pt}rc@{\hskip 5pt}r@{\hskip 10pt}rc@{\hskip 8pt}c}
   \hline
   \hline
   \noalign{\smallskip}
     &\multicolumn{2}{c}{Right Asc.}&& \multicolumn{2}{c}{Declination}\\
   \cline{2-3}\cline{5-6}
   \noalign{\smallskip}
   Sector\ \ & Min\ \ & Max\ \ && Min\ \ & Max\ \ && Source\\
   number & (h\ \ m) & (h\ \ m) && ($\degr$)\ \ \ & ($\degr$)\ \ \ && name\\
   \noalign{\smallskip}
   \hline
   \noalign{\smallskip}
   24 \hspace{5pt} & 01\ \ 20 & 02\ \ 40 && $-26.39$ & $-19.47$ && (0135$-$247)\\
   26 \hspace{5pt} & 01\ \ 20 & 02\ \ 40 && $-12.84$ & $ -6.38$ && (0138$-$097)\\
   35 \hspace{5pt} & 01\ \ 20 & 02\ \ 40 && $ 51.06$ & $ 62.73$ && (0144$+$584)\\
   44 \hspace{5pt} & 02\ \ 40 & 04\ \ 00 && $-12.84$ & $ -6.38$ && (0238$-$084)\\
   50 \hspace{5pt} & 02\ \ 40 & 04\ \ 00 && $ 26.39$ & $ 33.75$ && (0333$+$321)\\
   70 \hspace{5pt} & 04\ \ 00 & 05\ \ 20 && $ 41.81$ & $ 51.06$ && (0420$+$417)\\
  126 \hspace{5pt} & 08\ \ 00 & 09\ \ 20 && $ 62.73$ & $ 90.00$ && (0836$+$710)\\
  143 \hspace{5pt} & 09\ \ 20 & 10\ \ 40 && $ 51.06$ & $ 62.73$ && (0917$+$624)\\
  161 \hspace{5pt} & 10\ \ 40 & 12\ \ 00 && $ 51.06$ & $ 62.73$ && (1038$+$528)\\
  171 \hspace{5pt} & 12\ \ 00 & 13\ \ 20 && $ -6.38$ & $  0.00$ && (1253$-$055)\\
  178 \hspace{5pt} & 12\ \ 00 & 13\ \ 20 && $ 41.81$ & $ 51.06$ && (1216$+$487)\\
  230 \hspace{5pt} & 16\ \ 00 & 17\ \ 20 && $ 26.39$ & $ 33.75$ && (1600$+$335)\\
  239 \hspace{5pt} & 17\ \ 20 & 18\ \ 40 && $-33.75$ & $-26.39$ && \\
  263 \hspace{5pt} & 18\ \ 40 & 20\ \ 00 && $  6.38$ & $ 12.84$ && (1947$+$079)\\
  269 \hspace{5pt} & 18\ \ 40 & 20\ \ 00 && $ 51.06$ & $ 62.73$ && (1954$+$513)\\
  272 \hspace{5pt} & 20\ \ 00 & 21\ \ 20 && $-62.73$ & $-51.06$ && \\
  275 \hspace{5pt} & 20\ \ 00 & 21\ \ 20 && $-33.75$ & $-26.39$ && (2000$-$330)\\
  287 \hspace{5pt} & 20\ \ 00 & 21\ \ 20 && $ 51.06$ & $ 62.73$ && (2037$+$511)\\
  304 \hspace{5pt} & 21\ \ 20 & 22\ \ 40 && $ 41.81$ & $ 51.06$ && (2200$+$420)\\
  309 \hspace{5pt} & 22\ \ 40 & 24\ \ 00 && $-51.06$ & $-41.81$ && (2326$-$477)\\
  318 \hspace{5pt} & 22\ \ 40 & 24\ \ 00 && $ 12.84$ & $ 19.47$ && (2251$+$158)\\
   \noalign{\smallskip}
   \hline
   \end{tabular}
   \label{Tab:empty_sectors}
   \end{table}

   \begin{table}
   \caption{Identification of the 62 sectors over the celestial sphere where the related ICRF3 defining sources, also specified in the table, show moderately extended (category B) structures. Each sector is numbered and defined by a range in right ascension and a range in declination.}
   \small
   \renewcommand{\arraystretch}{0.98}
   \begin{tabular}{r@{\hskip 16pt}r@{\hskip 14pt}rc@{\hskip 5pt}r@{\hskip 10pt}rc@{\hskip 8pt}c}
   \hline
   \hline
   \noalign{\smallskip}
     &\multicolumn{2}{c}{Right Asc.}&& \multicolumn{2}{c}{Declination}\\
   \cline{2-3}\cline{5-6}
   \noalign{\smallskip}
   Sector\ \ & Min\ \ & Max\ \ && Min\ \ & Max\ \ && Defining\\
   number & (h\ \ m) & (h\ \ m) && ($\degr$)\ \ \ & ($\degr$)\ \ \ && source\\
   \noalign{\smallskip}
   \hline
   \noalign{\smallskip}
   2 \hspace{5pt} & 00\ \ 00 & 01\ \ 20 && $-62.73$ & $-51.06$ && 0047$-$579\\
  14 \hspace{5pt} & 00\ \ 00 & 01\ \ 20 && $ 26.39$ & $ 33.75$ && 0046$+$316\\
  16 \hspace{5pt} & 00\ \ 00 & 01\ \ 20 && $ 41.81$ & $ 51.06$ && 0110$+$495\\
  19 \hspace{5pt} & 01\ \ 20 & 02\ \ 40 && $-90.00$ & $-62.73$ && 0230$-$790\\
  31 \hspace{5pt} & 01\ \ 20 & 02\ \ 40 && $ 19.47$ & $ 26.39$ && 0149$+$218\\
  36 \hspace{5pt} & 01\ \ 20 & 02\ \ 40 && $ 62.73$ & $ 90.00$ && 0159$+$723\\
  46 \hspace{5pt} & 02\ \ 40 & 04\ \ 00 && $  0.00$ & $  6.38$ && 0305$+$039\\
  53 \hspace{5pt} & 02\ \ 40 & 04\ \ 00 && $ 51.06$ & $ 62.73$ && 0302$+$625\\
  54 \hspace{5pt} & 02\ \ 40 & 04\ \ 00 && $ 62.73$ & $ 90.00$ && 0346$+$800\\
  59 \hspace{5pt} & 04\ \ 00 & 05\ \ 20 && $-33.75$ & $-26.39$ && 0400$-$319\\
  66 \hspace{5pt} & 04\ \ 00 & 05\ \ 20 && $ 12.84$ & $ 19.47$ && 0507$+$179\\
  72 \hspace{5pt} & 04\ \ 00 & 05\ \ 20 && $ 62.73$ & $ 90.00$ && 0454$+$844\\
  73 \hspace{5pt} & 05\ \ 20 & 06\ \ 40 && $-90.00$ & $-62.73$ && 0530$-$727\\
  80 \hspace{5pt} & 05\ \ 20 & 06\ \ 40 && $-12.84$ & $ -6.38$ && 0605$-$085\\
  81 \hspace{5pt} & 05\ \ 20 & 06\ \ 40 && $ -6.38$ & $  0.00$ && 0539$-$057\\
  90 \hspace{5pt} & 05\ \ 20 & 06\ \ 40 && $ 62.73$ & $ 90.00$ && 0615$+$820\\
  91 \hspace{5pt} & 06\ \ 40 & 08\ \ 00 && $-90.00$ & $-62.73$ && 0738$-$674\\
  99 \hspace{5pt} & 06\ \ 40 & 08\ \ 00 && $ -6.38$ & $  0.00$ && 0743$-$006\\
 100 \hspace{5pt} & 06\ \ 40 & 08\ \ 00 && $  0.00$ & $  6.38$ && 0736$+$017\\
 101 \hspace{5pt} & 06\ \ 40 & 08\ \ 00 && $  6.38$ & $ 12.84$ && 0748$+$126\\
 105 \hspace{5pt} & 06\ \ 40 & 08\ \ 00 && $ 33.75$ & $ 41.81$ && 0641$+$392\\
 107 \hspace{5pt} & 06\ \ 40 & 08\ \ 00 && $ 51.06$ & $ 62.73$ && 0749$+$540\\
 109 \hspace{5pt} & 08\ \ 00 & 09\ \ 20 && $-90.00$ & $-62.73$ && 0842$-$754\\
 111 \hspace{5pt} & 08\ \ 00 & 09\ \ 20 && $-51.06$ & $-41.81$ && 0809$-$493\\
 112 \hspace{5pt} & 08\ \ 00 & 09\ \ 20 && $-41.81$ & $-33.75$ && 0826$-$373\\
 115 \hspace{5pt} & 08\ \ 00 & 09\ \ 20 && $-19.47$ & $-12.84$ && 0818$-$128\\
 121 \hspace{5pt} & 08\ \ 00 & 09\ \ 20 && $ 19.47$ & $ 26.39$ && 0834$+$250\\
 125 \hspace{5pt} & 08\ \ 00 & 09\ \ 00 && $ 51.06$ & $ 62.73$ && 0800$+$618\\
 127 \hspace{5pt} & 09\ \ 20 & 10\ \ 40 && $-90.00$ & $-62.73$ && 1022$-$665\\
 133 \hspace{5pt} & 09\ \ 00 & 10\ \ 40 && $-19.47$ & $-12.84$ && 1027$-$186\\
 139 \hspace{5pt} & 09\ \ 00 & 10\ \ 40 && $ 19.47$ & $ 26.39$ && 1012$+$232\\
 150 \hspace{5pt} & 10\ \ 40 & 12\ \ 00 && $-26.39$ & $-19.47$ && 1143$-$245\\
 160 \hspace{5pt} & 10\ \ 40 & 12\ \ 00 && $ 41.81$ & $ 51.06$ && 1150$+$497\\
 173 \hspace{5pt} & 12\ \ 00 & 13\ \ 20 && $  6.38$ & $ 12.84$ && 1236$+$077\\
 185 \hspace{5pt} & 13\ \ 20 & 14\ \ 40 && $-33.75$ & $-26.39$ && 1406$-$267\\
 186 \hspace{5pt} & 13\ \ 20 & 14\ \ 40 && $-26.39$ & $-19.47$ && 1435$-$218\\
 197 \hspace{5pt} & 13\ \ 20 & 14\ \ 40 && $ 51.06$ & $ 62.73$ && 1418$+$546\\
 199 \hspace{5pt} & 14\ \ 40 & 16\ \ 00 && $-90.00$ & $-62.73$ && 1448$-$648\\
 202 \hspace{5pt} & 14\ \ 40 & 16\ \ 00 && $-41.81$ & $-33.75$ && 1451$-$400\\
 205 \hspace{5pt} & 14\ \ 40 & 16\ \ 00 && $-19.47$ & $-12.84$ && 1443$-$162\\
 206 \hspace{5pt} & 14\ \ 40 & 16\ \ 00 && $-12.84$ & $ -6.38$ && 1510$-$089\\
 209 \hspace{5pt} & 14\ \ 40 & 16\ \ 00 && $  6.38$ & $ 12.84$ && 1502$+$106\\
 213 \hspace{5pt} & 14\ \ 40 & 16\ \ 00 && $ 33.75$ & $ 41.81$ && 1504$+$377\\
 216 \hspace{5pt} & 14\ \ 40 & 16\ \ 00 && $ 62.73$ & $ 90.00$ && 1448$+$762\\
 228 \hspace{5pt} & 16\ \ 00 & 17\ \ 20 && $ 12.84$ & $ 19.47$ && 1717$+$178\\
 233 \hspace{5pt} & 16\ \ 00 & 17\ \ 20 && $ 51.06$ & $ 62.73$ && 1623$+$578\\
 234 \hspace{5pt} & 16\ \ 00 & 17\ \ 20 && $ 62.73$ & $ 90.00$ && 1642$+$690\\
 241 \hspace{5pt} & 17\ \ 20 & 18\ \ 40 && $-19.47$ & $-12.84$ && 1730$-$130\\
 244 \hspace{5pt} & 17\ \ 20 & 18\ \ 40 && $  0.00$ & $  6.38$ && 1725$+$044\\
 253 \hspace{5pt} & 18\ \ 40 & 20\ \ 00 && $-90.00$ & $-62.73$ && 1935$-$692\\
 254 \hspace{5pt} & 18\ \ 40 & 20\ \ 00 && $-62.73$ & $-51.06$ && 1925$-$610\\
 256 \hspace{5pt} & 18\ \ 40 & 20\ \ 00 && $-41.81$ & $-33.75$ && 1954$-$388\\
 257 \hspace{5pt} & 18\ \ 40 & 20\ \ 00 && $-33.75$ & $-26.39$ && 1921$-$293\\
 260 \hspace{5pt} & 18\ \ 40 & 20\ \ 00 && $-12.84$ & $ -6.38$ && 1937$-$101\\
 276 \hspace{5pt} & 20\ \ 00 & 21\ \ 20 && $-26.39$ & $-19.47$ && 2037$-$253\\
 284 \hspace{5pt} & 20\ \ 00 & 21\ \ 20 && $ 26.39$ & $ 33.75$ && 2113$+$293\\
 289 \hspace{5pt} & 21\ \ 20 & 22\ \ 40 && $-90.00$ & $-62.73$ && 2142$-$758\\
 294 \hspace{5pt} & 21\ \ 20 & 22\ \ 40 && $-26.39$ & $-19.47$ && 2210$-$257\\
 297 \hspace{5pt} & 21\ \ 20 & 22\ \ 40 && $ -6.38$ & $  0.00$ && 2216$-$038\\
 306 \hspace{5pt} & 21\ \ 20 & 22\ \ 40 && $ 62.73$ & $ 90.00$ && 2229$+$695\\
 312 \hspace{5pt} & 22\ \ 40 & 24\ \ 00 && $-26.39$ & $-19.47$ && 2331$-$240\\
 315 \hspace{5pt} & 22\ \ 40 & 24\ \ 00 && $ -6.38$ & $  0.00$ && 2335$-$027\\
   \noalign{\smallskip}
   \hline
   \end{tabular}
   \label{Tab:category_B_sectors}
   \end{table}

   Aside from acquiring further VLBI data along the lines described above, the evolution of the frame will also depend at some level on the refinements of the modeling. One area where improvement is foreseen relates to the determination of the solar system acceleration vector. As the data accumulate, the time span will also extend, leading to an estimate of the vector amplitude with ever increased accuracy. Future data should also reveal whether that vector is directed entirely toward the Galactic center or is subject to some offset. Additionally, the Gaia mission is expected to deliver estimates of those parameters in future data releases, which will further contribute to improving such determination. Another question to be tackled relates to the source structure, which manifests itself through systematics in the VLBI delay measurements \citep{Xu2019} and instabilities in the individual source positions \citep{Gattano2018}. In the future, those effects will become more prominent as the precision of the measurements continues to increase, making it necessary to consider them in the modeling to further increase the quality of the frame. While the corresponding theoretical framework has been put forward long ago \citep{Charlot1990}, the practical implementation has not been straightforward due to the requirement to have multi-epoch images available for all the sources. While producing such series of images is an enormous task, the advent of several image databases \citep{Collioud2019, Hunt2019} now offers interesting prospects for the coming years in this area.

   Finally, another route to investigate that might possibly help to enhance the frame lies in exploring alternate analysis configurations. While the three catalogs that form ICRF3 have been derived independently, the corresponding data sets could be processed together. The observing system is largely similar at the three frequency bands and such a joint analysis should benefit from the strengths of each data set, most likely resulting in improvements of the overall frame. For example, systematics in the X/Ka~band catalog (see Sect.~\ref{Sec:intercomparison}) might end up reduced after incorporation of the S/X band and K~band data sets in the analysis. There are also reasons to expect an even closer alignment of the three catalogs with this scheme. It should be emphasized, though, that the source positions should be kept as separate parameters at the three frequencies, as otherwise any actual physical offsets, such as those suspected here for a fraction of the sources (see Sect.~\ref{Sec:comp_3-frequency-positions}), would affect the resulting source position estimates. It is worth noting that such a multi-frequency analysis has been recently tried, though not through using a unique software package but instead through a combination of normal equations derived at each frequency band, demonstrating that there would be some value in investigating further such a combination for future realizations of the ICRF \citep{Karbon2019}. Going further, one may even consider incorporating data from other space geodetic techniques in the combination. Those techniques include satellite and lunar laser ranging, global navigation satellite systems, and the ``doppler orbitography and radiopositioning integrated by satellite'' system, which, together with VLBI, contribute to the realization of the ITRF. While the impact on the celestial frame may be limited, such a multi-technique combination would have the advantage to deliver unified terrestrial and celestial frames \citep{Seitz2014}.

\section{Conclusion}
\label{Sec:conclusion}
   A new realization of the ICRF, denoted as ICRF3, has been produced based on VLBI data acquired over nearly 40~years in three frequency bands (S/X, K, and X/Ka band). This new realization is the first multi-frequency celestial reference frame ever generated. Another new feature incorporated in ICRF3 is the modeling of the galactocentric acceleration of the solar system, which considers an acceleration vector pointing toward the Galactic center with an amplitude of 5.8~$\mu$as/yr. The latter was derived by an adjustment directly to the S/X~band data. Source positions are reported for epoch 2015.0 and the above amplitude value should be used to propagate those positions to other epochs if necessary. The new frame includes positions for 4536~sources at S/X~band, 824~sources at K~band, and 678~sources at X/Ka~band, for a total number of 4588 sources, where 600 of these sources have independent positions available at the three frequencies.

   The noise floor in the individual source coordinates is at the level of 30~$\mu$as. At S/X~band, the median uncertainty is 127~$\mu$as in right ascension and 218~$\mu$as in declination, more than a factor of three improvement over the previous realization, ICRF2. This improvement reflects the efforts that have been accomplished during the past decade to re-observe all VCS-type sources (as identified in ICRF2) with the VLBA, while taking the opportunity to augment the number of observed sources by one-third at the same time. Most notably, the S/X band catalog includes also a pool of 500~sources, observed as part of IVS programs, which have highly accurate positions, with uncertainties in the range of 30--60~$\mu$as. The positional accuracy at K~band and X/Ka~band approaches that at S/X~band but remains a factor of 1.5--2 lower. A subset of 303 sources among the most observed ones at S/X~band has been identified for defining the frame based on their sky distribution, position stability, and the amount of structure.

   ICRF3 is aligned onto~ICRF2 to the accuracy of the latter. Comparing the S/X~band frame with the recently released Gaia-CRF2 frame in the optical domain, the two frames show no relative deformations above 30~$\mu$as. On the other hand, ICRF2 is found to have a significant dipolar deformation, approaching 100~$\mu$as, with respect to ICRF3, an effect that in large part results from not considering Galactic acceleration in the modeling for ICRF2. The K~band catalog shows no deformations above 50~$\mu$as with respect to the S/X~band frame, unlike the X/Ka~band catalog which suffers from dipolar and quadrupolar systematics. The latter is likely due to the limited geometry of the X/Ka~band data set. Comparisons of the individual source positions at the three frequency bands reveal significant offsets between frequencies for about 5\% of the sources, a percentage that increases to 22\% when comparing the ICRF3 positions to the Gaia-CRF2 optical positions. There are indications that those positional offsets may be the manifestation of extended source structures.

   ICRF3~was adopted by the IAU during its 30th General Assembly held in Vienna in August 2018 and is now the fundamental celestial reference frame to be used for all applications. Looking into the future, this first multi-frequency ICRF realization may be regarded as a first stage toward a fully integrated multi-band frame incorporating also the optical Gaia data.

\begin{acknowledgements}
      VLBI is a collaborative and cooperative endeavor involving many individuals and institutions around the world. This new celestial reference frame is the result of their efforts over the past 40 years. We wish to recognize and thank the designers and fabricators of VLBI instrumentation, from masers to receivers, data acquisition terminals and correlators, the schedule makers and session coordinators, the generations of model builders, software developers and analysts, and the national funding agencies who supported this work all along.

      All components of the International VLBI Service for Geodesy and Astrometry (IVS), which has organized this cooperation in the smoothest way over the past 20~years, deserve specific and deep acknowledgements, as do the Long Baseline Observatory (LBO) in the~US, formerly the Very Long Baseline Array (VLBA), for running survey observations that allowed for a considerable increase in the number of sources in the ICRF. The present work would not have been possible without the bulk of data acquired by these two VLBI arrays over the years. The US~Naval Observatory through a specific agreement with the LBO made available a large amount of VLBA observing time in 2017 and 2018 that allowed us to significantly strengthen the S/X~band and K~band frames, and is warmly thanked. The AuScope VLBI network, funded under the National Collaborative Research Infrastructure Strategy, an Australian Commonwealth Government Program, has been essential to increase VLBI observing in the Southern Hemisphere. Some of the AuScope sessions were supported by the Parkes radio telescope, a part of the Australia Telescope National Facility which is funded by the Australian Government for operation as a National Facility managed by CSIRO, allowing for observations of weaker sources. The group is also very much grateful to the Deep Space Network (DSN) for providing VLBI observing time on their three sites and making the X/Ka band frame a reality. Specific thanks are equally addressed to the European Space Agency (ESA) for occasionally making available the Malarg\"ue station in Argentina to observe jointly with the~DSN, which was essential to extend and strengthen the X/Ka~band frame in the south. The Hartebeesthoek Radio Astronomy Observatory and the University of Tasmania arranged unique K~band sessions with their telescopes and were instrumental to complete the sky coverage of the K~band frame in the south. A few sessions dedicated to densify the ICRF at S/X~band are the results of observations carried out by the European VLBI Network (EVN), a joint facility of independent European, African, Asian, and North American radio astronomy institutes, which is also very much thanked. The corresponding data were acquired under the EVN project codes EC013 and EC017.

      We are indebted to Fran\c cois Mignard and the Gaia Science Team for providing us with a comparison of the ICRF3 prototype catalog with the Gaia Data Release~2 celestial reference frame (Gaia-CRF2) prior to public release. We are also grateful to Arnaud Collioud for producing the world map in Fig.~{\ref{Fig:ICRF3_network_map}} that pictures the geographical location of the radio telescopes involved in the observations used for ICRF3. Portions of this research were carried out at the Jet Propulsion Laboratory, California Institute of Technology, under a contract (80NM0018D0004) with the National Aeronautics and Space Administration (NASA). The French contribution to this work was supported by the Programme National GRAM of CNRS/INSU with INP and IN2P3 co-funded by CNES. PC and GB wish also to acknowledge support from the ``Observatoire Aquitain des Sciences de l'Univers''. DG and DSM acknowledge support from NASA contracts NNG12HP00C and NNG17HS00C. ZM was partially supported by the Russian Government Program of Competitive Growth of Kazan Federal University.

      This research has made use of the Bordeaux VLBI Image Database (\url{http://bvid.astrophy.u-bordeaux.fr}) and Radio Reference Frame Image Database (\url{https://www.usno.navy.mil/USNO/astrometry/vlbi-products/rrfid}). It has also made use of Earth Orientation Parameters series from the IVS and IERS. The Gaia-CRF2 frame used in the comparisons results from data from the ESA mission Gaia (\url{https://www.cosmos.esa.int/gaia}), processed by the Gaia Data Processing and Analysis Consortium (DPAC, \url{https://www.cosmos.esa.int/web/gaia/dpac/consortium}). Funding for the DPAC has been provided by national institutions, in particular the institutions participating in the Gaia Multilateral Agreement.
\end{acknowledgements}

\bibliographystyle{aa} 
\bibliography{icrf3} 
\end{document}